\documentclass[twocolumn,twocolappendix]{aastex631}

\usepackage{amsmath}

\usepackage{orcidlink}

\newcommand{\AuthorORCID}[2]{%
  \href{https://orcid.org/#2}{\mbox{#1~{\Large\orcidlink{#2}}\kern-0.5em}}%
}

\begin{document}

\title{The Optical to X-ray Luminosity and Spectrum of Supernova Wind Breakouts}

\shorttitle{Spectrum of Supernova Wind Breakouts}
\shortauthors{Wasserman et al.}

\author{\AuthorORCID{Tal Wasserman}{0009-0005-7414-3965}}
\affiliation{Dept. of Particle Phys. \& Astrophys., Weizmann Institute of Science, Rehovot 76100, Israel}

\author{\AuthorORCID{Nir Sapir}{0000-0003-0655-3688}}
\affiliation{Dept. of Particle Phys. \& Astrophys., Weizmann Institute of Science, Rehovot 76100, Israel}
\affiliation{Plasma Physics Department, Soreq Nuclear Research Center, Yavne 81800, Israel}

\author{Peter Szabo} % (no ORCID provided)
\affiliation{Dept. of Particle Phys. \& Astrophys., Weizmann Institute of Science, Rehovot 76100, Israel}

\author{\AuthorORCID{Eli Waxman}{0000-0002-9038-5877}}
\affiliation{Dept. of Particle Phys. \& Astrophys., Weizmann Institute of Science, Rehovot 76100, Israel}

\begin{abstract}
Observations indicate that optically thick circum-stellar medium (CSM) at radii of $10^{14}-10^{15}~$cm around Type II core-collapse supernovae (SN) progenitors is common (and may be present in other types of massive star explosions). The breakout of the SN radiation-mediated shock (RMS) through such CSM leads to the formation of a collisionless shock (CLS). We analyze the evolution of the shock structure and associated radiation field during and after the RMS-CLS transition for non-relativistic shock breakout velocity ($v_{\rm bo}=10^9v_9~{\rm cm/s}<0.1c$) through a hydrogen-rich CSM ``wind" density profile, $\rho\propto r^{-2}$, with breakout radius $R_{{\rm bo}}=10^{14}R_{14}~$cm much larger than the progenitor radius. An analytic description of the key properties of the emitted optical to X-ray radiation is provided, supported by numeric radiation-hydrodynamics calculations self-consistently describing the time-dependent spatial distribution of the plasma and radiation, governed by Bremsstrahlung emission/absorption and inelastic Compton scattering. The characteristic energy of the photons carrying most of the luminosity, $\approx10^{43}R_{14}v_9^2~$erg/s, shifts from UV to X-ray, reaching 1~keV as the shock reaches $\approx3R_{\rm bo}$, in $\approx3R_{14}/v_9~$d. The X-ray signal is not suppressed by propagation through the upstream wind, and its absence may suggest that the dense CSM does not extend much beyond $R_{\rm bo}$. Our results provide the basis for a quantitative calculation of the high energy $\gamma$-ray and neutrino emission that is expected from particles accelerated at the CLS, and will allow using data from upcoming surveys that will systematically detect large numbers of young SNe, particularly ULTRASAT, to infer the pre-explosion mass loss history of the SN progenitor population.
\end{abstract}

\section{Introduction} \label{sec:intro}

The earliest emission of radiation from a SN explosion is associated with a ``shock breakout". As the RMS \citep{weaver_structure_1976} that drives the ejection of the envelope propagates outwards, the optical depth of the plasma lying ahead of it decreases. When the optical depth drops to $\approx c/v$, where $v$ is the shock velocity, radiation escapes, and the shock dissolves. In the absence of an optically thick CSM, the breakout occurs as the shock reaches the edge of the star \citep{lasher_simple_1975,klein_x-ray_1978}, producing a bright X-ray/UV flash with a typical duration of hours, followed by a UV/optical emission from the expanding cooling shocked-envelope on a days timescale. Existing theoretical analyses \citep[e.g.][]{rabinak_early_2011,nakar_early_2010,piro_shock_2010,katz_non-relativistic_2012,sapir_non-relativistic_2013,sapir_uvoptical_2017,piro_shock_2021,morag_shock_2023,morag_shock_2024} provide a good understanding, and tools for accurate description, of the radiation emitted during breakout and envelope cooling \citep[see, e.g.][for review]{waxman_shock_2017}. Observations of the envelope (shock-)cooling phase, which is well understood theoretically, can constrain the radius and composition of the progenitor star, as well as the explosion energy \citep[e.g.][]{rabinak_early_2011,nakar_early_2010,irani_early_2024}. Recent work applying the most advanced models \citep{morag_shock_2023,morag_shock_2024} to a large set of Type II SNe with early optical-UV observations \citep{irani_early_2024} shows that CSM-free models account very well for the observations of about $50\%$ of the population; the progenitor radii derived by the models are consistent with the radii distribution measured locally \citep{irani_early_2024}. The light curves of the other 50\% are inconsistent with shock cooling and indicate the presence of an optically thick CSM shell.

The prevalence of a compact distribution of dense CSM around many, and perhaps most core-collapse SN progenitors is supported by growing observational evidence (see \S~\ref{subsec:CSM}). In the presence of an optically thick CSM with $\tau>c/v$,  shock breakout occurs within the CSM, potentially extending its duration to days. CSM breakouts are accompanied by the conversion of the RMS to a collisionless shock (CLS) \citep[][\S~\ref{subsec:Wind+CLS} below]{katz_x-rays_2011}. The RMS-CLS transition implies a dramatic increase in the temperature of the shock-heated plasma, from 10's of eV to 10's of keV \citep{katz_x-rays_2011}, shifting the peak of the spectrum from the UV to the X-ray band on the breakout timescale $R_{\rm bo}/v$. In addition, the CLS is expected to produce a ``tail" of high energy particles accelerated to energies well above the $\sim10$~keV temperature that characterizes the bulk of the shock-heated electrons and protons. While the high energy particles are not expected to modify significantly the shock dynamics and the optical-X-ray emission \citep[][\S~\ref{subsec:This_Work} below]{katz_x-rays_2011}, they are expected to produce high energy ($h\nu\gg m_ec^2$) photons and  (multi-TeV) neutrinos \citep{katz_x-rays_2011,murase_new_2011,zirakashvili_type_2016,li_pev_2019,sarmah_high_2022,pitik_optically_2023,kimura_high-energy_2025,cosentino_high-energy_2025}.

CSM breakouts are interesting both because their observations provide information on the progenitors and their pre-explosion evolution as shown recently for SN 2023ixf (see \S~\ref{subsec:CSM}), and also because they may be the sources of several classes of powerful transients: Non-relativistic (NR)-CSM breakouts are considered as possible explanations of (at least part of) the super-luminous SN class \citep[e.g.][]{ofek_supernova_2010,chevalier_shock_2011,ginzburg_superluminous_2012,moriya_analytic_2013,rest_fast-evolving_2018}, of “double peak” SNe \citep[e.g.][]{piro_using_2015,nicholl_lsq14bdq_2015}, and of the early part of the emission of SNe of type IIn \citep[e.g.][]{ofek_interaction-powered_2014,drout_rapidly_2014,gezari_galex_2015,ibik_ps1-11aop_2025}; Fast CSM breakouts have been proposed as the sources of X-ray flashes and low-luminosity $\gamma$-ray bursts associated with Type Ib/c SNe  \citep[e.g.][]{tan_trans-relativistic_2001,campana_association_2006,waxman_grb_2007,soderberg_extremely_2008, balberg_supernova_2011}; Rare SNe with prominent X-ray emission lasting from days to years have also been linked to dense CSM \citep[e.g.][]{fox_x-ray_2000,chandra_radio_2012,ofek_x-ray_2013,levan_superluminous_2013,ofek_sn_2014,dwarkadas_x-ray_2016,chandra_circumstellar_2018,pellegrino_x-ray_2024}. 

Furthermore, if compact dense CSM is indeed common, then the resulting neutrino flux may account for a significant fraction of the observed $>10$~TeV neutrino background \citep{sarmah_high_2022,waxman_shock_2025}. Note that the neutrino production mechanism we discuss is different than the ``choked-jet" mechanism \citep{meszaros_tev_2001,murase_tev-pev_2013,denton_bright_2018,guetta_low-_2023,zegarelli_towards_2024}.
The neutrinos are expected to be produced over a few days timescale, preceding or coincident with the bright UV (possibly followed by bright X-ray) breakout emission\footnote{The emission of neutrinos by SN CLS driven into the low-density winds surrounding massive stars at large radii was discussed by several authors \citep[e.g.][]{murase_new_2011,petropoulou_point-source_2017,sarmah_high_2022,murase_interacting_2024}. The neutrino luminosity produced by shocks driven into typical wind/ISM extends over hundreds of days and is too low to account for the neutrino background \citep[e.g.][]{waxman_shock_2025}.}. SNe producing $>1$ neutrino-induced muon events are expected at a rate of $\sim0.1/~$yr in the $\simeq1$~km$^2$ IceCube detector\footnote{An association of a single muon-induced 10~TeV neutrino with a nearby SN within a few days of the explosion would be significant at approximately 99.9\% confidence level \citep{waxman_shock_2025}.} \citep{icecube_collaboration_first_2006}, and in the KM3Net \citep{adrian-martinez_letter_2016} and Baikal-GVD \citep{malyshkin_baikal-gvd_2023} detectors under construction, and at a rate of $\sim1/~$yr in larger detectors under planning and construction (IceCube Gen. 2, \cite{grant_neutrino_2019}; P-One, \cite{agostini_pacific_2020}; TRIDENT, \cite{ye_multi-cubic-kilometre_2023}).

As the capabilities of rapid transient searches improve, both from the ground (e.g. GOTO, \cite{steeghs_gravitational-wave_2022}, and LAST, \cite{ofek_large_2023}) and from space with the expected launch of the wide-field UV space telescope ULTRASAT \citep{shvartzvald_ultrasat_2024}, a systematic detection of many SNe of all types at early, $<1$\,d time will be possible. A quantitative theory describing the evolution of the electromagnetic spectrum during a CSM breakout, as the RMS transforms to CLS, is not yet available (see below and \S~\ref{subsec:Previous_Works}) and is needed to enable using early UV measurements, early spectra, and possibly early X-ray measurements, to determine the pre-explosion mass loss history of the SN progenitor population and provide constraints on the yet unknown mass ejection mechanism (see \S~\ref{subsec:CSM}), as well as to derive quantitative estimates of the high energy $\gamma$-ray and neutrino luminosity and spectrum produced by CSM breakouts \citep{waxman_shock_2025}.

The calculations of NR, $v/c<0.1$ CSM breakout spectra face several challenges.
\\ \noindent[i] Steady-state shock structure solutions are not applicable due to the non-steady nature of the shock structure at breakout, as it changes from an RMS to a CLS in a complicated manner on a timescale comparable to the dynamical time, $R_{\rm bo}/v$.
\\ \noindent[ii] During and following the breakout, the radiation spectrum is far from thermal.
\\ \noindent[iii] Inelastic Compton scattering, which is challenging to include (and is not included in all time-dependent) radiation-hydro codes, plays a crucial role. For example, in determining the post-CLS electron temperature profile, and in shaping the optical-X-ray photon spectrum through Comptonization. It is important to note that at breakout, the photon diffusion time across the shocked CSM shell is comparable to the dynamical time. Hence, the escaping photon spectrum is determined by the entire temperature and density profiles of the shocked plasma.
\\ \noindent[iv] The CLS heats electrons over a length scale that is many orders of magnitude smaller than $R_{\rm bo}$, and the electrons cool in the post-shock region on a length scale which may be orders of magnitude smaller than $R_{\rm bo}$. This implies that a very high resolution (prohibitively large without adaptive mesh refinements) is required in order to obtain the correct electron temperature profile.
\\ \noindent[v] The interaction of the escaping UV--X-ray radiation with the initially cold and neutral upstream CSM is challenging to calculate, as it requires following the evolution of ionization and heating of the CSM.

In this paper, we construct a quantitative description, addressing all the above challenges, of the evolution of the shock structure and of the electromagnetic, optical--X-ray low-resolution spectrum (``spectral energy distribution”) during and following the breakout. We consider a spherically symmetric breakout from a hydrogen-rich CSM with a ``wind" density profile, $\rho\propto r^{-2}$, a shock driven either by a constant velocity piston or by an expanding shocked stellar envelope with an initial polytropic structure, and breakout radius much larger than the initial stellar radius. We consider the parameter range \{$10^{12}\mathrm{cm}<R_\mathrm{bo}<10^{15}\mathrm{cm}$,~ $0.2<v_9<2$\}, and track the evolution up to the time at which the shock reaches the wind Thomson photosphere.

We choose to analyze this simplest form of the problem, which is mostly determined by two parameters, the breakout radius $R_{\rm bo}$, and velocity $v_{\rm bo}$, for two reasons. First, it enables a clear understanding of the governing physical processes, which, in turn, allows the construction of an analytic description of the key quantities that describe the time-dependent shock profile structure and radiation field. These results may be used for constructing semi-analytic solutions for a wide range of density profiles and non-spherically symmetric breakouts. Second, the solutions obtained for this ``simple" setup are expected to describe well the key features of observed breakouts for cases where deviations from spherical symmetry are not very large, the length scale for density variation at $R_{\rm bo}$ is not much smaller than $R_{\rm bo}$, and the mass of the shocked CSM is small compared to the envelope mass. This is due to the following reasons.
\\ \noindent[i] When the characteristic length scale for density variation at $R_{\rm bo}$ is comparable to $R_{\rm bo}$, the width of the RMS is comparable to $R_{\rm bo}$ and a CLS is formed due to the radiation flux reduction with radius (see \S~\ref{subsec:Wind+CLS}). A qualitatively different behavior is expected in cases where the CSM density is sharply 'truncated', i.e. decreasing significantly at $R_{\rm bo}$ over a length scale $\ll R_{\rm bo}$, in which case the RMS width is $\ll R_{\rm bo}$, a CLS does not necessarily form, and the breakout resembles a breakout from a stellar edge with a large radius with duration $\ll R_{\rm bo}/v$. Such ``edge breakouts" \citep[see, e.g.][]{chevalier_shock_2011,khatami_landscape_2024} may be obtained in cases where the dense CSM has an ``edge", i.e. it is sharply truncated at some radius, and its optical depth is $\gg c/v$, implying an ejected mass $M_\mathrm{bo}\gg 0.01R_{14}^2v_9^{-1} M_\odot$ out to $R_{\rm bo}$. As discussed in \S~\ref{subsec:CSM}, while observations do not provide stringent constraints on the density structure of the dense CSM and the mechanisms leading to its ejection are not clear, the observed long luminosity rise times and the inferred mass ejection rates suggest that ``wind breakouts" (rather than ``edge breakouts") are common. Finally, we note that the results of numeric simulations of mass ejection following energy deposition in the envelope of a giant star yield density distributions, which are not very different from $r^{-2}$ \citep[e.g.][]{kuriyama_radiation_2020,tsuna_analytical_2021,tsang_et_al_3d_2022}.
\\ \noindent[ii] The shock-accelerated outer part of the envelope acts as a ``piston" driving a shock through the CSM. Since the dependence of the velocity of the outer envelope shells on the shell mass is weak, e.g. $v\propto m^{-0.1}$ for a convective $n=3/2$ polytrope (where $m$ is the mass measured from the stellar surface inwards), the velocity of the ``piston" driving the shock into the CSM does not vary significantly during breakout as long as the accumulated mass of the shocked CSM up to $R_\mathrm{bo}$ ($\sim0.01R_{14}^2v_9^{-1} M_\odot$) is small compared to the envelope mass. A more detailed explanation is given in \S~\ref{sec:ejecta_results}.

In this work, we consider the limit of a breakout radius much larger than the stellar radius and much smaller than the radius out to which the dense CSM extends, $R_\star\ll R_\mathrm{bo}\ll R_\mathrm{dCSM}$. The ratio $R_\mathrm{bo}/R_\star$ below which the finite stellar radius significantly affects the breakout is derived in Appendix \ref{sec:luminosity_contributions}, and the effect of a finite $R_\mathrm{dCSM}/R_\mathrm{bo}$ is discussed qualitatively in \S~\ref{sec:discussion}. Observations suggest that while the separation of these radii may be significant, it is not necessarily very large, see Figure \ref{fig:density}. This is the case, e.g., for SN\,2023ixf (see \S~\ref{subsec:CSM} below), where a coincidence is apparent - the optical depth of the compact, order $\sim0.1 M_\odot$ CSM is comparable to $c/v$ at $R_\mathrm{dCSM}\sim10^{14.5}$~cm. As discussed in \S~\ref{subsec:CSM}, the ejection of a fraction of a solar mass to $\sim10^{14.5}$~cm, and hence this type of coincidence, may be common.

\begin{figure}
\label{fig:density}
    \centering
    \includegraphics[height=6.5cm]{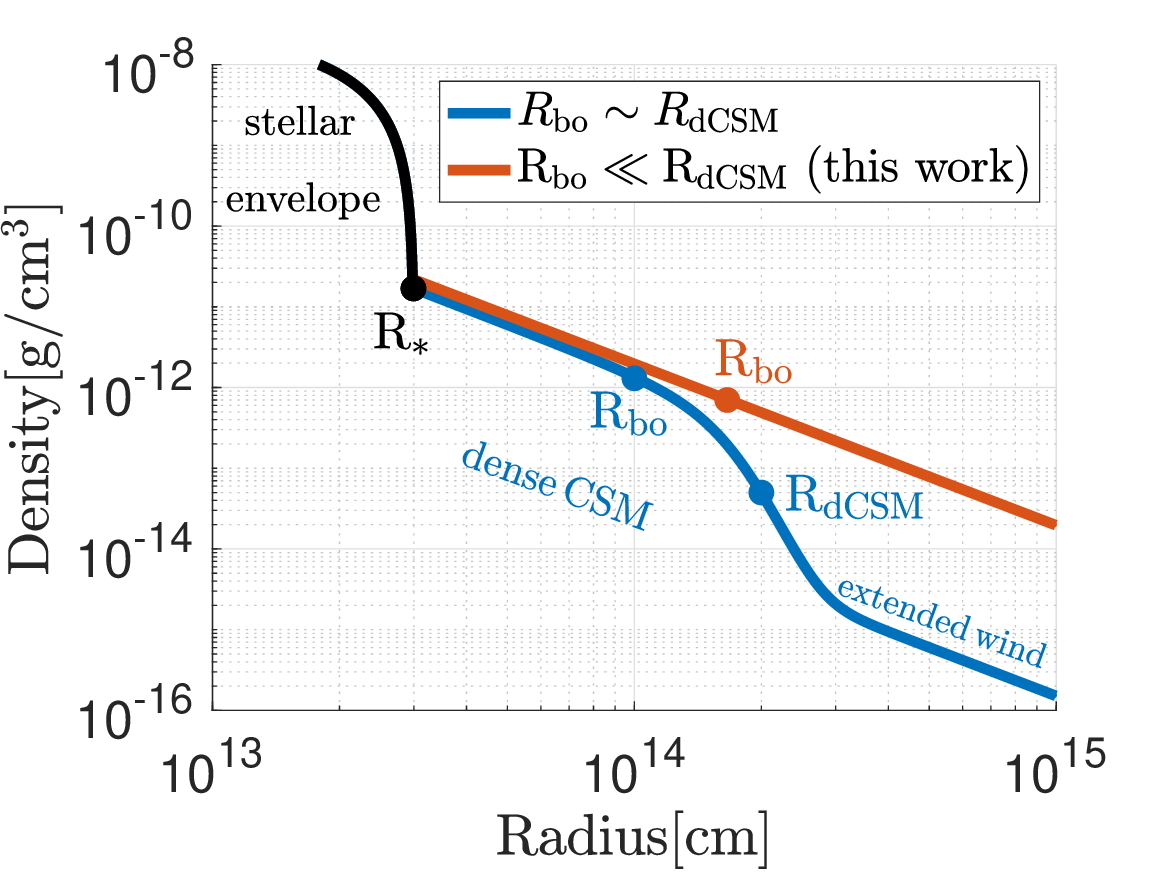}
    \caption{A CSM density profile with a finite $R_\mathrm{dCSM}$, similar to the one inferred from observations of the nearby SN 2023ixf (see \S~\ref{subsec:CSM} below), and the wind density profile considered in this paper. A qualitative discussion of the impact of a finite $R_\mathrm{dCSM}$ is given in \S~\ref{sec:discussion}.}
\end{figure}

Radiation-hydrodynamics solutions are obtained using an adapted version (see \S~\ref{subsec:This_Work}) of our 1D NR radiation-hydrodynamics code \citep{sapir_numeric_2014,morag_shock_2023}, solving radiation transport in the multi-group diffusion approximation (which is valid for the $v/c\ll 1$ shocks considered; see \S~\ref{subsec:This_Work}). This code successfully reproduced a wide range of analytic results, including those describing planar shock breakout \citep{sapir_numeric_2014}, and was successfully used to describe the early shock-cooling multi-band light curves of a large number of Type II SNe \citep{irani_early_2024}. Since the high energy particles tail accelerated by the CLS is not expected to modify significantly the shock dynamics and the optical to X-ray radiation field  \citep[][\S~\ref{subsec:This_Work} below]{katz_x-rays_2011}, we solve the dynamics of the flow and the optical--X-ray radiation field neglecting the presence of the high energy particles.

The rest of the paper is organized as follows. The observational evidence for compact, optically thick CSM around core-collapse SN progenitors is discussed in \S~\ref{subsec:CSM}, including a concise reference to theoretical work addressing the possible mechanisms leading to such CSM ejection. In \S~\ref{subsec:Wind+CLS}, we repeat the derivation of \cite{katz_x-rays_2011}, showing that a CLS is formed during wind breakout, and introduce notations that are used throughout the paper. Earlier theoretical work on wind breakouts is discussed in \S~\ref{subsec:Previous_Works}. The approximations used in the calculations presented in this paper are described and justified in \S~\ref{subsec:This_Work}. The radiation-hydro equations and the numeric code are described in \S~\ref{sec:equations}. Numeric results and analytic approximate results are given in \S~\ref{sec:constant_results} for the case of a constant velocity piston-driven shock and in \S~\ref{sec:ejecta_results} for a polytropic expanding envelope driven shock. In \S~\ref{sec:discussion}, we summarize our results, provide a qualitative discussion of the impact of a finite dense CSM radius, and explain how the time-dependent spectral luminosity tables that are given in the supplementary material may be used for obtaining model predictions.

\subsection{Pre-Explosion Mass Loss of Core-Collapse SN Progenitors}
\label{subsec:CSM}

Systematic analyses \citep{ofek_precursors_2014,strotjohann_bright_2021} of precursor emission (preceding the SN explosion) in a large sample of SNe of type IIn, i.e., showing narrow line spectra indicative of CSM interaction \citep{schlegel_new_1990, smith_mass_2014,gal-yam_observational_2017}, find that significant precursor emission, with optical photons' energy exceeding $10^{47}~$erg, is common during the $\sim90$ days preceding the SN explosion. Various suggestions have been made for the precursor emission mechanism, which is not yet well understood, including: pair instability pulsations \citep[e.g.][]{rakavy_instabilities_1967,woosley_pulsational_2007}, binary interaction \citep[e.g.][]{chevalier_common_2012,soker_explaining_2013}, radiation-driven instability \citep[e.g.][]{suarez-madrigal_local_2013}, unstable late-stage nuclear burning \citep[e.g.][]{smith_preparing_2014,woosley_remarkable_2015}, dissipation of internal gravity waves driven by core burning \citep[e.g.][]{shiode_setting_2013,fuller_pre-supernova_2017,fuller_pre-supernova_2018}, and core magnetic activity \citep[][]{cohen_pre-supernova_2024}.

Although not well understood, the precursor probably reflects the deposition of energy (exceeding the observed optical photons' energy) at the stellar envelope. Since $10^{47}$\,erg corresponds to the binding energy of $\approx1M_\odot$ at the envelope of a RSG, $GM_\star M/R_\star$ with $M_\star=10M_\odot,\,M=1\,M_\odot,R_\star=10^{13.5}\,{\rm cm}$, and since the energy deposition duration is shorter than or comparable to the dynamical time of the envelope, $1/\sqrt{G\rho}\approx100$\,d, the precursor is expected to be associated with the ejection of a significant fraction of a solar mass \citep[consistent with the results of recent analytic and numeric studies,][]{kuriyama_radiation_2020,linial_partial_2021,matzner_wave-driven_2021,tsuna_analytical_2021,ko_eruption_2022,tsang_et_al_3d_2022,corso_mass_2024}. The ejected mass is expected to expand at a velocity comparable to the escape velocity, $\sqrt{2GM_\star/R_\star}\approx100\,{\rm km{\rm/s}}$, implying that a shell ejected at time $t_{\rm pr}$ preceding the explosion expands to a radius of $\approx10^{14.5}(t_{\rm pr}/1{\rm yr})$~cm by the explosion time. Indeed, \citet{strotjohann_bright_2021} infer a dense 0.1--1\,$M_\odot$ CSM out to $10^{14}-10^{15}$\,cm radii for most of their sample. Interestingly, \citet{jacobson-galan_final_2022} report the detection of a precursor prior to a spectroscopically-regular Type II SN.

Independent evidence for the prevalent presence of optically thick CSM shells around spectroscopically regular Type II SN progenitors is obtained from early, 1\,day timescale observations of optical-UV SN light curves. \citet{irani_early_2024} carried out the first systematic analysis of early ($\sim1$\,d) optical-UV light curves of a large sample of Type II SNe \citep[see][for a discussion of analyses of individual SNe]{morag_shock_2023}. They find that while the early light curves of $\approx50\%$ of Type II SN are consistent with the emission from the expanding shocked stellar envelope, so-called ``shock cooling" emission \citep[see][for review]{waxman_shock_2017}, the light curves of the other 50\% are inconsistent with shock cooling and indicate the presence of an optically thick CSM shell. The extended, days-long rise of the luminosity to high values of $10^{43}-10^{44}$\,erg/s and the high color temperature, with most emission in the UV, are consistent with shock breakout from such a shell \citep[][]{ofek_supernova_2010,waxman_shock_2017}: The shock breakout radius is inferred from the duration of the luminosity rise, which is given by the SN driven shock crossing time, $R_{\rm bo}=vt\approx10^{14.5}(t/3\,{\rm d})(v/10^9{\rm cm{\rm~s^{-1}}})$; The shell mass at $R_{\rm bo}$ is determined by the requirement that the optical depth at $R_{\rm bo}$ equals $c/v$, $M_\mathrm{bo}=(c/v)4\pi R_{\rm 14.5}^2/\kappa_T\approx0.05M_\odot$ (where $\kappa_T$ is the Thomson opacity); The observed breakout energy, $\approx10^{49}$\,erg is consistent with $M_\mathrm{bo}v^2/2\approx5\times10^{49}$\,erg; The high, $\sim10$\,eV temperature is consistent with a blackbody radiation carrying the breakout energy at $R_\mathrm{bo}$ \citep[see \S~\ref{subsec:Wind+CLS} below; We note, however, that this temperature is higher than the $\sim3$~eV temperatures inferred from the optical-UV bands in][]{irani_early_2024}. This work supports earlier suggestions from analyses of SN Type II rise times \citep[e.g.][]{forster_delay_2018,morozova_measuring_2018} that many of these explosions occur within a compact CSM distribution.

For a given CSM shell mass $M_{\rm bo}$ within $10^{14.5}$\,cm, the corresponding wind mass loss is given by $\dot{M}/(v_{\rm w}/100\,{\rm km{\rm~s^{-1}}})=M_{\rm bo} {\rm /yr}$, where $v_{\rm w}$ is the wind velocity. Note that the mass loss rates corresponding to $M_{\rm bo}=0.01-0.1\,M_\odot$, $\dot{M}/(v_{\rm w}/100{\rm km{\rm~s^{-1}}})=0.01-0.1M_\odot$/yr are orders of magnitude higher than the mass loss rates typically observed for RSGs, $\dot{M}/(v_{\rm w}/100\,{\rm km{\rm~s^{-1}}})\le10^{-4}M_\odot/$yr \citep{de_jager_mass_1988,marshall_asymptotic_2004,van_loon_empirical_2005}.

Additional independent evidence for the existence of an optically thick $\sim10^{14.5}$\,cm CSM shell is provided by ``flash spectroscopy" \citep{gal-yam_wolfrayet-like_2014} which revealed the presence of strong narrow spectral lines from high ionization species that disappear a few days following the SN explosion \citep{gal-yam_wolfrayet-like_2014,khazov_flash_2016, yaron_confined_2017, zhang_sn_2020,bruch_large_2021,terreran_early_2022,jacobson-galan_final_2024}. These lines are most naturally explained as due to the ionization and excitation by the breakout UV emission of a compact CSM shell that is swept up by the SN shock on a few days timescale (hence extending to $\sim10^{14.5}$\,cm) with optical depth corresponding to mass loss rates of  $\dot{M}/(v_{\rm w}/100{\rm km{\rm\, s^{-1}}})=10^{-3}-10^{-2}M_\odot/$yr \citep{yaron_confined_2017,dessart_explosion_2017,boian_diversity_2019}. \cite{bruch_prevalence_2023} find that $>50\%$ of SNe type II likely show such CSM features.

Finally, recent observations of the nearby 6.4\,Mpc distance SN\,2023ixf enabled an unprecedentedly detailed study of the CSM structure around a super-giant SN progenitor. Early multi-wavelength (optical--UV--X-ray) and spectra measurements, as well as late extensive and broad-band X-ray and radio monitoring, were successfully carried out thanks to the early detection and relatively short distance, providing stringent constraints on the CSM at the progenitor's vicinity \citep[e.g.][]{bostroem_early_2023,jacobson-galan_sn_2023,grefenstette_early_2023,hiramatsu_discovery_2023,chandra_chandras_2024,zimmerman_complex_2024,a_j_dinosaur_2025}. The observations are consistent with a shock breakout through a dense CSM shell, with $\dot{M}/(v_{\rm w}/100{\rm km{\rm\, s^{-1}}})\approx0.03M_\odot/$yr extending to $\approx2\times10^{14}$~cm, surrounded by a much lower density extended wind, $\dot{M}/(v_{\rm w}/100{\rm km{\rm\, s^{-1}}})\approx10^{-4}M_\odot/$yr, at larger radii (see Figure \ref{fig:density}).

The discussion of this sub-section focused on the most common SN progenitors, the hydrogen-rich super-giant progenitors of Type II SNe, for which observations allow one to draw initial constraints on the population as a whole. It should be noted that recent evidence shows that other types of massive star explosions also occur within compact CSM, including some Type Ic events lacking hydrogen and helium \citep{irani_sn_2024}, as well as rare populations of events where the CSM is dominated by helium \citep[Type Ibn,][]{pastorello_giant_2007,pastorello_massive_2008, karamehmetoglu_luminous_2021,hosseinzadeh_type_2017}, or carbon and oxygen \citep[Type Icn,][]{gal-yam_wcwo_2022,perley_type_2022,pellegrino_diverse_2022}.

\subsection{Wind Breakout \& CLS Formation}
\label{subsec:Wind+CLS}

For a wind density profile
\begin{equation}
    \label{eq:rho-r}
    \rho(r)=\frac{\dot{M}}{4\pi r^2 v_{\rm w}},
\end{equation}
the breakout radius, where the optical depth is $c/v_\mathrm{bo}$, $v_\mathrm{bo}$ being the shock velocity at the breakout radius, is
\begin{equation}
    \label{eq:Rbr}
    R_{\rm bo}=\frac{v_\mathrm{bo}}{c}\frac{\kappa\dot{M}}{4\pi v_{\rm w}}\approx5.7\times10^{13}\kappa_{.34}\dot{M}_{-2}v_{\rm w,7}^{-1}v_9\,{\rm cm},
\end{equation}
where $\dot{M}=10^{-2}\dot{M}_{-2}M_\odot/$yr, $v_{\rm w}=10^7v_{\rm w,7}{\rm~cm/s}$, and $\kappa=0.34\kappa_{.34}{\rm cm^2/g}$ is the opacity, which is dominated at the relevant temperature and density ranges by electron (Thomson) scattering (see \S~\ref{subsec:This_Work}). Correspondingly, the wind $\tau=1$ Thomson photosphere is located at $R_\mathrm{ph}=(c/v_\mathrm{bo})R_\mathrm{bo}$.
The wind density at the breakout radius is
\begin{equation}
\label{eq:density}
\begin{split}
\rho_\mathrm{bo}&\equiv\rho\left(r=R_{\mathrm{bo}}\right)=\frac{c/v_{\mathrm{bo}}}{\kappa_\mathrm{T} R_{\mathrm{bo}}}\\
&\approx8.8\times10^{-13}\kappa_{.34}^{-1}R_{14}^{-1}v_9^{-1}\mathrm{g/cm^3}.
\end{split}
\end{equation}

Throughout the paper, we denote quantities normalized to their value at $R_{\rm bo}$ by a tilde, e.g.
\begin{equation}
\label{eq:normalization}
    \begin{split}
        \tilde{r}&\equiv r/R_\mathrm{bo},\\
        \tilde{v}&\equiv v/v_\mathrm{bo},\\
        \tilde{\rho}&\equiv \rho/\rho_\mathrm{bo},\\
        \tilde{\tau}&\equiv \tau/\tau_\mathrm{bo}.
    \end{split}
\end{equation}
In these notations, $\tilde{\rho}=\tilde{r}^{-2}$, and the optical depth of the plasma lying beyond radius $r$ is $\tilde{\tau}=\tilde{r}^{-1}$ (and the photosphere is at $\tilde{r}_\mathrm{ph}\approx30v_9^{-1}$). The diffusion time at radius $r$, $t_\mathrm{diff}\equiv \rho\kappa r^2/c=R_\mathrm{bo}/v_\mathrm{bo}$, is everywhere equal to the breakout time.

The characteristic breakout rise time, luminosity, energy, and color temperature are \citep{ofek_supernova_2010}
\begin{equation}
    \begin{split}
        t_\mathrm{rise}&\approx t_\mathrm{bo}= R_\mathrm{bo}/v_\mathrm{bo} =1R_{14}/v_9~\mathrm{d},\\
        L_\mathrm{bo}&\approx4\pi R_\mathrm{bo}^2\times\frac{1}{2}\rho_\mathrm{bo} v_\mathrm{bo}^3=10^{43.5}R_{14}v_9^2~\mathrm{erg/s},\\
        E_\mathrm{bo}&\approx L_\mathrm{bo}\times t_\mathrm{bo}=10^{48.5}R_{14}^2v_9~\mathrm{erg},\\
        T_\mathrm{color}&\approx\left(\frac{1}{2}\rho_\mathrm{bo}v_\mathrm{bo}^2/a_\mathrm{BB}\right)^{1/4}=10\left(v_9/R_{14}\right)^{1/4}~\mathrm{eV}.
    \end{split}
\end{equation}

\cite{katz_x-rays_2011} showed that during the wind breakout, the shock wave cannot be supported by the radiation momentum transfer, leading to the conversion of the RMS to a CLS. We repeat their argumentation here. In an RMS, the plasma lying ahead of the shock is accelerated by the radiation scattering off the electrons\footnote{Bound-free absorption does not contribute significantly to the plasma acceleration as the equilibrium ionization fractions are high (\ref{sec:ionization}). However, if ``metals" are only partially ionized by the radiation, then their bound-bound absorption opacity may be large (the de-excitation timescales are very short). The bound-bound contribution to the opacity is significant for solar abundance only for plasma temperature around $\sim15~$eV, and vanishes for $>20~$eV. For shock velocities $v_9>0.4$, the plasma temperatures exceed $20~$eV (\ref{sec:ionization}), rendering a domination of Thomson opacity for the radiation force. Note that since the plasma is optically thick to photons emitted by de-excitations, the net number of unexcited atoms is anyway significantly suppressed. Further analysis is required to calculate the modification of the opacity for lower shock velocities.}. The velocity to which the plasma at radius $r$ may be accelerated by radiation is
\begin{equation}
\label{eq:radacc}
    v=\frac{\kappa}{c}\int dt j=\frac{\kappa}{c}\frac{E(r)}{4\pi r^2},
\end{equation}
where $j$ is the radiation energy flux, and $E(r)$ is the energy carried by radiation propagating across a sphere of radius $r$. Assuming that the radiation energy is dominated by the energy produced as the shock propagates through the wind, which is valid for $R_\star\ll R_\mathrm{bo}$, the maximal available radiation energy is $E(r)=M(r)v^2/2\propto r$. As this energy increases linearly with radius, but the fraction of momentum delivered to the plasma decreases as $\propto r^{-2}$ (assuming the plasma does not expand considerably during the passage of the radiation), we see that there exists a radius, which in this analysis is $\tilde{r}\approx 1/2$, beyond which the radiation will no longer be able to accelerate the plasma to velocity $v$. Beyond this radius, the shock can no longer be mediated by radiation, and it must transform into a collisional or collisionless shock. Since the plasma frequency, $\omega_\mathrm{p}\approx 10^9 R_{14}^{-1/2}v_9^{-1/2}\mathrm{s}^{-1}$, is much larger than the protons Coulomb collision rate $\nu_\mathrm{C}\approx10^{-2}R_{14}^{-1}v_9^{-4}\mathrm{s}^{-1}$, a narrow width of a few skin depth, $\sim c/\omega_\mathrm{p}\approx 100~$cm CLS will form \citep{waxman_tev_2001}. The plasma is heated by the CLS to a temperature comparable to the kinetic energy of the protons, $\approx 100v_9^2~$keV, producing an X-ray-dominated radiation spectrum \citep{katz_x-rays_2011}.

\subsection{Earlier Work}
\label{subsec:Previous_Works}

In many analytic and numeric studies \citep[e.g.][]{chevalier_shock_2011,ginzburg_superluminous_2012, morozova_measuring_2018,tsuna_type_2019,takei_numerical_2020,margalit_analytic_2022,khatami_landscape_2024}, the radiation was assumed to be in thermal equilibrium with the plasma. While these studies provide information about the bolometric light curve, they do not provide a description of the evolution of the spectrum, which is far from thermal at and beyond the breakout radius. 

Some numeric studies \citep[e.g.][and subsequent studies of these groups]{moriya_supernovae_2011,dessart_numerical_2015} addressed this limitation using multi-group radiation hydrodynamics codes. These calculations do not, however, account for two processes that play a significant role in shaping the radiation spectrum. First, they do not include inelastic Compton scatterings, which (as shown below) are important for determining the optical--X-ray spectrum. Second, their grid resolution is insufficient for correctly capturing the temperature to which the plasma is heated by the CLS. Since the shock transition is spread in the numeric calculations over a few grid points, its numeric width is orders of magnitude larger than the physical $c/\omega_\mathrm{p}\approx 100~$cm width. As a result, the plasma numeric heating rate is many orders of magnitude smaller than the physical heating rate, such that the plasma is incorrectly limited by the plasma cooling (radiation emission) processes. This prevents the plasma from reaching the correct high temperatures of $\approx100$~keV, suppressing the emission of X-rays and yielding a UV-dominated spectrum. \citet{suzuki_supernova_2019} achieved sufficient grid resolution and obtained the correct plasma temperature by using an adaptive mesh refinement technique. However, they employed a two-temperature approximation for the plasma and radiation, which limits their coupling and results, again, in a too low radiation temperature.

The effect of inelastic Compton scatterings is also challenging to incorporate analytically. e.g. \citet{svirski_optical_2012}, estimated the X-ray emission based on the Bremsstrahlung cooling contribution to total emission, neglecting Comptonization effects. We show that upscattering of soft photons can yield significant X-ray emission, resulting in dominant X-ray luminosity much earlier than predicted by \citet{svirski_optical_2012}. While progress has been made \citep[e.g.][]{svirski_sn_2014,margalit_optical_2022,irwin_unexplored_2024} in analytic estimates of the effects of Compton scattering, a complete self-consistent calculation of the time-dependent shock structure and radiation spectrum during and following the RMS-CLS transition is still lacking.

The interaction of the escaping radiation with the upstream plasma is also often inconsistently treated, as the heating of the upstream plasma is neglected, leading to the conclusion that the X-ray luminosity is significantly absorbed \citep[e.g.][]{svirski_optical_2012,chevalier_x-rays_2012}. We demonstrate that the radiation efficiently heats the upstream plasma, resulting in a state of negligible absorption for most shock velocities.

Finally, let us comment on the numeric method used in \citet{ioka_spectrum_2019} \citep[and in subsequent papers of this group, e.g.][]{ito_monte_2020} for breakout calculations. This method approximates the temporal evolution as an ``adiabatic" transition between planar steady-state shock structures with a finite time-dependent fraction of radiation escaping upstream (to account for the escape of radiation as the shock approaches the breakout radius). The ``adiabatic" approximation is not a valid approximation since the shock structure changes on a dynamical timescale during which the upstream density varies significantly\footnote{ Particularly, it does not capture the effect of photon diffusion from the higher density regions downstream, which leads to the significant suppression of the electron temperature compared to that obtained in steady solutions \citep{sapir_non-relativistic_2013}.}. A more severe limitation of the method is the use of planar geometry, for which a CLS does not form. This method does not allow, therefore, to describe the RMS-CLS transition and the resulting plasma heating and X-ray emission, predicting a persistent soft spectrum with decreasing color temperature.

\subsection{Our Approximations}
\label{subsec:This_Work}

Our numeric code solves the 1D spherically symmetric multi-group radiation-hydrodynamics equations using diffusion approximation for radiation transport, including Bremsstrahlung emission/absorption and inelastic Compton scattering. The main approximations adopted are described and justified below.
\\ \noindent [i] Using the diffusion approximation is justified for the NR velocities considered, $v/c\ll1$. During the RMS phase, the width of the shock transition region is $\approx c/v$ times larger than the photon mean free path. During the CLS phase, the shock width is much smaller, implying a $\approx v/c$ deviation from isotropy in the radiation field in the plasma frame. However, the deviation is expected to be small for $v/c\ll 1$. We have verified that our results are not sensitive to changing the diffusion approximation to the ``P$_1$ diffusion approximation" \citep{castor_radiation_2004}, and to the choice of the Eddington factor used in the Eddington approximation for the outer boundary conditions (see \S~\ref{sec:equations}) (the luminosity is determined deep in the wind, where the optical depth is $c/v\gg 1$).
\\ \noindent [ii] The electrons and ions are assumed to move as a single fluid with the same velocity, also in regions, particularly within the RMS, where acceleration is dominated by Compton scattering of electrons due to the strong coupling provided by collective plasma instabilities, see \S~2.3.1 of \citet{budnik_relativistic_2010}: the ratio between the plasma time and the time between Compton scatterings of an electron is $\approx 10^{-12}(n_e/10^{12}{\rm cm^{-3}})^{1/2}$. The plasma instabilities developed in these regions due to the negligible velocity separation may be interesting to study \citep{vanthieghem_role_2022}, but carry negligible energy and do not affect the shock structure.
\\ \noindent [iii] An artificial viscosity term captures the CLS. This is a valid approximation as the CLS width, of the order of the plasma skin depth, is very small compared to all other length scales of the problem, e.g., the photon scattering mean free path. A ``cooling limiter" is introduced to capture the correct shock-heated electron temperature and post-shock cooling profile with acceptable grid resolution. The algorithm is described, and its validity is demonstrated in Appendix \ref{sec:cooling_limiter}.
\\ \noindent [iv] A complete first-principles understanding of the CLS structure, particularly of the fractions $\epsilon_\mathrm{pl},\epsilon_\mathrm{B}$ of post-shock internal energy carried by plasma particles and magnetic fields, is not yet available \citep[see e.g.][for reviews]{gupta_ab-initio_2023,sironi_relativistic_2015}. We adopt here $\epsilon_\mathrm{pl}=\epsilon_\mathrm{B}=1/2$ and equipartition between electrons and protons. The dependence of the results on $\epsilon_\mathrm{B}$ is expected to be small, as long as $\epsilon_\mathrm{pl}$ is of order unity (synchrotron emission is unimportant since the thermal electrons are NR and the synchrotron photon density is limited by strong self-absorption). The electron-proton equipartition assumption is justified since for the breakout velocities considered, the electrons are heated rapidly by proton collisional heating relative to their radiative (Compton) cooling, such that the (quasi-thermal) electrons and protons are close to equipartition at the post-shock flow (the electron temperature will be limited by cooling to values lower than the proton temperature only at high, $v_9>2$ velocities; see Appendix \ref{sec:electron-proton} for details).

We assume that the energy density of the magnetic field evolves adiabatically in the post-shock flow, with an adiabatic index of $4/3$ (appropriate for a non-ordered field), since we expect the characteristic time for magnetic field dissipation to be longer than the dynamical time. The Ohmic dissipation timescale of the magnetic fields is $t_B\gtrsim(m_p/m_e)(l\omega_p/c)^2t_{ee}$ \citep{krall_principles_1973, waxman_tev_2001}, where $l$ is the field coherence length, $\omega_p$ is the plasma frequency, and $t_{ee}$ is the electron-electron collision timescale. If the upstream plasma is unmagnetized and the field is generated by plasma instabilities at the shock, its initial coherence length is $\sim10 c/\omega_p$ \citep{sironi_relativistic_2015}, for which the dissipation time is short, $t_B/t_\mathrm{bo}\approx10^{-4}v_9^2(T_e/100~\mathrm{eV})^{3/2}$. However, both theoretical results and observations \citep{gruzinov_gamma-ray_2001,keshet_magnetic_2009} suggest that the coherence length of the field grows to the scale of the Larmor radius of accelerated protons. For the highest energy to which protons may be accelerated by the CLS, $>100~$TeV \citep{katz_x-rays_2011,waxman_shock_2025}, this yields $t_B/t_\mathrm{bo}>10^5(\varepsilon_\mathrm{max}/100~\mathrm{TeV})^2(T_e/100~\mathrm{eV})^{3/2}$.
\\ \noindent [v] As the plasma temperatures $<m_e c^2$ for $v_9<2$, we use the NR Bremsstrahlung emission/absorption and the Kompaneets approximation to describe Compton scatterings. At the higher end of the considered shock velocities, these are accurate up to $\sim10\%$, which is the same order of correction for electron-electron Bremsstrahlung and double Compton emission.
\\ \noindent [vi] Although the code includes the contribution of ``metals" to radiative processes (Bremsstrahlung, bound-bound, and bound-free; approximating the ionization and excitation distributions as thermal), we neglect these contributions in the present calculations. This is justified since in the high-temperature low-density range of interest, the opacity is dominated by electron scattering, the ions are highly ionized, and their impact on the shock structure and downstream flow is small \citep[see][where it is also shown that the photon production rate is dominated by Bremsstrahlung rather than double Compton or recombination, even for a low abundance of metals]{sapir_non-relativistic_2013}.
\\ \noindent[vii] We show in Appendix \ref{sec:ionization} that absorption of the escaping $>0.1$~keV X-ray photons by the initially cold and neutral upstream plasma is not significant for $v_9\gtrsim0.4$, due to the rapid ionization and heating of the upstream plasma. We, therefore, approximate in our numeric calculations the upstream plasma as fully ionized. 
\\ \noindent [viii] We neglect in the current analysis the effects of high energy, ``cosmic ray" (CR), protons and electrons accelerated by the CLS. CRs may affect the dynamics through escape of neutrinos produced in inelastic $pp/p\gamma$ collisions, through generation of $h\nu\gg m_ec^2$ photons for which the scattering cross section differs from Thomson's, and through modification of the optical depth by e$^\pm$ pair production. These effects are, however, small. For efficient CR acceleration, $\epsilon_\mathrm{CR}\sim0.2$, and conversion of a large fraction of the CR energy to pions, $f_\pi\sim0.5$, neutrinos carry only $0.5 f_\pi\epsilon_{CR}\sim0.05$ of the energy.
The dynamical effect of high energy photons is small since only a fraction of the CR energy is radiated at $h\nu\gg m_ec^2$ and since the pair production optical depth is significant only for $h\nu\gg m_ec^2$ photons, limiting their propagation (the resulting pairs lose their energy rapidly radiating lower energy photons). The contribution of pairs to the optical depth is small for $v/c\le0.1$ \citep{katz_x-rays_2011}.
Finally, the electron downstream temperature, determined by the balance of radiative cooling to proton collisional heating, is not significantly modified since the energy density of $h\nu<m_ec^2$ photons is not significantly modified.
\\ \noindent [ix] We neglect the wind velocity and the star's gravity, the effects of which are small for the fast $v_\mathrm{bo}$ considered.

\section{Radiation-Hydrodynamics Equations \& Numeric Methods} \label{sec:equations}

\subsection{Equations}\label{sec:eqs}
The temporal evolution of the plasma density and velocity profiles are determined by the mass and momentum conservation equations, 
\begin{equation}
\label{eq:mass}
    \dot{\rho}=-\rho \nabla_r v,
\end{equation}
\begin{equation}
\label{eq:vdot}
    \dot{v}=-\frac{1}{\rho}\partial_r P_\mathrm{tot}=-\frac{1}{\rho}\partial_r\left(\frac{1}{3}e_\mathrm{r}+\frac{2}{3}e_\mathrm{pl}+\frac{1}{3}e_\mathrm{B}+q\right).
\end{equation}
Here, over-dot represents a Lagrangian derivative, $\dot{f}\equiv(\partial_t+v\partial_r)f$, $\nabla_r$ is the radial part of the divergence operator, $\{e_\mathrm{r},e_\mathrm{pl},e_\mathrm{B}\}$ are the \{radiation, thermal plasma, magnetic field\} energy densities, and $q$ is the artificial viscosity pressure, for which we use the von Neumann-Richtmyer method with an additional linear term \citep{wilkins_use_1980}.

The radiation energy density is an integral over photon energies $\varepsilon$, of the spectral energy density $e_{\mathrm{r},\varepsilon}$
\begin{equation}
    e_\mathrm{r}=\int d\varepsilon e_{\mathrm{r},\varepsilon}.
\end{equation}
The energy conservation equations are
\begin{equation}
\label{eq:e}
\begin{split}     
\dot{e}_{\mathrm{r},\varepsilon}&=\dot{e}_{\mathrm{r},\varepsilon}^{\mathrm{mech}}+\dot{e}_{\mathrm{r-pl},\varepsilon}+\dot{e}_{\mathrm{r},\varepsilon}^{\mathrm{diff}},\\
\dot{e}_\mathrm{pl}&=\dot{e}_\mathrm{pl}^{\mathrm{mech}}-\dot{e}_\mathrm{r-pl},\\
\dot{e}_\mathrm{B}&=\dot{e}_\mathrm{B}^{\mathrm{mech}},
\end{split}
\end{equation}
with mechanical, radiation-plasma interaction, and diffusion terms defined as follows. 

The mechanical work terms are given by
\begin{equation}
\begin{split}
\dot{e}_{\mathrm{r},\varepsilon}^{\mathrm{mech}}&=-\left(\frac{4}{3}e_{\mathrm{r},\varepsilon}-\frac{1}{3}\partial_\varepsilon(\varepsilon e_{\mathrm{r},\varepsilon})\right)\nabla_r v,\\
\dot{e}_\mathrm{pl}^{\mathrm{mech}}&=-\left(\frac{5}{3}e_\mathrm{pl}+\tilde{\epsilon}_\mathrm{pl}q\right)\nabla_r v,\\
\dot{e}_\mathrm{B}^{\mathrm{mech}}&=-\left(\frac{4}{3}e_\mathrm{B}+\tilde{\epsilon}_\mathrm{B}q\right)\nabla_r v.
\end{split}
\end{equation}
The radiation frequency-dependent compression term follows \citet{castor_radiation_2004}. At the CLS stage, the viscosity at the shock converts kinetic energy into specific energy fractions of plasma and magnetic field energy. We use $\tilde{\epsilon}_\mathrm{pl}=0.42$ to obtain $\epsilon_\mathrm{pl}=0.5$ at the post-shock plasma\footnote{The value of $\tilde{\epsilon}_\mathrm{pl}$ was determined from numeric calculations of planar non-radiative viscous shocks in a uniform mixture of ideal gases with adiabatic indices of $5/3$, representing the plasma, and $4/3$, representing the magnetic field component.}.

The radiation-plasma coupling term is
\begin{equation}
\begin{split}
\label{eq:radmat}
    \dot{e}_{\mathrm{r-pl},\varepsilon}=&\rho \kappa_\varepsilon c\left(B_\varepsilon-e_{\mathrm{r},\varepsilon}\right)+\\
    &\rho \kappa_\mathrm{T} c\frac{\varepsilon}{m_e c^2}\partial_\varepsilon\left[T\partial_\varepsilon\left(\varepsilon e_{\mathrm{r},\varepsilon}\right)+\left(\varepsilon-4T\right)e_{\mathrm{r},\varepsilon}\right].
\end{split}
\end{equation}
The first term is the Bremsstrahlung emission and absorption, where $\kappa_\varepsilon\left(\rho,T\right)$ is the Bremsstrahlung opacity, $B_\varepsilon\left(T\right)$ is the Planck spectral energy density, and  the plasma temperature is
\begin{equation}
    T=\frac{e_\mathrm{pl}}{\frac{3}{2}\frac{1+Z}{A}\frac{\rho}{m_p}},
\end{equation}
where $Z$($A$) is the atomic(mass) number. In the numeric calculations, we use $Z=A=1$ and the electron Thomson scattering opacity is $\kappa=\kappa_T\approx0.4~\mathrm{cm^2/g}$. For hydrogen/helium-dominated plasmas, the results have limited dependence on this choice, with a straightforward scaling of different quantities derived analytically (see, for example, the bolometric luminosity eq. (\ref{eq:luminosity})). The second term of eq. (\ref{eq:radmat}) is the Kompaneets approximation for (inelastic) Compton scattering \citep{kompaneets_establishment_1956}, excluding the negligible stimulated emission term (which was verified numerically to be insignificant). For what follows, it is useful to write the bolometric form of the two interactions
\begin{equation}
\begin{split}
\label{eq:bolradmat}
    \dot{e}_\mathrm{Brem}=&\rho \kappa_\mathrm{T} c\left(\frac{32}{\pi^3}\right)^{1/2}\alpha_e\frac{\rho}{m_p}\left(m_e c^2 T\right)^{1/2},\\
    \dot{e}_\mathrm{Comp}=&\rho \kappa_\mathrm{T} c\frac{4\left(T-T_\mathrm{r}\right)}{m_e c^2}e_\mathrm{r},
\end{split}
\end{equation}
where $\alpha_e$ is the fine structure constant, and the radiation temperature is defined as
\begin{equation}
\label{eq:T_rad}
T_\mathrm{r}\equiv\frac{1}{4}\frac{\int d\varepsilon\, \varepsilon  e_{\mathrm{r},\varepsilon}}{e_\mathrm{r}}\equiv\frac{1}{4}\Bar{\varepsilon},
\end{equation}
a quarter of the energy-weighted average photon energy.

The radiation diffusion term is\footnote{We use Thomson opacity for the diffusion coefficient since Thomson scattering opacity dominates over Bremsstrahlung opacity almost everywhere. Bremsstrahlung opacity is important in the dense shell produced downstream behind the shock (\S~\ref{sec:constant_results}, \S~\ref{sec:ejecta_results}). However, the radiation field is almost uniform across the shell (\S~\ref{sec:constant_results}), so the Bremsstrahlung correction to the diffusion opacity is not important. We have verified numerically that adding the Bremsstrahlung opacity yields a negligible impact on the results.}
\begin{equation}
    \label{eq:diff}
    \dot{e}_{\mathrm{r},\varepsilon}^{\mathrm{diff}}=-\nabla_r j_\varepsilon=\nabla_r\left(\frac{c}{3\rho \kappa_\mathrm{T}}\partial_r e_{\mathrm{r},\varepsilon}\right).
\end{equation}
The radiation flux at the outer boundary ($\tau=1$) is determined using the Eddington approximation $j_\varepsilon=f_E e_{\mathrm{r},\varepsilon}(\tau=1)$, where $f_E$ is the dimensionless order unity Eddington factor. We have verified that the results are not sensitive to the exact value of $f_E$ in the range 0.3-0.5, and show numeric results for $f_E=0.5$.

\subsection{Numeric Code, Validation \& Convergence}\label{sec:code}

The radiation-hydrodynamics equations (\ref{eq:mass})-(\ref{eq:diff}) are solved using an adapted version of our 1D NR radiation-hydrodynamics code \citep{sapir_numeric_2014,morag_shock_2023}. The mass and momentum conservation equations, (\ref{eq:mass}) and (\ref{eq:vdot}), are solved by the standard leap-frog on a staggered mesh method. The energy conservation set of equations (\ref{eq:e}) is solved using operator splitting. The equations are divided into three parts, radiation diffusion, radiation-plasma interaction, and radiation mechanical work, and these parts are solved consecutively. The radiation-plasma interaction, eq. (\ref{eq:radmat}), is solved implicitly in an iterative fashion for a given plasma temperature. The corresponding radiative energy transfer is introduced to the plasma in eq. (\ref{eq:e}), and the solution for the plasma temperature is then inserted again into the radiation-plasma interaction, until convergence is achieved. The resulting spectral radiation energy density and the plasma energy density solve the set of equations simultaneously. In this way, energy conservation in the radiative interactions between the radiation and the plasma is assured to numeric precision. This code was verified against simple benchmark problems that have bolometric and spectral analytical solutions \citep[see more details in][]{sapir_numeric_2014} and
was successfully used to describe the early shock-cooling multi-band light curves of a large number of type II SNe \citep{irani_sn_2024}.

A ``cooling limiter” is introduced to capture the correct CLS-heated plasma temperature and the following plasma cooling profile with acceptable grid resolution. The algorithm is described, and its validity is demonstrated in Appendix \ref{sec:cooling_limiter}.

The Courant condition in the dense shell produced downstream behind the shock (\S~\ref{sec:constant_results}, \S~\ref{sec:ejecta_results}) is the most restricting constraint in determining the numeric time step, which is of order $\sim10^{-5}$ of the breakout time. The spatial grid contains $\sim10^3$ logarithmically-spaced cells \citep[see][for the spatial grid of the stellar envelope]{sapir_numeric_2014}, and we use $\sim10^2$ logarithmically-spaced frequency bins covering the range $[10^{-1},10^{6}]$~eV. The hydrodynamic and spectral results (presented in \S~\ref{sec:constant_results}, \S~\ref{sec:ejecta_results}) were checked to be converged (with an error of typically $\lesssim1\%$) with increasing ($\times2,\times4$) the spatial, frequency, and temporal resolutions.

\section{Constant Velocity Piston} \label{sec:constant_results}

We analyze in this section the breakout of a shock wave driven by a constant velocity piston into the wind. The constant piston velocity implies a constant shock velocity, $v_\mathrm{bo}$, which is slightly, $(1-\Delta\tilde{v}_\mathrm{DS})(\gamma+1)/2=1.125$ times, higher than the piston velocity, $\Delta\tilde{v}_\mathrm{DS}$ being the downstream velocity difference between the shock and the piston (see below), and $\gamma$ the adiabatic index\footnote{The adiabatic index in the RMS stage is $4/3$, while for the two-gases mixture in the CLS stage, it is $\gamma=\frac{5}{3}\epsilon_\mathrm{pl}+\frac{4}{3}\epsilon_B=3/2$. We use the latter for setting the piston velocity to obtain a shock velocity of $v_\mathrm{bo}$ at the CLS stage. The small variations of the adiabatic index have only a small impact on our results.}. 

We define the shock ``position" to be at the location of the peak plasma temperature (at the RMS stage, the shock width corresponds to $\Delta\tau\sim c/v$, while in the CLS stage, the width is a few grid points). The time dependence of the shock position is simply a linear propagation, $\tilde{r}_\mathrm{sh}= \tilde{t}$, where the tilde sign denotes normalization to the breakout value, eq. (\ref{eq:normalization}), and we denote with subscript ``sh" quantities evaluated at the shock position. We describe the initial and inner-boundary conditions in \S~\ref{subsec:piston_initial}, present the resulting numeric plasma profiles and radiation spectrum in \S~\ref{subsec:piston_results}, and derive analytic approximations describing key numeric results in \S~\ref{subsec:piston_analytic}.

\subsection{Initial \& Inner-Boundary Conditions}
\label{subsec:piston_initial}

The plasma is initially at rest, with density given by a wind profile, eq. (\ref{eq:density}), and initial temperature $T_0=0.1$~eV. The radiation is initially set to be in thermal equilibrium with the plasma. The results are insensitive to the exact values of $T_0$.

The inner boundary conditions are a piston at rest at radius $\tilde{r}_\mathrm{p}=\tilde{r}_\mathrm{p0}=0.1$ at initial time $\tilde{t}=0.1$, accelerating with constant acceleration to the piston velocity at $\tilde{t}=\tilde{t}_a=1.01$, and remaining at a constant velocity at later time (the finite time of acceleration allows faster convergence with longer time steps). The results are not sensitive to the exact values of $\tilde{r}_\mathrm{p0}$ and $\tilde{t}_a$. The spectrum of the incoming flux associated with the acceleration \citep[see][for implementation details]{sapir_numeric_2014} is thermal. The results are not sensitive to the spectral shape of the incoming flux.

The artificial viscosity that is used to capture the CLS is a function of the velocity variation between grid points. During the RMS stage, the viscosity contribution vanishes with increasing grid resolution. In order to allow faster convergence, we activate the artificial viscosity only at $\tilde{r}_\mathrm{sh}> 0.2$.

\subsection{Numeric Results}
\label{subsec:piston_results}

\begin{figure}
    \centering
    \includegraphics[height=14cm]{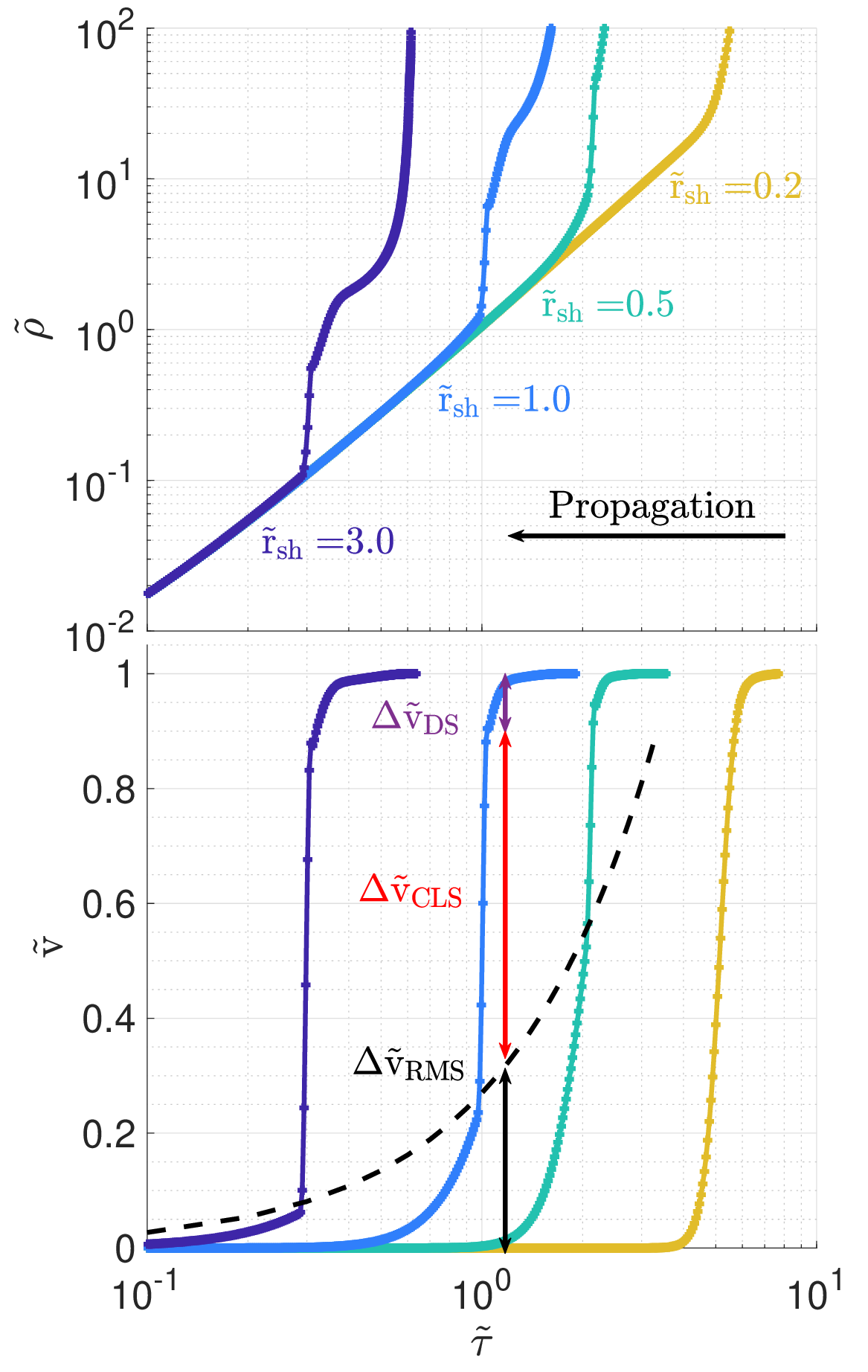}
    \caption{Plasma density (\textit{upper panel}) and velocity  (\textit{lower panel}) profiles as a function of optical depth at different shock radii (normalized by eq. (\ref{eq:normalization}), except the velocity which is normalized to the piston velocity). At small shock radii, the radiation diffusing upstream accelerates the plasma to the post-shock velocity. When the RMS approaches the breakout radius, the plasma is accelerated by the radiation diffusing upstream to a smaller velocity, $\Delta\tilde{v}_\mathrm{RMS}$  (see text) described analytically by eq. (\ref{eq:v_RMS}) (dashed line), and then further accelerated by $\Delta\tilde{v}_\mathrm{CLS}$ by the CLS.}
    \label{fig:constrhov}
\end{figure}

\begin{figure*}[ht]
\centering
    \hspace*{-2cm}
    \includegraphics[height=8cm]{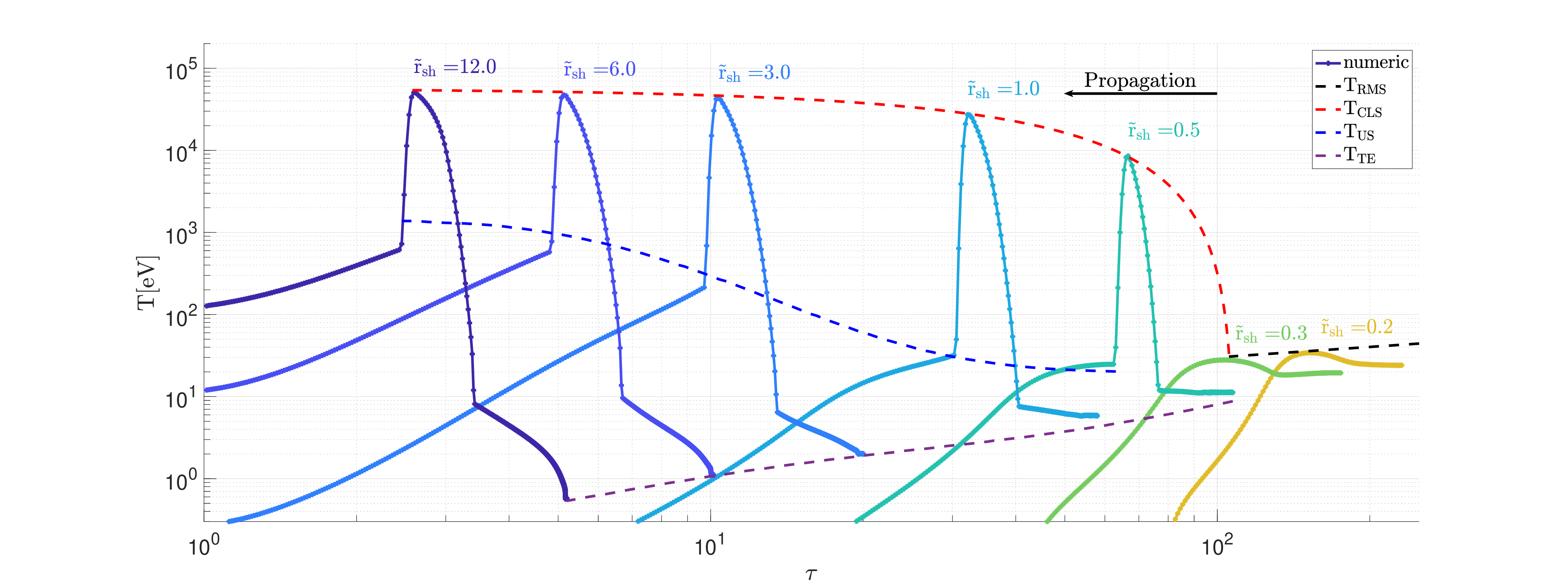}
    \hspace*{2cm}
    \caption{Solid lines show the numeric plasma temperature profiles as a function of optical depth at different shock radii, for $R_{14}=1$ and $v_9=1$. Analytic approximations for key temperature features that describe the profile structure are shown in dashed lines: The RMS temperature $T_\mathrm{RMS}$, eq. (\ref{eq:T_RMS}); the CLS temperature $T_\mathrm{CLS}$, eq. (\ref{eq:T_CLS}); the temperature to which upstream plasma is heated by the diffusing radiation $T_\mathrm{US}$, eq. (\ref{eq:T_US}); and the thermal equilibrium temperature achieved in the dense shell $T_\mathrm{TE}$, eq. (\ref{eq:T_TE}). The analytic approximations for $T_\mathrm{US}$ and $T_\mathrm{TE}$ are obtained using the numeric radiation temperature (see text).}
    \label{fig:consttemp}
\end{figure*}

\begin{figure}[ht]
    \centering
    \includegraphics[height=6.5cm]{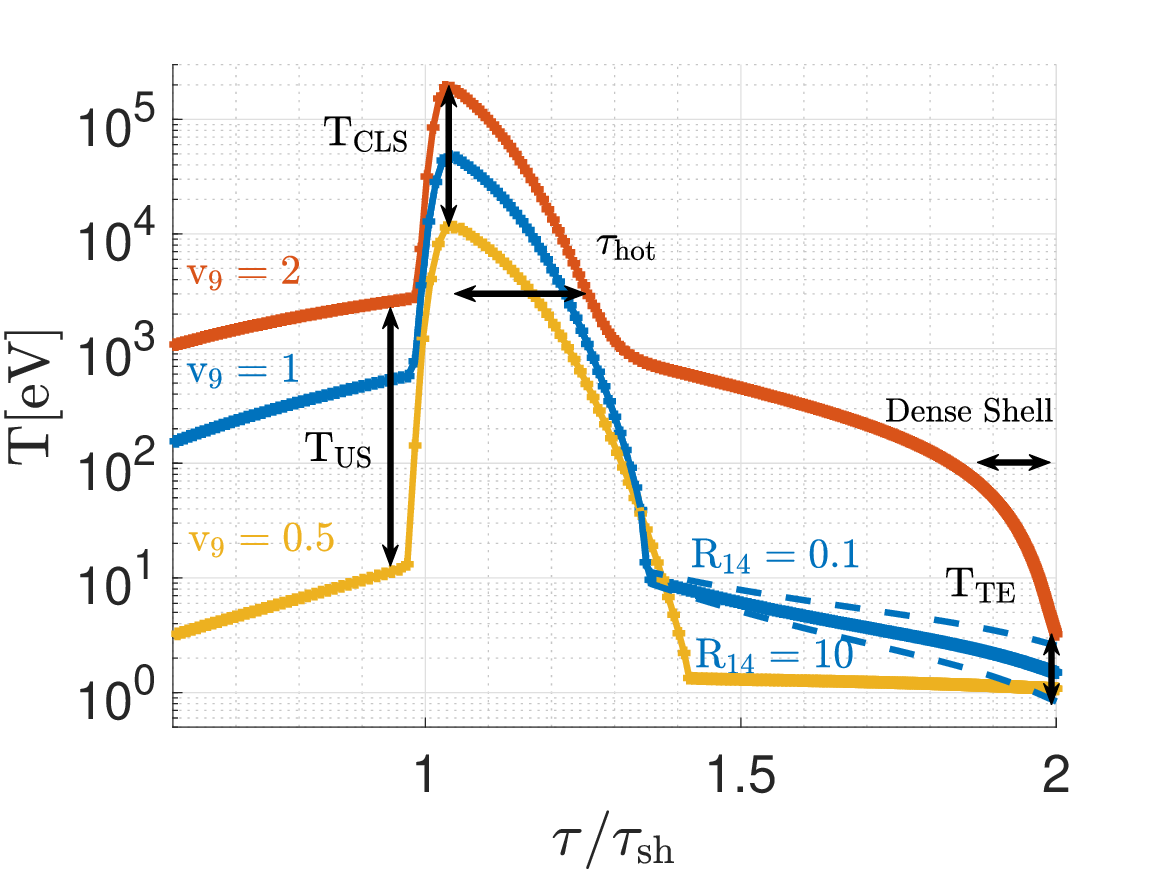}
    \caption{Plasma temperature profiles as a function of (normalized) optical depth, for $\tilde{r}_\mathrm{sh}=6$, $R_{14}=1$ and $v_9=0.5,1,2$. For $v_9=1$, we also show  $R_{14}=0.1,10$. Analytic approximations describe well the variations in $T_\mathrm{CLS}$, eq. (\ref{eq:T_CLS}), $T_\mathrm{US}$, eq. (\ref{eq:T_US}), $T_\mathrm{TE}$, eq. (\ref{eq:T_TE}) and $\tau_\mathrm{hot}$, eq. (\ref{eq:tau_CLS}).}
    \label{fig:T_comp}
\end{figure}

\begin{figure}[ht]
    \centering
    \includegraphics[height=6.5cm]{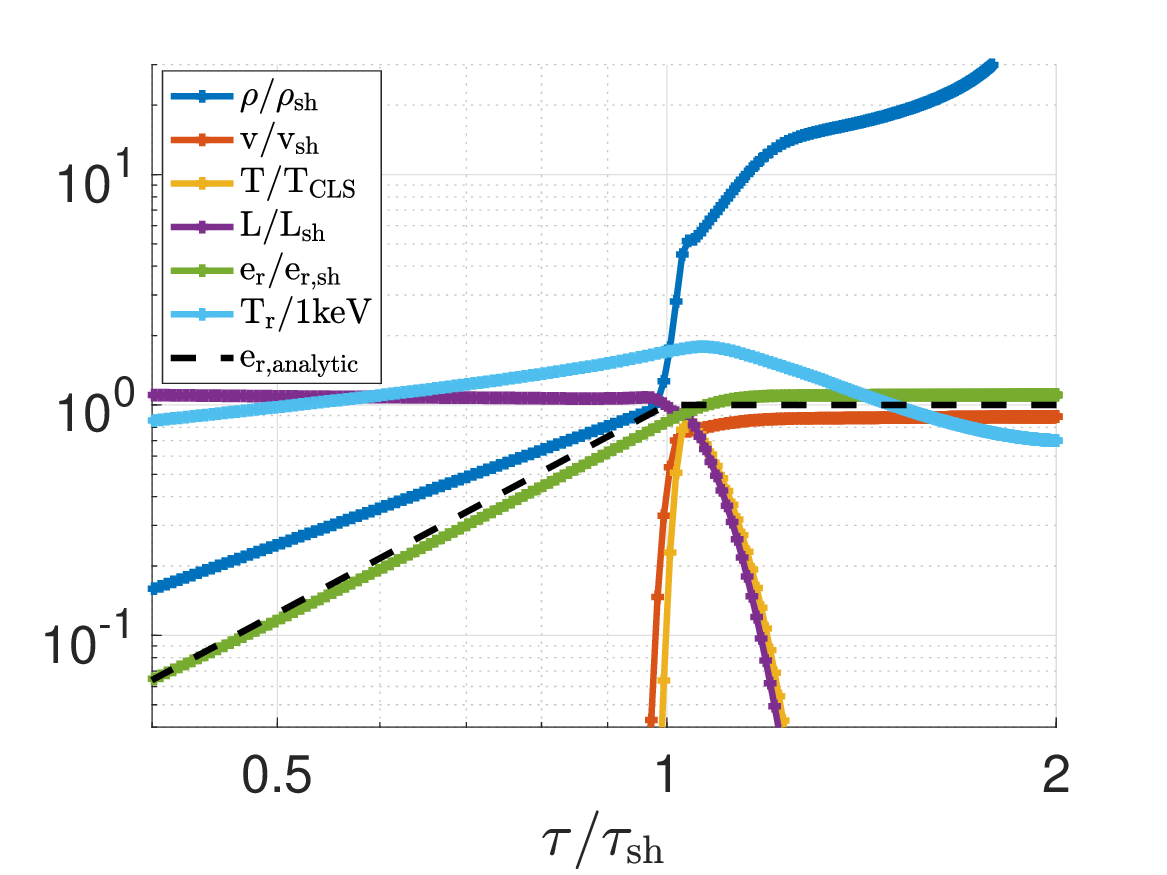}
    \caption{Radiation-hydrodynamic profiles at the CLS for $\tilde{r}_\mathrm{sh}=6$, $R_{14}=1$, and $v_9=1$: density, velocity, plasma temperature, bolometric luminosity, bolometric radiation energy density, and radiation temperature, as functions of optical depth, normalized to their analytic shock values. The analytic CLS temperature and luminosity are given by eqs. (\ref{eq:T_CLS}) and (\ref{eq:luminosity}) respectively. The analytic radiation energy density profile is given by eq. (\ref{eq:e_r}) (dashed line). All profiles presented are nearly independent of $\tilde{r}_\mathrm{sh}$, $R_\mathrm{bo}$, and $v_\mathrm{bo}$, with the exception of the evolving radiation temperature (which is normalized by $1~$keV), and of the thickness of the post-shock hot layer (see text), which depends on the shock radius according to eq. (\ref{eq:tau_CLS}).}
    \label{fig:profiles}
\end{figure}

\subparagraph{\textbf{Hydrodynamical Profiles}.} The plasma density and velocity profiles obtained at different shock radii $\tilde{r}_\mathrm{sh}$ are shown in Figure \ref{fig:constrhov}. When normalized to their breakout values, eq. (\ref{eq:normalization}), these profiles are nearly independent of $R_\mathrm{bo}$ and $v_\mathrm{bo}$. The plasma temperature profiles obtained at different $\tilde{r}_\mathrm{sh}$ for $R_\mathrm{14}=1$, $v_\mathrm{9}=1$ are shown in Figure \ref{fig:consttemp}. Figure \ref{fig:T_comp} compares 
the plasma temperature profiles for different $R_\mathrm{bo}$ and $v_\mathrm{bo}$ values, and Figure \ref{fig:profiles} shows the spatial dependence of several hydrodynamical quantities at a fixed time within the CLS stage.

At small radii, the radiation diffusing upstream accelerates the plasma close to the post-shock velocity, $2(\gamma+1)^{-1}v_\mathrm{bo}$, and a smooth velocity curve is obtained within the transition region (Figure \ref{fig:constrhov}). At larger radii, comparable to the breakout radius, the velocity evolution within the shock transition region can be divided into two parts: the plasma is smoothly accelerated by the radiation diffusing upstream to a smaller velocity, $\Delta\tilde{v}_\mathrm{RMS}$, and then abruptly by $\Delta\tilde{v}_\mathrm{CLS}$ by the CLS. The plasma is then further gradually accelerated in the downstream flow by $\Delta\tilde{v}_\mathrm{DS}$ to the piston velocity (these three quantities are normalized to the piston velocity, so their sum $=1$). As the radiation ``struggles" to accelerate the upstream plasma (\S~\ref{subsec:Wind+CLS}), a gradual increase of $\Delta\tilde{v}_\mathrm{CLS}$ at the expense of $\Delta\tilde{v}_\mathrm{RMS}$ is obtained. 

We define the time, and the corresponding shock radius, $\tilde{r}_\mathrm{CLS}$ of ``CLS onset", as the time after which the peak plasma temperature begins to increase. As can be seen in Figure \ref{fig:consttemp}, the transition to a rising peak plasma temperature is abrupt and occurs at $\tilde{r}_\mathrm{CLS}\approx0.3$, corresponding to $\tilde{\tau}_\mathrm{CLS}\approx 3.3$. The CLS onset radius is found to be nearly independent of $R_\mathrm{bo}$ and $v_\mathrm{bo}$. An analytic description of the evolution of $\Delta \tilde{v}_\mathrm{RMS}$ is given below, eq. (\ref{eq:v_RMS}), which describes well the ``break" between the smooth RMS-accelerated plasma and the CLS-accelerated plasma, see Figure \ref{fig:constrhov}. Beyond $\tilde{r}_\mathrm{CLS}$, we obtain $\Delta\tilde{v}_\mathrm{DS}\approx0.1$ (normalized to the piston velocity) which stays approximately constant at different $\tilde{r}_\mathrm{sh}$, $R_\mathrm{bo}$ and $v_\mathrm{bo}$. The density diverges near the piston and produces a dense plasma shell, as expected from the self-similar solution for a viscous shock driven by a constant velocity piston into a wind density profile (and is similar to the dense shell generated by the expanding envelope driven shock, see \S~\ref{sec:ejecta_results}). While the dense shell may have an impact on the resulting radiation spectrum (described analytically by eq. (\ref{eq:T_TE})), the peak density value, which depends on the grid resolution, does not affect the flow dynamics and radiation field.

Figure \ref{fig:consttemp} shows that the peak plasma temperature rises rapidly after the CLS onset radius, as mentioned above (for $v_9=1$, the CLS temperature $T_\mathrm{CLS}$ is $10^3$ times higher than the RMS temperature $T_\mathrm{RMS}$ within less than the breakout timescale). The upstream plasma temperature $T_\mathrm{US}$ also rises with shock propagation as it is heated by the diffusing radiation originating from the shocked plasma. However, this heating process is not as rapid and occurs after shock propagation of over a few breakout radii. In contrast, the downstream plasma in the dense shell behind the shock cools with shock propagation. We derive below (\S~\ref{subsec:piston_analytic}) analytic approximations for the key features that describe the temperature profile. While the temperature profile is sensitive to $v_\mathrm{bo}$ (which affects, for example, $T_\mathrm{CLS}$ and $T_\mathrm{US}$), it is weakly dependent on $R_\mathrm{bo}$, as can be seen in Figure \ref{fig:T_comp}.

The post-CLS hot layer of shocked plasma is evident in Figures \ref{fig:consttemp}-\ref{fig:profiles}. We derive below an analytic approximation for its (physical) width, eq. (\ref{eq:tau_CLS}). Note that when using the cooling limiter, the resulting cooling profile of the plasma in this layer at early times is smeared to be wider than the physical cooling width, but nevertheless yielding the correct and converged radiation emission; see Appendix \S~\ref{sec:cooling_limiter}.

\begin{figure*}[ht]
    \centering
    \includegraphics[height=5.5cm]{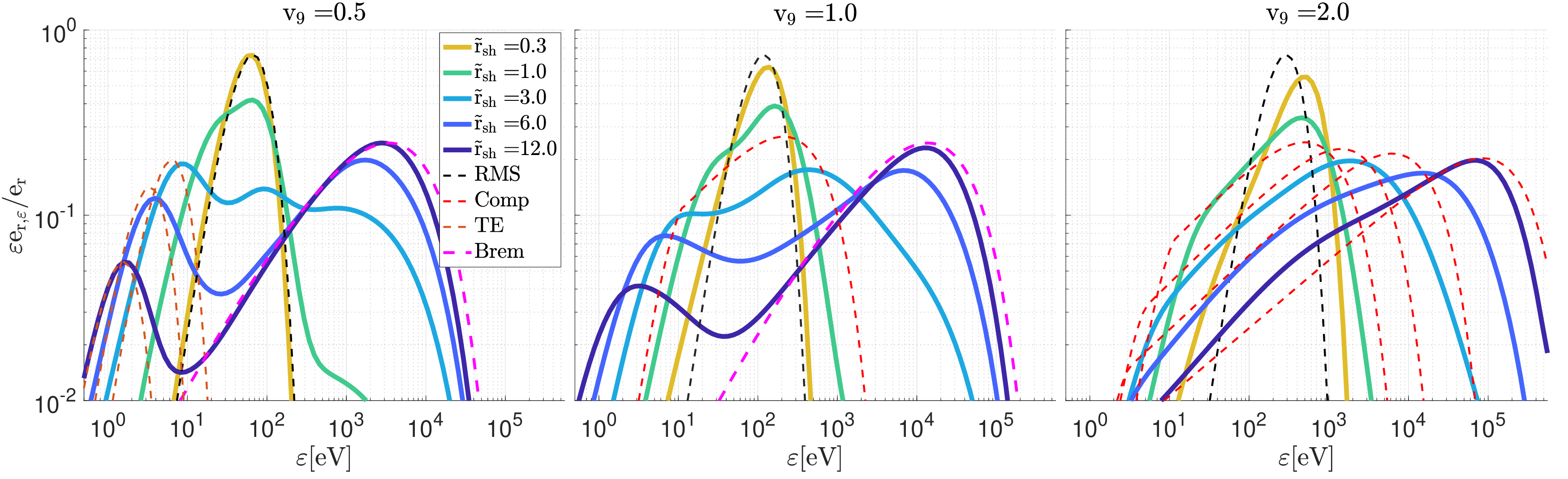}
    \caption{Solid lines show the numeric radiation spectral energy density at $\tilde{r}_\mathrm{sh}$, $\mathrm{e}_{{\rm r},\varepsilon}\equiv \partial_\varepsilon\mathrm{e}_{\rm r}$, as a function of photon energy, $\varepsilon$, normalized to the bolometric energy density $\mathrm{e}_{\rm r}$, eq. (\ref{eq:e_r}), for different shock radii, $R_{14}=1$ and $v_9=0.5,1,2$. Analytic approximations for various regimes are shown in dashed lines (corresponding to shock radii of their adjacent solid lines): Comptonized Wien (black, eq. (\ref{eq:T_RMS})), unsaturated Comptonization (red, eq. (\ref{eq:unsaturated})), thermal radiation originating from the dense shell (brown, eq. (\ref{eq:T_TE})), and Bremsstrahlung cooling (magenta, eq. (\ref{eq:brem})).}
    \label{fig:constspect}
\end{figure*}

\begin{figure}[ht]
    \centering
    \includegraphics[height=6.5cm]{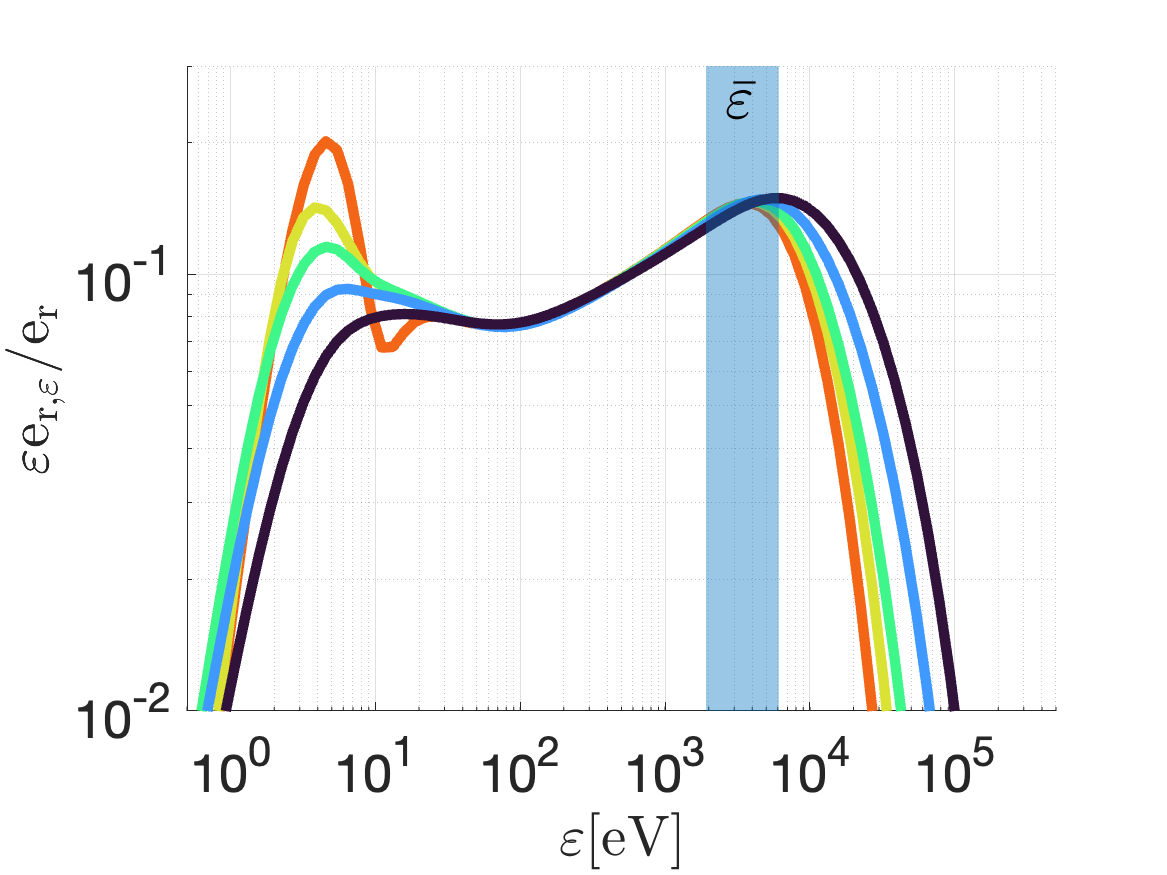}
    \caption{Radiation spectral energy density normalized to the bolometric energy density at different locations downstream, between the piston (red) and the shock (dark blue), for $\tilde{r}_\mathrm{sh}=6$, $R_{14}=1$, and $v_9=1$. The shaded area represents the variation of the (energy-weighted) average photon energy, $\Bar{\varepsilon}$, defined in eq. (\ref{eq:T_rad}).}
    \label{fig:downspect}
\end{figure}

\subparagraph{\textbf{Spectral Evolution}.} Figure \ref{fig:constspect} shows the radiation spectral energy density at $\tilde{r}_\mathrm{sh}$ for different shock radii, $R_{14}=1$ and $v_9=0.5,1,2$, and Figure \ref{fig:downspect} shows the spectrum variations along the downstream at a fixed time.

The spectral shape of the radiation energy density changes dramatically with shock propagation. Initially, at the RMS stage, the spectrum resembles a simple Comptonized-Wien distribution. During the transition to the CLS stage, the spectrum undergoes a hardening phase. Even after the shock is well into the CLS stage (e.g. $\tilde{r}_\mathrm{sh}=3$), the spectrum continues to harden significantly as the shock propagates. The average photon energy exceeds $1$~keV after propagation over a few $R_\mathrm{bo}$. For the higher velocities, $v_9\gtrsim 1.4$, the spectrum exhibits a single peak, while for lower velocities, an additional softer peak appears at UV energies, as shown in Figure \ref{fig:constspect}. The spectra depend only weakly on $R_\mathrm{bo}$, see \S~\ref{subsec:piston_analytic} and \S~\ref{subsec:envelope_spectrum}.

Figure \ref{fig:profiles} shows that in the upstream, the luminosity is uniform, and the bolometric radiation energy density decreases as $r^{-3}$. This behavior is consistent with that predicted by the analytic analysis below; see eq. (\ref{eq:luminosity}) and (\ref{eq:e_r}). In the downstream, the radiation energy density is roughly uniform, and the variations in the spectrum and its average energy (or temperature, eq. (\ref{eq:T_rad})) are limited, with a small decrease in the radiation temperature approaching the piston; see Figures \ref{fig:profiles} and \ref{fig:downspect}, and the explanation below.

\subsection{Analytic Approximations}
\label{subsec:piston_analytic}

\subparagraph{\textbf{Bolometric Luminosity}.} After the breakout, the radiation reaches all the way to the photosphere. Assuming that all the energy deposited in plasma particles is efficiently radiated away (as demonstrated below, eq. (\ref{eq:rad_eff})), the upstream luminosity (at $\tilde{r}>\tilde{r}_\mathrm{sh}$) remains nearly constant in time and uniform spatially. For strong shock waves, the shock produces an internal energy density of $2(\gamma^2-1)^{-1}\rho v_\mathrm{sh}^2$ at a velocity $(\gamma-1)/(\gamma+1)v_\mathrm{sh}$ relative to the shock (and a fraction $\epsilon_\mathrm{pl}$ is deposited in plasma particles and radiated away), so the luminosity is
\begin{equation}
\begin{split}
\label{eq:luminosity}
L_\mathrm{sh}&=4\pi r_\mathrm{sh}^2\times\epsilon_\mathrm{pl}2(\gamma+1)^{-2}\rho(r_\mathrm{sh})v_\mathrm{sh}^3\\
&=8\pi\epsilon_\mathrm{pl}\left(\gamma+1\right)^{-2}\frac{c}{\kappa_T}R_\mathrm{bo}v_\mathrm{bo}^2\approx1.5\times10^{43}R_{14}v_9^2~\mathrm{erg/s},
\end{split}
\end{equation}
where we used the wind density profile eq. (\ref{eq:density}). The uniform luminosity and its value are consistent with the numeric result, as demonstrated in Figure \ref{fig:profiles}.

\subparagraph{\textbf{Downstream Uniformity and Bolometric Radiation.}} Since the diffusion time is shorter than the dynamical time after breakout (\S~\ref{subsec:Wind+CLS}), photons diffuse quickly in the downstream region between the shock and the piston (relative to the shock propagation timescale). As a result, the downstream bolometric radiation energy density is expected to be roughly uniform, as well as the radiation spectrum and temperature. For the uniform luminosity upstream, the radiation energy density can be directly inferred from the flux, eq. (\ref{eq:diff}), and follows $\tilde{r}^{-3}$. We, therefore, approximate
\begin{equation}
\label{eq:e_r}
\begin{split}  
e_\mathrm{r}(r,t)&\approx e_\mathrm{bo}
\begin{cases} \tilde{r}^{-3}& \tilde{r}_\mathrm{sh}<\tilde{r}<\tilde{r}_\mathrm{ph}\\
\tilde{r}^{-3}_\mathrm{sh}& \tilde{r}_\mathrm{p}<\tilde{r}<\tilde{r}_\mathrm{sh}
\end{cases},\\
e_\mathrm{bo}&=\frac{L_\mathrm{sh}}{4\pi R_\mathrm{bo}^2v_\mathrm{bo}}\approx 1.2\times10^5 R_{14}^{-1}v_9~\mathrm{erg/cm}^3.
\end{split}
\end{equation}
This structure provides an excellent approximation to the numeric results, as shown in Figure \ref{fig:profiles}. Note that near the photosphere ($\tau=1$), the diffusion approximation no longer holds, and a transition to free-streaming, $e_\mathrm{r}\sim \tilde{r}^{-2}$, occurs.

The downstream radiation spectrum (and its average energy) does exhibit some limited variations, see Figure \ref{fig:downspect} (and \ref{fig:profiles}), which are a result of thermal-reprocessing of hard photons in the dense shell (see below).

\subparagraph{\textbf{RMS Velocity Difference ($\Delta\tilde{v}_\mathrm{RMS}$)}.} The numeric result of $\tilde{r}_\mathrm{CLS}\approx 0.3$ aligns with the analytic prediction of eq. (\ref{eq:radacc}), $\tilde{r}_\mathrm{CLS}\approx 1/2$. Based on the argument leading to this equation, we expect $\Delta\tilde{v}_\mathrm{RMS}$ to decrease as $\tilde{r}_\mathrm{sh}^{-1}$
\begin{equation}
\label{eq:v_RMS}
    \Delta\tilde{v}_\mathrm{RMS}(\tilde{r}_\mathrm{sh}>\tilde{r}_\mathrm{CLS})\approx(1-\Delta\tilde{v}_\mathrm{DS}) \tilde{r}_\mathrm{CLS}/\tilde{r}_\mathrm{sh}, 
\end{equation}
normalized to the piston velocity. This is in good agreement with the numeric results shown in Figure \ref{fig:constrhov}. Correspondingly, $\Delta\tilde{v}_\mathrm{CLS}\approx1-\Delta\tilde{v}_\mathrm{DS}-\Delta\tilde{v}_\mathrm{RMS}$.

\subparagraph{\textbf{RMS Temperature ($T_\mathrm{RMS}$).}} For $v_9\gtrsim1$, the shock transition region is expected to be far from thermal equilibrium \citep{katz_fast_2010}. The RMS peak plasma temperature can be estimated using the interpolation formula of \cite{sapir_non-relativistic_2013}, for a wind density profile, eq. (\ref{eq:density})
\begin{equation}
\label{eq:T_RMS}
    \begin{split}        \mathrm{log}_{10}\frac{T_\mathrm{RMS}}{\mathrm{eV}}\approx&0.08+1.18v_9^{1/2}\\
        &-\left(0.26-0.05v_9^{1/2}\right)\mathrm{log}_{10}\left(\tilde{r}_\mathrm{sh}^2 R_{14}v_9\right).
    \end{split}
\end{equation}
This formula is derived from numeric calculations of the number of photons contributing to the Comptonized-Wien spectrum that was shown to be generated in such conditions and can be approximated with a blackbody shape shifted to temperature $T_\mathrm{RMS}$ \citep[see][for details]{katz_fast_2010}. While originally fitted to the density profile of a stellar envelope, it performs well in the wind case, with errors up to tens of percent at the high end of our shock velocities, $v_9\sim2$, see the numeric result in Figures \ref{fig:consttemp}, \ref{fig:constspect}. Notably, for $v_9=1$, $T_\mathrm{RMS}$ is $\sim1.5$ times higher than the blackbody temperature predicted under the thermal equilibrium assumption.

Following the CLS onset radius, $\tilde{r}_\mathrm{sh}>\tilde{r}_\mathrm{CLS}$, three key temperatures characterize the temperature profile (and the generated radiation spectrum)
\begin{enumerate}
    \item \textbf{CLS Temperature (\(T_\mathrm{CLS}\))}: The peak temperature of the CLS-heated plasma;
    \item \textbf{Upstream Temperature (\(T_\mathrm{US}\))}: The temperature to which the immediate upstream plasma is heated by the diffusing radiation;
    \item \textbf{Thermal Equilibrium Temperature (\(T_\mathrm{TE}\))}: The temperature of thermal equilibrium achieved in the dense shell behind the shock.
\end{enumerate}
We derive below analytic approximations for these temperatures (Figure \ref{fig:consttemp} shows a comparison of the analytic and numeric results).

\subparagraph{\textbf{CLS Temperature ($T_\mathrm{CLS}$).}} The CLS peak temperature can be estimated at each shock radius using the shock internal energy density corresponding to the instantaneous value of the CLS velocity difference $\Delta\tilde{v}_\mathrm{CLS}$
\begin{equation}
    \label{eq:T_CLS}
    T_\mathrm{CLS}=\epsilon_\mathrm{pl}\frac{2}{3(\gamma+1)^2}m_p (\frac{\Delta\tilde{v}_\mathrm{CLS}}{1-\Delta\tilde{v}_\mathrm{DS}}v_\mathrm{bo})^2 \overset{\tilde{r}_\mathrm{sh}\gg 1}{\rightarrow} 56v_9^2~\mathrm{keV}.
\end{equation}
The CLS temperature increases quickly, and at breakout, it reaches about half of its terminal value, as shown in Figure \ref{fig:consttemp}. It is independent of $R_\mathrm{bo}$.

\subparagraph{\textbf{Upstream Temperature ($T_\mathrm{US}$).}} The upstream equilibrium temperature at each radius is determined by balancing the Compton heating with the Bremsstrahlung cooling. Equating the two expressions in eq. (\ref{eq:bolradmat}), using the radiation energy density eq. (\ref{eq:e_r}), and solving the resulting quadratic equation yields \citep{liedahl_x-ray_1999}
\begin{equation}
\label{eq:T_US}
\begin{split}
    \frac{T_\mathrm{eq}}{T_\mathrm{r}}&=1+\frac{T_\mathrm{c1}}{2 T_\mathrm{r}}\left(1-\sqrt{1+\frac{4 T_\mathrm{r}}{T_\mathrm{c1}}}\right),\\
    T_\mathrm{c1}&\equiv\frac{2\alpha_e^2}{\pi^3}\frac{(m_e c^2)^3}{(e_\mathrm{r}/n_e)^2}\approx
    \begin{cases}
    16\tilde{r}^2v_9^{-4}~\mathrm{eV}&\tilde{r}<\tilde{r}_\mathrm{ph}\\
    14\times10^3v_9^{-6}~\mathrm{eV}&\tilde{r}_\mathrm{ph}<\tilde{r}\\
    \end{cases}.
\end{split}
\end{equation}
The plasma temperature at the immediate upstream, $T_\mathrm{US}$, is obtained by evaluating the equilibrium temperature at $\tilde{r}_\mathrm{sh}$.
If the radiation temperature exceeds the critical temperature $T_\mathrm{c1}$, the plasma equilibrium temperature is close to the radiation temperature $T_\mathrm{eq}=T_\mathrm{r}(1-\sqrt{T_\mathrm{c1}/T_\mathrm{r}})$ (although the spectrum is far from thermal). Conversely, if the radiation temperature is below the critical value, the plasma equilibrium temperature is much lower, approaching $T_\mathrm{eq}=T_\mathrm{r}^2/T_\mathrm{c1}$. We show these two limits of eq. (\ref{eq:T_US}) in Figure (\ref{fig:T_eq}) of Appendix \S~\ref{sec:ionization} (where we also show that for low radiation temperatures, photo-ionization heating dominates over Compton heating, resulting in a different $T_\mathrm{eq}$). 

The evolution of the radiation spectrum and temperature is discussed below, where we provide an analytic description only in limiting regimes. To estimate the upstream temperature during the whole evolution (shown in Figure \ref{fig:consttemp}), we use the numeric radiation temperature.

\subparagraph{\textbf{Thermal Equilibrium Temperature ($T_\mathrm{TE})$.}}

The plasma density at the dense shell formed downstream is sufficiently high for Bremsstrahlung to dominate both absorption and emission, yielding a thermal radiation spectrum at low photon energies, for which the dense shell's effective absorption optical depth is larger than unity (recall that the Bremsstrahlung opacity decreases with photon energy). The dense shell temperature $T_\mathrm{TE}$ is therefore approximately given by the temperature for which the Bremsstrahlung opacity for photons with energy $\sim T_\mathrm{TE}$ is $\sim\tau^2$ times smaller than the Thomson opacity (where $\tau$ is the Thomson optical depth between the shell and the shock), enabling efficient absorption within the dense shell, before escaping the system. This yields, for $T_\mathrm{r}\gg T_\mathrm{TE}$
\begin{equation}
\label{eq:T_TE}
    \begin{split}
    T_\mathrm{TE}=& f_\mathrm{TE} \tau^{1/2}\left(\frac{T_\mathrm{r}}{m_e c^2}h^3c^3e_\mathrm{r,sh}\right)^{1/4}\\
    \approx&5\left(\frac{T_\mathrm{r}}{25~\mathrm{eV}}\right)^{1/4}\tilde{r}_\mathrm{sh}^{-5/4}R_{14}^{-1/4}v_9^{-1/4}\mathrm{eV}.
    \end{split}
\end{equation}
Here, $f_\mathrm{TE}$ is an order unity dimensionless parameter calibrated numerically, incorporating the required ratio of Compton heating to Bremsstrahlung absorption of the plasma for an approximated thermal equilibrium (corresponding to some required high density). Using  $f_\mathrm{TE}=0.35$ yields a good agreement with the numeric results for different $\tilde{r}_\mathrm{sh}$, $R_\mathrm{bo}$, and $v_\mathrm{bo}$ (see Figures \ref{fig:consttemp}, \ref{fig:T_comp}).

The dense shell reprocesses some fraction of the high energy photon luminosity generated near the CLS into lower energy, $\approx T_\mathrm{TE}$ photon luminosity. This is significant mostly for lower shock velocities, $v_9\lesssim0.8$ (see Figure \ref{fig:constspect}, and the analytic explanation of the spectral evolution below in this subsection). We note that the dense shell may be subject to 3D instabilities, which are not captured by our 1D calculations and may reduce the shell's density, thus also reducing the efficiency of high energy radiation reprocessing. Analytic stability analyses \citep[e.g.][]{chevalier_hydrodynamic_1995,duffell_one-dimensional_2016}, which do not include the contribution of radiation and magnetic fields to the pressure and entropy, find instability development that significantly reduces the shell's density, while multidimensional numeric calculations including radiation \citep[e.g.][]{suzuki_supernova_2019} report limited density reduction with unaffected light curves.

\subparagraph{\textbf{Post CLS Cooling Mechanism}.} Plasma heated by the CLS to $T_\mathrm{CLS}$, eq. (\ref{eq:T_CLS}), cools through Bremsstrahlung emission and inverse Compton scattering. It is useful to define the emissivity ratio of these processes at the shock radius using eq. (\ref{eq:bolradmat}), with $T_\mathrm{CLS}\gg T_\mathrm{r}$ \citep[see also][]{chevalier_x-rays_2012,svirski_optical_2012}
\begin{equation}
\label{eq:Q}
Q_\mathrm{Brem}\equiv\dot{e}_\mathrm{Brem,sh}/\dot{e}_\mathrm{Comp,sh}\approx0.1 \tilde{r}_\mathrm{sh}v_9^{-3}.
\end{equation}
Compton cooling dominates at high velocities and near breakout, while Bremsstrahlung cooling becomes increasingly dominant at larger radii. $Q$ is related to the critical temperature defined in eq. (\ref{eq:T_US}) by $Q_\mathrm{Brem}=\sqrt{T_\mathrm{c1}/T_\mathrm{CLS}}(\gamma+1)/(\gamma-1)$, and is independent of $R_\mathrm{bo}$.

\subparagraph{\textbf{
Shock Cooling Efficiency}.} The cooling time is
\begin{equation}
\label{eq:t_cool}
t_\mathrm{cool}\equiv\frac{e_\mathrm{pl,sh}}{\dot{e}_\mathrm{Comp,sh}+\dot{e}_\mathrm{Brem,sh}}\approx 10^{2.5}\frac{\tilde{r}_\mathrm{sh}^3 R_{14}/v_9}{1+Q_\mathrm{Brem}}~\mathrm{s},
\end{equation}
and is a small fraction of the dynamical time
\begin{equation}
\label{eq:rad_eff}
    t_\mathrm{cool}/t_\mathrm{dyn}\approx 10^{-2.5}\frac{\tilde{r}_\mathrm{sh}^2}{1+Q_\mathrm{Brem}}.
\end{equation}
Thus, the shock is radiative up to the photosphere for all velocities considered, justifying eq. (\ref{eq:luminosity}).

\subparagraph{\textbf{Hot Layer Optical Depth ($\tau_\mathrm{hot}$).}} The Thomson optical depth of the hot layer downstream of the CLS, which we define as extending inward to the location where the plasma temperature drops by an order of magnitude, can be estimated using the cooling time\footnote{A better treatment would track the evolution of the relevant quantities (e.g. plasma density and temperature, and the background radiation density and temperature) which change during cooling. The resulting corrections are more significant at larger radii, where the cooling time exceeds the timescale over which these quantities evolve significantly. However, this approximation suffices and does not alter the qualitative picture.}
\begin{equation}
\label{eq:tau_CLS}
\tau_\mathrm{hot}\approx f_\tau\rho_\mathrm{sh}\kappa_\mathrm{T} v_\mathrm{bo}t_\mathrm{cool}
        \approx
        0.2\frac{\tilde{r}_\mathrm{sh}/v_9}{1+Q_\mathrm{Brem}}.
\end{equation}
Here, $f_\tau$ is a dimensionless parameter calibrated by the numeric calculations. Using $f_\tau=2$ yields a good agreement with the numeric results for different $\tilde{r}_\mathrm{sh}$, $R_\mathrm{bo}$, and $v_\mathrm{bo}$. The optical depth of the hot layer increases linearly with shock radius at high velocities while remaining approximately constant at low velocities (where $Q_\mathrm{Brem}$ is significant). It is independent of $R_\mathrm{bo}$.

\subparagraph{\textbf{Spectral Evolution}.} In Appendix \ref{sec:ionization}, we show that photons escaping far upstream are not significantly absorbed or degraded in energy as they propagate through the wind, such that the spectrum at the shock position determines the emitted spectrum. While the bolometric radiation energy density is independent of the details of the radiation processes (as long as they remain efficient), the spectral shape depends on how the shock-heated plasma cools and on subsequent interactions of the emitted radiation with the downstream plasma. The Compton cooling spectrum depends on the background radiation spectrum, whereas the Bremsstrahlung cooling spectrum is independent of it, eq. (\ref{eq:radmat}). Thus, we may expect a different spectral shape at different $\tilde{r}_\mathrm{sh}$, $R_\mathrm{bo}$, and $v_\mathrm{bo}$. We now describe the overall spectral evolution at the considered parameter range, using approximations that are derived below for the limits of pure Compton/Bremsstrahlung cooling.

At the RMS stage ($\tilde{r}_\mathrm{sh}<\tilde{r}_\mathrm{CLS}$), the spectrum is well described by a thermal/Comptonized-Wien distribution with temperature $T_\mathrm{RMS}$, eq. (\ref{eq:T_RMS}), see $\tilde{r}_\mathrm{sh}=0.3$ of Figure \ref{fig:constspect}. At the CLS stage, qualitatively different behavior is obtained for $v_9\gtrsim1.3$ and $v_9\lesssim1.3$.

For $v_9\gtrsim1.3$, $Q_\mathrm{Brem}<1$, eq. (\ref{eq:Q}) holds up to the photosphere, implying that the post-shock cooling is always dominated by Compton scatterings and the spectrum is approximately described by unsaturated Comptonization with energies increasing significantly on a few breakout timescales. An analytic approximation for the resulting spectrum in this case is derived below, eq. (\ref{eq:unsaturated}), and compared to the $v_9=2$ numeric results in Figure \ref{fig:constspect} (the validity of the approximation is further supported by static numeric calculations discussed in Appendix \ref{sec:static}).

For $v_9\lesssim1.3$, $Q_\mathrm{Brem}$ increases with shock radius, leading to a transition from an unsaturated Comptonization spectrum to a Bremsstrahlung cooling spectrum. An analytic approximation for the Bremsstrahlung cooling spectrum is derived below, eq. (\ref{eq:brem}).\footnote{The transition occurs before $Q_\mathrm{Brem}$ exceeds unity for three main reasons: (1) When $Q_\mathrm{Brem}$ is not much smaller than unity upscattered photons already carry much less energy; (2) The plasma density increases downstream of the shock, causing $Q_\mathrm{Brem}$ to be larger during part of the cooling process, resulting in ``early" Bremsstrahlung cooling at intermediate temperatures; (3) The number of thermal photons originating from the dense shell that is formed behind the shock, eq. (\ref{eq:T_TE}) and which ``serves" as the soft photon input of the unsaturated Comptonization, decreases, implying that the spectrum resembles the Bremsstrahlung cooling shape even if the shocked plasma predominantly cools via Compton scattering (Compton cooling occurs predominantly by scattering photons emitted from the hot shock region since most of the energy is deposited in these more energetic photons).} This transition is evident in the $v_9=1$ numeric results shown in Figure \ref{fig:constspect}. We lack an analytic approximation for the spectral shape during the transition. After the transition, the spectral shape is dominated by Bremsstrahlung cooling, with a thermal radiation component originating in the dense shell, $B_\varepsilon(T_\mathrm{TE})$, eq. (\ref{eq:T_TE}). This thermal radiation diminishes with increasing shock radius, and the spectrum approaches a pure Bremsstrahlung cooling shape after a few breakout times. Figure \ref{fig:constspect} shows a good agreement between the analytic approximations for the Bremsstrahlung cooling and thermal components of the spectrum with the numeric results (the validity of the approximation is further supported by static numeric calculations discussed in Appendix \ref{sec:static})\footnote{The ``leftover" Bremsstrahlung cooling component deviates initially from the derived shape because energetic photons lose more energy via Compton downscatterings near the dense shell. Only at larger shock radii does the spectrum converge to the derived Bremsstrahlung cooling shape. The thermal component shape also deviates from a Planck distribution when Compton cooling is not entirely negligible.}.

In both regimes, the spectrum becomes X-ray-dominated after a few breakout times. While the rise time and luminosity scales are dependent on $R_\mathrm{bo}$ and $v_\mathrm{bo}$ (\S~\ref{subsec:Wind+CLS}), we find that the spectral shape is primarily sensitive to $v_\mathrm{bo}$, and is only weakly dependent on $R_\mathrm{bo}$ (see also \S~\ref{subsec:envelope_spectrum}) through $T_\mathrm{RMS}$ and $T_\mathrm{TE}$.

Finally, we derive below the approximations for the spectral shapes in the limiting regimes of Bremsstrahlung and Compton cooling dominance.

\subparagraph{\textbf{Bremsstrahlung Cooling Spectrum}.} We consider the regime where Bremsstrahlung dominates the cooling, and Compton energy loss is negligible ($Q_\mathrm{Brem}\gg1$). A plasma at temperature $T_\mathrm{CLS}$, cooling solely via Bremsstrahlung, eq. (\ref{eq:radmat}) and (\ref{eq:bolradmat}), to a much lower temperature\footnote{The exact low-temperature limit can be derived similarly to eq. (\ref{eq:T_US}). The spectral shape is not sensitive to its value since most of the emission takes place at high temperatures.}, emits photons with a spectral shape
\begin{equation}
\label{eq:brem}
\begin{split}
e_{\mathrm{Brem},\varepsilon}&\approx\int_{T_\mathrm{CLS}}^0\frac{dT}{\dot{T}}\dot{e}_{\mathrm{Brem},\varepsilon}\approx\int_{T_\mathrm{CLS}}^0dT\frac{\dot{e}_{\mathrm{Brem},\varepsilon}}{\dot{e}_\mathrm{Brem}}\\
    &\approx\int_{\varepsilon/2T_\mathrm{CLS}}^\infty dx\frac{e^{-x}K_0(x)}{x},
\end{split}
\end{equation}
where the spectral Bremsstrahlung Gaunt factor is calculated in the Born approximation, and $K_0$ is the zeroth modified Bessel of the second kind. The average photon energy, eq. (\ref{eq:T_rad}) can be solved analytically and is $\bar{\varepsilon}=T_\mathrm{CLS}/3$. Note that the spectral shape and average differ from the thermal Bremsstrahlung spectrum emitted from a plasma of constant temperature. When Bremsstrahlung cooling dominates, and $T_\mathrm{CLS}$ is given by $\approx56v_9^2~$keV, eq. (\ref{eq:T_CLS}), the average photon energy is
\begin{equation}
\label{eq:T_brem}
    \bar{\varepsilon}_\mathrm{brem}\approx 19v_9^2~\mathrm{keV}.
\end{equation}
The spectral shape agrees well with the numeric results for the spectrum at intermediate and low velocities ($v_9\lesssim 1.2$) at late times, as shown in Figure \ref{fig:constspect} (the validity of the approximation is further supported by static numeric calculations discussed in Appendix \ref{sec:static}).

\subparagraph{\textbf{Unsaturated Comptonization Spectrum}.} For $v_9\gtrsim1.3$, we have $Q_\mathrm{Brem}<1$ all the way to the photosphere, and cooling is always dominated by Compton emission. However, this does not necessarily imply that Compton cooling dominates the spectral shape, as this depends on the Compton parameter of the medium. 

We adapt the solution for unsaturated Comptonization of a soft photon input \citep{shapiro_two-temperature_1976,rybicki_radiative_1979,fransson_x-ray_1982} to our problem. For a steady-state input of soft photons, up to energy $\varepsilon_\mathrm{soft}$, injected into a finite optically thick uniform medium with a moderate $y\equiv\frac{4T}{m_e c^2}\tau^2$ Compton parameter (not high enough to saturate to a Comptonized-Wien spectrum), an analytic solution of the Kompaneets equation, eq. (\ref{eq:radmat}) is available in two regions\footnote{The exact solution near $\varepsilon\sim T$ is available analytically only for the case $y=1$ \citep{shapiro_two-temperature_1976}.}
\begin{equation}
\begin{split}
e_{\mathrm{Comp},\varepsilon}&\propto
    \begin{cases} \varepsilon^{-\alpha(y)}&\varepsilon_\mathrm{soft}\ll \varepsilon\ll T\\
  e^{-\varepsilon/T} & T\ll\varepsilon
\end{cases},\\
\alpha(y)&=\sqrt{9/4+4/y}-3/2.
\end{split}
\end{equation}
For $y>1$, we have $0<\alpha<1$, implying that most of the energy, $\varepsilon e_\varepsilon$, is carried by a few ``lucky" photons that were upscattered many times. For $y<1$, the power law is steeper, and Compton scatterings do not boost the photons energy significantly (For $y\gg1$, the saturated regime, a different solution applies).

In our case, the medium is not uniform, and the steady-state assumption does not hold (Figure \ref{fig:consttemp}). However, we can make some crude approximations to understand the physical picture. We treat the medium roughly as a narrow hot layer with a temperature $T_\mathrm{CLS}$ eq. (\ref{eq:T_CLS}) and optical depth $\tau_\mathrm{hot}$ eq. (\ref{eq:tau_CLS}), embedded in a larger optical depth $\tau_\mathrm{sh}$ of colder plasma. We further assume that scatterings in the hot layer are approximately $\tau_\mathrm{hot}/\tau_\mathrm{sh}$ less likely\footnote{This is a rough estimate that neglects correlations between scatterings. More detailed treatments exist \citep[e.g.][and subsequent works]{sobolev_number_1966} but are beyond the scope of this paper.}. Instead of scattering $\tau_\mathrm{sh}^2$ times, photons scatter in the hot layer $\tau_\mathrm{sh}\tau_\mathrm{hot}$ times.

Considering the thermal photons emitted at $\varepsilon_\mathrm{soft}\approx 3T_\mathrm{TE}$ as the source for Comptonization, the spectrum can be approximated as
\begin{equation}
\label{eq:unsaturated}
\begin{split}
    e_{\mathrm{Comp},\varepsilon}&\approx
    \begin{cases}
        B_\varepsilon\left(T_\mathrm{TE}\right)& \varepsilon<3T_\mathrm{TE}\\
        B_{3T_\mathrm{TE}}\left(T_\mathrm{TE}\right)\left(\frac{\varepsilon}{3T_\mathrm{TE}}\right)^{-\alpha(y)}e^{-\frac{1-\alpha(y)}{4}\frac{\varepsilon}{T_\mathrm{r}}}&3T_\mathrm{TE}<\varepsilon
    \end{cases},
    \\
        y&=\frac{4T_\mathrm{CLS}}{m_e c^2}\tau_\mathrm{sh}\tau_\mathrm{hot}\approx\frac{2}{1+Q_\mathrm{Brem}}.
    \end{split}
\end{equation}
$y$ depends only on $Q_\mathrm{Brem}$ and is of order unity for small $Q_\mathrm{Brem}$. This implies that for Compton-dominated cooling, most of the energy is carried by soft photons upscattered in the hot layer. $v_9=2$ of Figure \ref{fig:constspect} shows agreement between the numeric and analytic spectral shapes (the validity of the approximation is further supported by static numeric calculations discussed in Appendix \ref{sec:static}).

The spectral cut-off energy, beyond which the energy density drops exponentially, determines the radiation temperature $T_{\rm r}$ (ignoring the soft photon part). Since a finite time is required for photons to upscatter in the hot layer and approach the steady-state energy distribution, we expect the temperature to increase as
\begin{equation}
\label{eq:T_unsat}
    T_\mathrm{r}(t)\approx T_\mathrm{RMS} e^{f_\mathrm{r}y \tilde{t}},
\end{equation}
during the RMS-CLS transition. Here, $f_\mathrm{r}$ is an order unity dimensionless factor calibrated by the numeric results. Using $f_\mathrm{r}=0.3$ yields a good agreement with the numeric results (see Figure \ref{fig:constspect}) for different $R_\mathrm{bo}$ and $v_\mathrm{bo}$. It takes a few breakout times for the radiation spectrum to shift to high energy since $y$ is close to unity. At longer times, the steady-state cut-off energy would be, in fact, smaller than $T_\mathrm{CLS}$, as the shock nears the photosphere, where photons escape more easily, limiting their maximum achievable temperature. The number of $T_\mathrm{TE}$ photons that were not significantly upscattered is determined by the total radiation energy density, eq. (\ref{eq:e_r}).

\section{Polytropic Envelope Driven Shock}
\label{sec:ejecta_results}

In this section, we investigate the case of a shock wave driven into the wind by an expanding envelope with a polytropic pre-shock density profile. The initial and inner-boundary conditions are described in \S~\ref{subsec:envelope_initial}. The forward-reverse shock structure resulting from the envelope-wind ``collision" is described analytically in \S~\ref{subsec:envelope_forward}. We show that the forward shock dominates the emission and decelerates slowly, implying that the main characteristics of the flow and radiation field are similar to those obtained for the constant velocity piston case. This enables conclusions to be drawn from the constant velocity piston results and encourages us to adopt the same methods and interpretation regarding the processes that shape the resulting radiation spectrum. The numeric plasma profiles and radiation spectral evolution are presented in \S~\ref{subsec:envelope_plasma} and in \S~\ref{subsec:envelope_spectrum}, respectively.

\subsection{Initial \& Inner-Boundary Conditions}
\label{subsec:envelope_initial}

For a polytropic stellar envelope, the density near the stellar edge (where the enclosed mass is nearly constant) can be approximated as a power law of the distance from the edge \citep[e.g.][]{matzner_expulsion_1999}. The initial density profile of the envelope-wind is
\begin{equation}
\label{eq:rho_env}
\rho_0(r)=
\begin{cases} \frac{M_\star}{R_\star^3}\left(\frac{R_\star-r}{R_\star}\right)^{3/2}&r<\hat{R}_\star\\
  \rho_\mathrm{bo}\tilde{r}^{-2} & \hat{R}_\star<r
\end{cases},
\end{equation}
where $M_\star$ and $R_\star$ are the star mass and radius, and $\hat{R}_\star$ is the radius of an equal density of the envelope and the wind, in practice very close to $R_\star$.

The inner boundary is set within the envelope at a depth beyond which the envelope mass is 4 times the wind mass up to the photosphere, $M_\mathrm{wind}=c/v_\mathrm{bo}M_\mathrm{bo}\approx0.14R_{14}^2v_9^{-1}M_\odot$. We have verified that the results are not sensitive to increasing the envelope layer mass included in the calculations to 20 times the wind mass. The RMS shock is initiated at 30 shock widths ($\Delta\tau\sim c/v$) inward the envelope, with an initial shock velocity $\tilde{v}_\mathrm{sh,0}=f_v(M_\star/M_\mathrm{bo})^{\lambda/(n+1)}$ corresponding to an eventual shock velocity $v_\mathrm{bo}$ at $R_\mathrm{bo}$, see \S~\ref{subsec:envelope_forward}. $f_v=0.7$ is a dimensionless parameter calibrated by the numeric calculation and was tested to be insensitive to different $R_\mathrm{bo}$, $v_\mathrm{bo}$. The initial RMS hydrodynamical profiles are taken from \cite{sapir_non-relativistic_2011}, and the acceleration of the inner envelope boundary is taken from the hydrodynamical post-shock adiabatic expansion solution of \cite{matzner_expulsion_1999}. For the $R_{\rm bo}\gg R_\star$ limit considered here, the radiation diffusing from the expanding shock-heated envelope does not affect the plasma and radiation field evolution at breakout; see Appendix \ref{sec:luminosity_contributions}.

The rest of the initial conditions are similar to those used in the constant velocity piston case, \S~\ref{subsec:piston_initial}.

\subsection{Forward Shock Domination \& Deceleration}
\label{subsec:envelope_forward}

As the shock approaches the stellar surface (or, in this case, the envelope-wind boundary), it accelerates and the flow approaches the planar self-similar solutions of \citet{gandelman_shock_1956} and \citet{sakurai_problem_1960}. The shock velocity diverges near the edge as
\begin{equation}
\label{eq:sakurai}
    v_\mathrm{sh}(r)\approx v_\star\left(\frac{R_\star-r}{R_\star}\right)^{-\lambda},
\end{equation}
where $v_\star$ is the velocity scale set by the explosion energy and envelope mass, and $\lambda\approx0.19n$ \citep{grasberg_self-similar_1981}.

As the shock propagates into the wind, it decelerates due to the increasing accumulated mass. This leads to the formation of a reverse shock that penetrates into the expanding envelope and decelerates it. Before the reverse shock re-shocks a mass shell of the ejecta, the shell continues to expand at roughly twice  \citep[see][]{matzner_expulsion_1999} the velocity it had when the original shock had reached the edge, eq. (\ref{eq:sakurai}). The expanding envelope is thus described, prior to being re-shocked, by a homologous expansion with a velocity distribution that is constant in time as a function of mass, $v_\mathrm{ej}\propto m^{-\lambda/(n+1)}$ (integrating the envelope density eq. (\ref{eq:rho_env})).

An envelope shell of a certain mass $m$ is decelerated once the wind mass accumulated by the forward shock is comparable to $m$. The forward shock's velocity is thus approximately given by \citep{chevalier_self-similar_1982}
\begin{equation}
\label{eq:v_ej}
    v_\mathrm{sh}(m)\approx v_\mathrm{ej}(m)\approx v_\mathrm{bo}\left(\frac{m}{M_\mathrm{bo}}\right)^{-\lambda/(n+1)}.
\end{equation}
Since the wind mass is linear with shock radius, the deceleration with radius or time is\footnote{When the reverse shock penetrates deeper into the ejected envelope, there will be corrections to the velocity and luminosity scaling due to deviations from the planar self-similar solution. These corrections would be small, provided that the wind mass is small compared to the envelope mass.}
\begin{equation}
\label{eq:_v_sh}
    \tilde{v}_\mathrm{sh}\approx\tilde{r}_\mathrm{sh}^{-\lambda/(n+1)}\approx\tilde{t}^{-\lambda/(\lambda+n+1)}\approx\tilde{t}^{-0.1}.
\end{equation}
The weak dependence of the envelope shells' velocity on their mass implies a weak dependence of the forward shock velocity on radius. The decreasing shock velocity implies a decreasing luminosity produced by the forward shock, $L_\mathrm{sh}\propto\tilde{t}^{-0.3}$.

The rate at which internal energy is generated by the reverse shock is much smaller than that generated by the forward shock. Using the thin shell approximation \citep[e.g.][]{chevalier_supernova_2003}, the ratio of energy production rates by the reverse and forward shocks is $(p-4)/2(p-3)^2\approx5\%$, where $p\approx(3\lambda+n+1)/\lambda\approx12$ is the power-law of the ejecta density as a function of radius, $\rho_\mathrm{ej}\propto r^{-p}$.

\subsection{Plasma Profiles}
\label{subsec:envelope_plasma}

\begin{figure}
    \centering
    \includegraphics[height=14cm]{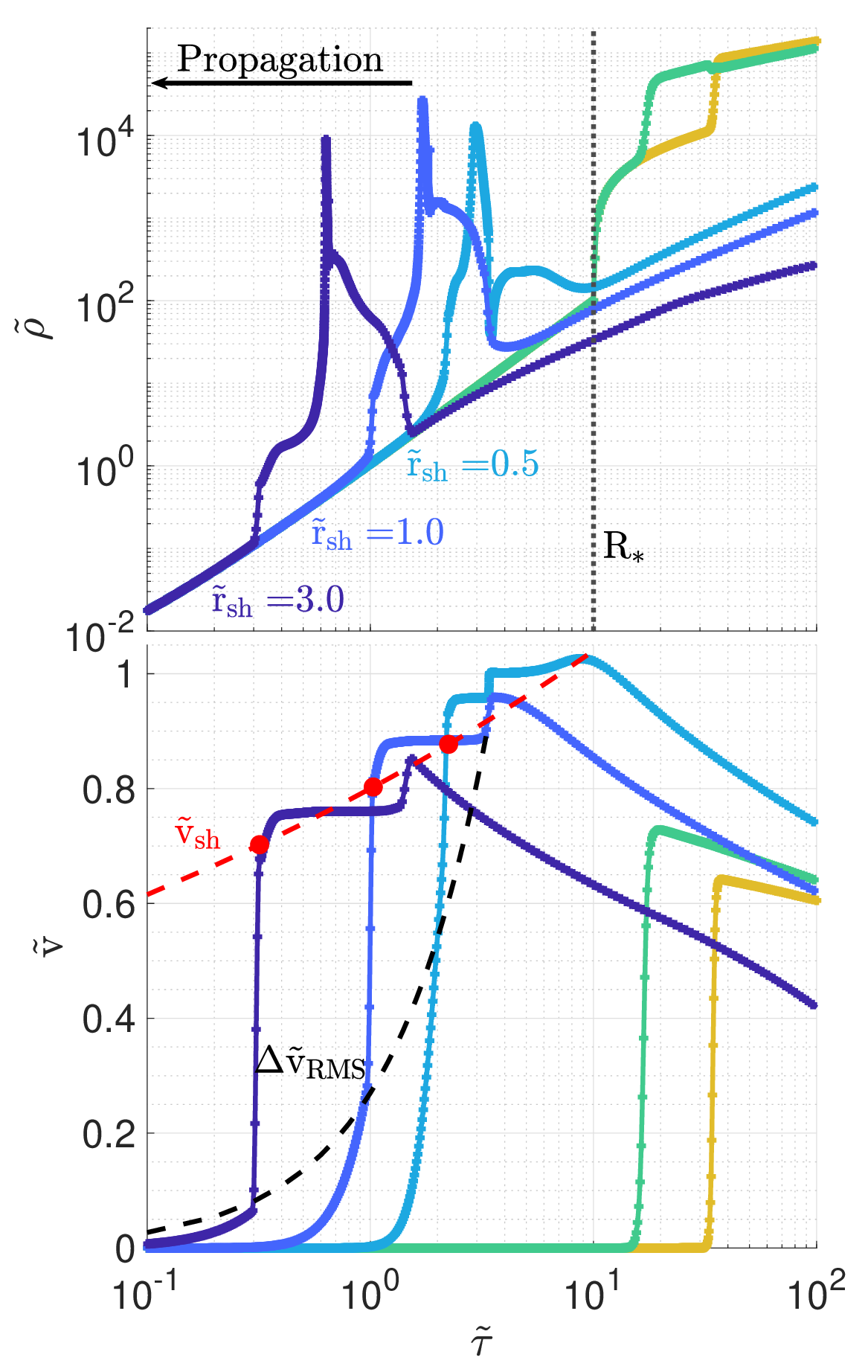}
    \caption{Plasma density (\textit{upper panel}) and velocity (\textit{lower panel}) profiles as a function of optical depth at different shock radii, normalized to their breakout value. We present calculations for star radius $R_\star=0.1R_\mathrm{bo}$ and mass $M_\star=10M_\odot$, which become independent of these parameters as the shock propagates beyond a few $R_\star$. The shock deceleration is then well-fitted by eq. (\ref{eq:_v_sh}) (red dashed). During the RMS-CLS transition, the RMS velocity difference, $\Delta\tilde{v}_\mathrm{RMS}$, is described analytically by eq. (\ref{eq:v_RMS}) (black dashed).}
    \label{fig:rhov}
\end{figure}

The plasma density and velocity profiles at different shock radii are shown in Figure \ref{fig:rhov}. When normalized to their breakout value, eq. (\ref{eq:normalization}), these profiles are nearly independent of $R_\mathrm
{bo}$ and $v_\mathrm{bo}$. The transition from RMS to CLS follows a path similar to that of the constant velocity piston case (see the definition and derivation of $\Delta\tilde{v}_\mathrm{RMS}$, eq. (\ref{eq:v_RMS}), in \S~\ref{subsec:piston_results}). The velocity profile is nearly uniform within the shocked wind and re-shocked ejecta, with the velocity deceleration well described by eq. (\ref{eq:_v_sh}). 

The density structure (Figure \ref{fig:rhov}) is consistent with the one obtained in the self-similar analysis of the forward-reverse shock of a uniformly expanding power-law density profile gas that moves into a stationary power-law density profile \citep{chevalier_self-similar_1982}, including the compression at the ejecta-wind interface. As in the constant velocity piston case, the exact peak density value depends on grid resolution but does not affect the flow dynamics and the radiation field. 

\subsection{Radiation Evolution}
\label{subsec:envelope_spectrum}

\begin{figure}[ht]
    \centering
    \includegraphics[height=6.5cm]{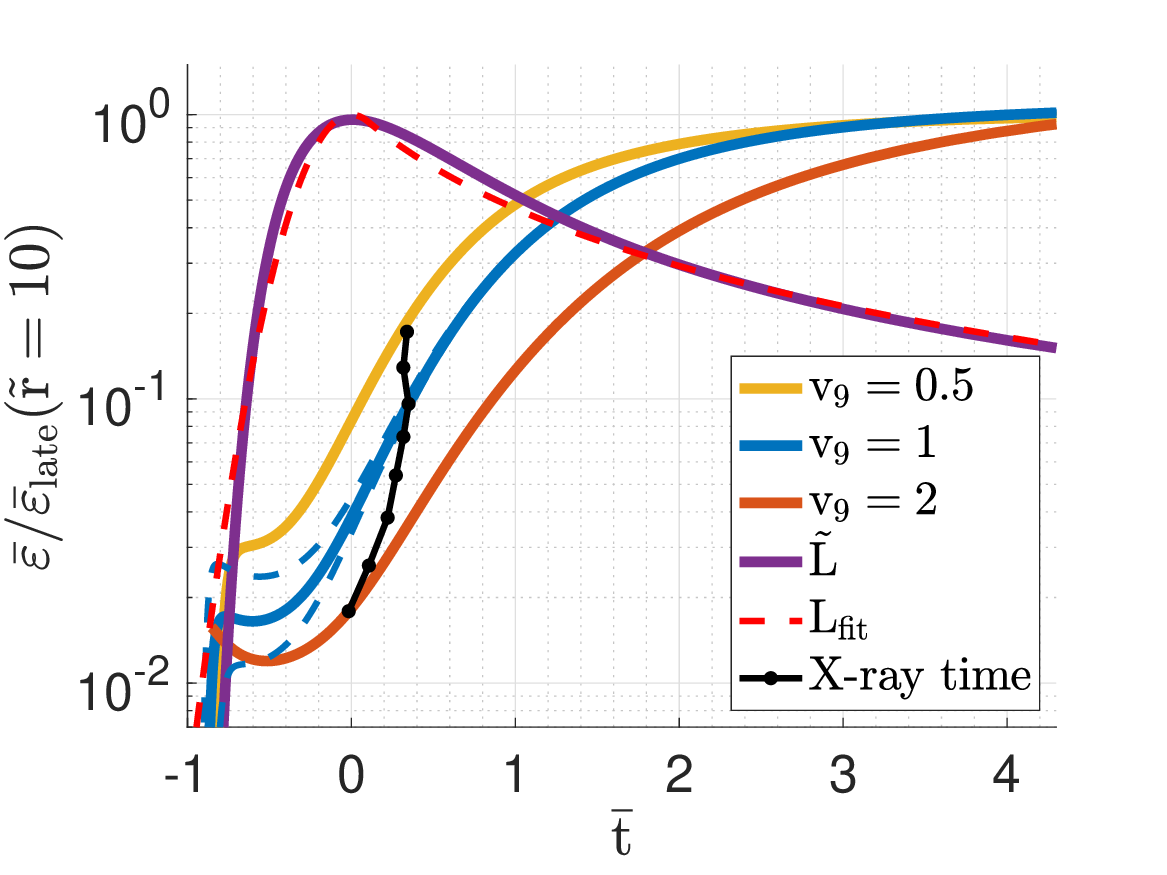}
    \caption{Average photon energy (energy-weighted, eq. (\ref{eq:T_rad})) emitted from the photosphere, normalized to $\bar{\varepsilon}_\mathrm{sh,late}(\tilde{r}_\mathrm{sh}=10)$ of eq. (\ref{eq:typical_energy}), as a function of photospheric time, defined by eq. (\ref{eq:photospheric_time}), for $R_{14}=1$ and $v_9=0.5,1,2$. For $v_9=1$, we also show the results for $R_{14}=0.1,10$ (dashed blue lines). The purple line shows the bolometric luminosity curve, normalized to $L$ given by eq. (\ref{eq:luminosity}), with a red dashed line of an analytic fit of a Crystal Ball function (see eq.~(\ref{eq:photospheric_time}) and the text following it). The black line shows the ``X-ray time", after which $>1~$keV photons compose a significant ($>10\%$) part of the luminosity, where the black dots correspond to  $v_9=0.6-2$ with a step size of $0.2$.}
    \label{fig:t_ph}
\end{figure}

\begin{figure*}[ht]
    \centering
    \includegraphics[height=5.5cm]{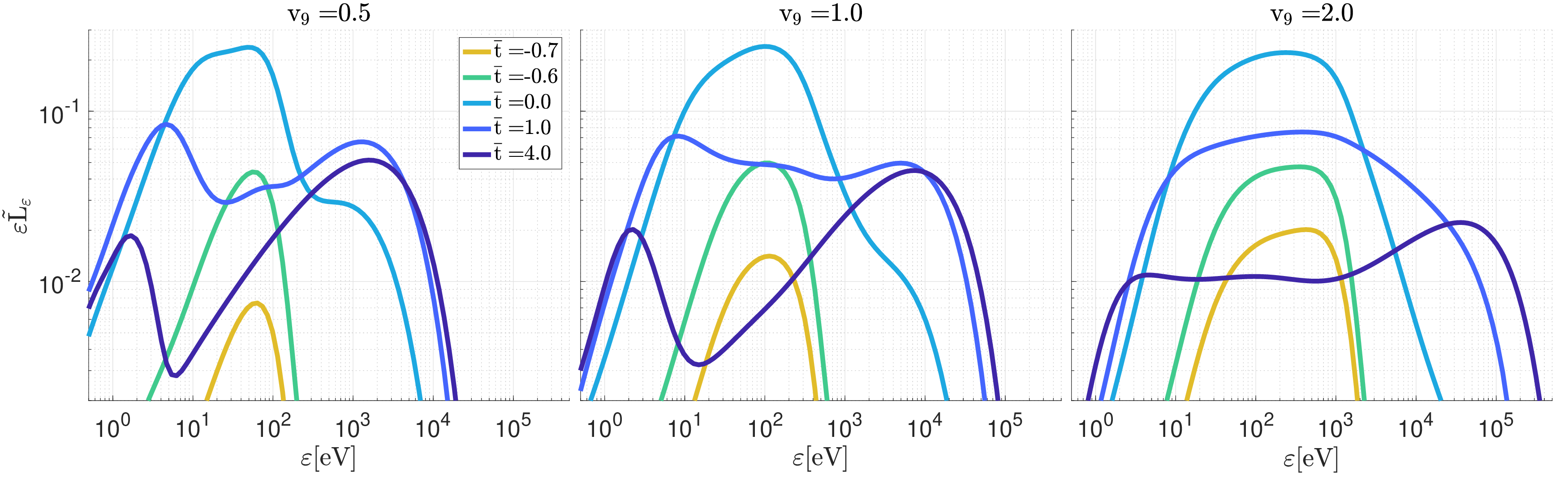}
    \caption{Spectral luminosity, $L_\varepsilon\equiv\partial L/\partial \varepsilon$, emitted from the photosphere, normalized to $L$ given by eq. (\ref{eq:luminosity}), at different photospheric times, defined by eq. (\ref{eq:photospheric_time}). Results are shown for $R_{14}=1$ and $v_9=0.5,1,2$.}
    \label{fig:luminosity}
\end{figure*}

Figure $\ref{fig:t_ph}$(\ref{fig:luminosity}) shows the average photon energy and the bolometric (spectral) luminosity, emitted from the photosphere, as a function of photospheric time (defined by eq. (\ref{eq:photospheric_time}) below), for $v_9=0.5,1,2$. Figure \ref{fig:xray} shows the average photon energy at the shock at $\tilde{r}_\mathrm{sh}=10$, and the shock radius beyond which $>1~$keV photons compose a significant, $>10\%$, part of the shock energy spectrum, termed ``X-ray radius", for different $v_\mathrm{bo}$.

\begin{figure}[ht]
    \centering
    \includegraphics[height=6.5cm]{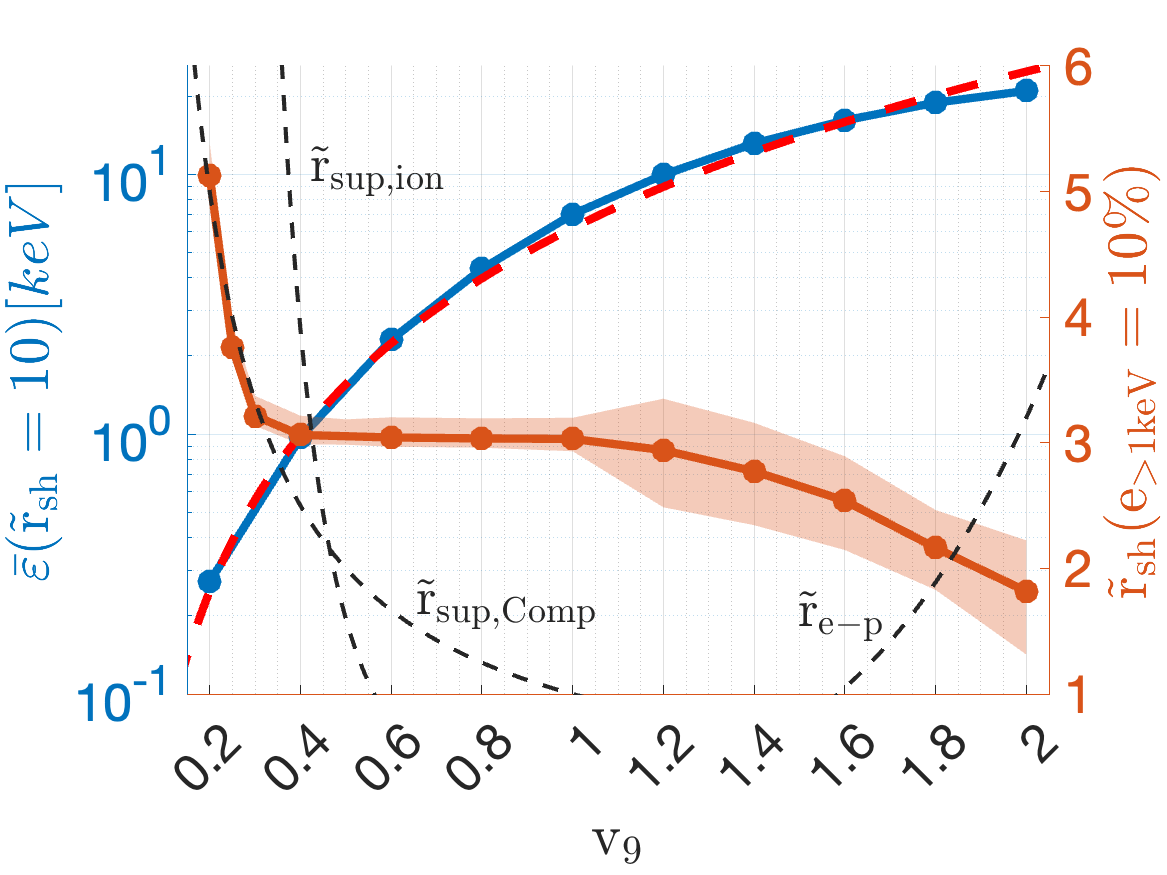}
    \caption{Solid lines show the average photon energy at the shock $\bar{\varepsilon}_\mathrm{sh}$ (energy-weighted, eq. (\ref{eq:T_rad})) at $\tilde{r}_\mathrm{sh}=10$ (\textit{left axis}), and the ``X-ray radius", defined in the text (\textit{right axis}), as a function of $v_\mathrm{bo}$. $\bar{\varepsilon}_\mathrm{sh}(10)$ is insensitive to $R_\mathrm{bo}$, with variations of the order of the plotted line width, and the X-ray radius is weakly sensitive to $R_\mathrm{bo}$, with the shaded area representing a spread of $10^{13}-10^{15}~$cm. $\bar{\varepsilon}_\mathrm{sh}(10)$ is well described by eq. (\ref{eq:typical_energy}) (red dashed curve). Also shown in black dashed lines (corresponding to right axis): the shock radii beyond which there is negligible upstream UV--X-ray energy absorption (by photo-ionization or Compton scattering, eqs.~(\ref{eq:r_rep,ion}), (\ref{eq:r_rep,comp})) and the radius beyond which there is electron-proton equipartition at the shock, eq. (\ref{eq:r_ep}).}
    \label{fig:xray}
\end{figure}

The average time it takes photons to diffuse from the shock to the wind Thomson photosphere ($\tilde{r}\approx30v_9^{-1}$) depends on the shock position. At $R_\mathrm{bo}$, it is extended from the diffusion time by a logarithmic correction $\approx t_\mathrm{bo}\ln(c/v_\mathrm{bo})$ \citep{ginzburg_superluminous_2012}. Note that photons emitted from the photosphere at some instance may originate from emission of the shock at different radii, and so the spectrum evolution is smeared over the diffusion time through the wind (the light-crossing time of the photosphere does not introduce another smearing as it is of the same order of the breakout or diffusion time, $R_\mathrm{ph}/c=R_\mathrm{bo}/v_\mathrm{bo}$). To describe a temporal emission from the photosphere, we thus use a ``photospheric time" coordinate shifted relative to peak luminosity time $t_\mathrm{pl}$ and normalized
\begin{equation}
\label{eq:photospheric_time}
    \Bar{t}\equiv\frac{t-t_\mathrm{pl}}{t_\mathrm{bo}\ln(c/v_\mathrm{bo})}.
\end{equation}
The bolometric luminosity as a function of the photospheric time, normalized to $L$ of eq. (\ref{eq:luminosity}), is universal: it is nearly independent of $R_\mathrm{bo}$ and $v_\mathrm{bo}$ (Figure \ref{fig:t_ph}, with variations of the order of $10\%$ for different $R_\mathrm{bo}$, $v_\mathrm{bo}$). It can be usefully fitted by a Crystal Ball function (a continuously differentiable function of a Gaussian core and a power-law tail), with parameters $\alpha=0.33,~n=1.1,~\sigma=0.3$. The universal bolometric luminosity peak is found at $t_\mathrm{pl}/(t_\mathrm{bo}\ln(c/v_\mathrm{bo}))=1.2$, with left width at half maximum of $0.42$, and right width at half maximum of $1.06$.

Unlike radiation breakout from a stellar surface \citep{sapir_non-relativistic_2013} where the radiation temperature increases along with the quick rise in emitted luminosity, in the wind breakout the spectrum evolution is smeared over the diffusion time, and the temperature does not increase by much during the luminosity rise, as can be seen in Figures \ref{fig:t_ph} and \ref{fig:luminosity}. 

As in the constant velocity piston case, the spectral shape changes dramatically with time, hardening during a few breakout times from a thermal/Comptonized-Wien distribution at 10's of eV to a 10's of keV dominated spectrum (Figures \ref{fig:t_ph}-\ref{fig:xray}). We now derive estimations for the average photon energy at late times and for the X-ray radius (or time).

As the shock decelerates slowly, the analytic approximations derived for the constant velocity piston (\S~\ref{subsec:piston_analytic}) remain approximately valid for the polytropic envelope driven shock, provided they are adjusted for the shock velocity decay. Since the shock is decelerating, the Bremsstrahlung to Compton emission ratio $Q_\mathrm{Brem}$, eq. (\ref{eq:Q}) increases faster, and the thermal radiation originating from the dense shell at $T_\mathrm{TE}$ decreases faster, compared to the constant velocity piston case. As a result, even for the highest breakout velocity considered ($v_9=2$), the unsaturated Comptonization spectrum, eq. (\ref{eq:unsaturated}) is present only at early times, with a faster transition to the Bremsstrahlung spectrum, eq. (\ref{eq:brem}). This transition is, however, not complete (see Figure \ref{fig:luminosity}); for high velocities, some Comptonization effects remain, while for lower velocities, the thermal radiation of the dense shell is not as negligible as in the piston case. The average photon energy at the shock can thus be estimated from the analytic result for Bremsstrahlung cooling, eq. (\ref{eq:T_brem}). Taking into account the shock deceleration up to e.g. $\tilde{r}_\mathrm{sh}=10$, the average photon energy should be reduced by a factor $(10^{-0.1})^2\approx0.6$, eq. (\ref{eq:_v_sh}), resulting in $\approx12v_9^2~$keV. Since the pure Bremsstrahlung cooling limit is not reached (as explained above), this is a bit of an overestimate, and the fitted numeric result (Figure \ref{fig:xray}) is consistent with average photon energy at the shock converging at late times to
\begin{equation}
    \label{eq:typical_energy}    \bar{\varepsilon}_\mathrm{sh,late}\approx 6\tilde{r}_\mathrm{sh,10}^{-0.2}v_9^2~\mathrm
    {keV}.
\end{equation}
This is also the asymptote of the average energy of photons emitted from the photosphere (see Figure \ref{fig:t_ph}).

As can be seen in Figure \ref{fig:xray}, X-rays compose a sizable fraction of the energy already at $3R_\mathrm{bo}$. This X-ray radius is weakly sensitive to $R_\mathrm{bo}$; For smaller $R_\mathrm{bo}$, the RMS temperature, eq. (\ref{eq:T_RMS}) increases, reducing the X-ray radius. However, at lower velocities, the thermal radiation temperature in the dense shell also rises such that the reprocessing is stronger, approximately canceling the RMS temperature change effect. The overall effect is small, as changing $R_\mathrm{bo}$ by orders of magnitude modifies the X-ray radius by only tens of percent. We see that the X-ray photospheric time is typically one-third, corresponding to approximately one breakout time, past peak luminosity time, see Figure \ref{fig:t_ph}. $\bar{\varepsilon}_\mathrm{sh,late}(\tilde{r}_\mathrm{sh}=10)$ is reached after another $\sim7$ breakout times.

\section{Discussion}
\label{sec:discussion}

In this paper, we solved the evolution of the shock structure and its associated radiation field in non-relativistic wind breakouts. We confirmed numerically (\S~\ref{subsec:piston_results}) that the RMS is converted to a CLS beyond $\tilde{r}_\mathrm{CLS}\approx0.3$, and derived analytically the evolution of the shocked plasma temperature, that reaches $T_\mathrm{CLS}\approx56v_9^2~$keV at the end of the RMS-CLS transition (eq. (\ref{eq:T_CLS}), Figure \ref{fig:consttemp}). We showed that a hot plasma layer with Thomson optical depth $\tau_\mathrm{hot}\approx0.2$, eq. (\ref{eq:tau_CLS}) is formed behind the CLS, with electrons and protons near equipartition (for $v_9\lesssim2$, Appendix \ref{sec:electron-proton}). The hot plasma cools down by Bremsstrahlung and inverse Compton emission. Compton cooling dominates at high shock velocities and small shock radii, eq. (\ref{eq:Q}), resulting in an unsaturated Comptonized spectrum given by eq. (\ref{eq:unsaturated}). Bremsstrahlung cooling dominates at low velocity and large shock radii, yielding a spectrum given by eq.~(\ref{eq:brem}). The resulting radiation field is quite uniform across the downstream shocked plasma (Figure \ref{fig:downspect}). A dense low-temperature shell is formed at the inner edge of the shocked wind (\S~\ref{subsec:piston_results}, \S~\ref{subsec:envelope_plasma}), reprocessing part of the high energy photon radiation generated at the CLS to lower, $\sim5~$eV energies, eq. (\ref{eq:T_TE}). 

In both the Compton and Bremsstrahlung cooling regimes, the characteristic photon energy shifts from 10's of eV during the RMS phase to X-ray energies during the CLS phase, reaching $\approx1$~keV at $3R_\mathrm{bo}$. The observed flux is dominated by X-rays at $\approx1R_{14}/v_9~$d past peak bolometric luminosity, and the characteristic photon energy increases to $\approx6v_9^2$~keV after $\approx7R_{14}/v_9~$d (eq. (\ref{eq:T_brem}), (\ref{eq:T_unsat}), (\ref{eq:typical_energy}), Figures \ref{fig:constspect}, \ref{fig:t_ph}-\ref{fig:xray}). The bolometric luminosity light curve is determined by $R_\mathrm{bo}$ and $v_\mathrm{bo}$ (\S~\ref{subsec:Wind+CLS}; a universal analytic light curve is given in \S~\ref{subsec:envelope_spectrum}, see Figure \ref{fig:t_ph}). The spectral shape is determined primarily by $v_\mathrm{bo}$. For $v_9>0.4$, the absorption of escaping X-ray radiation by the (initially cold and neutral) upstream wind plasma is negligible (and for $v_9>0.8$ this is true also for very large radii and mass of CSM, Appendix \ref{sec:ionization}).

Tables tabulating our numeric results for the time-dependent normalized local shock spectrum, as well as the escaping radiation's spectral luminosity, are given in \url{https://github.com/talwas/WindBreakout}. We provide spectra over photon energy range of $1-10^{5.5}~$eV, for $R_{14}\in\{0.01,0.1,1,10\}$, $v_{9}\in\{0.5,1,2\}$. Results are given for both the constant velocity piston case and the shocked polytropic envelope case. The spectral luminosity energy distribution can be directly compared against observations, and extrapolated for different $R_\mathrm{bo}$, $v_\mathrm{bo}$, as it exhibits a smooth dependence on these parameters.

We presented solutions for the limit of a breakout radius much larger than the stellar radius and much smaller than the radius out to which the dense CSM extends,
$R_\star\ll R_\mathrm{bo}\ll R_\mathrm{dCSM}$. As noted (\S~\ref{sec:intro}), observations suggest that while the separation of these radii may be significant, it is not necessarily very large. 
\\ \noindent-\textit{Finite $R_\star/R_\mathrm{bo}$}. We show in Appendix \ref{sec:luminosity_contributions} that the radiation diffusing from the expanding envelope up to the shock comprises at breakout a fraction  $\sim R_\star/R_\mathrm{bo}$ of the energy density. This implies that for cases where $R_\star$ is a significant fraction of $R_\mathrm{bo}$, the spectrum of the optical-UV part of the escaping radiation would be significantly softer than obtained in the $R_\star\ll R_\mathrm{bo}$ limit (this may partially account for the lower, $T_\mathrm{color}\sim3~$eV compared to $10~$eV temperatures inferred for some cases considered to be CSM breakouts, see \S~\ref{sec:intro}). 
\\ \noindent-\textit{Finite $R_\mathrm{bo}/R_\mathrm{dCSM}$}. In cases where the dense CSM does not extend much beyond the breakout radius, we expect the radiation to be approximately described by our results up to when the shock reaches a small optical depth near $R_\mathrm{dCSM}$. At later times, X-ray radiation will be produced by the CLS propagating into the lower-density extended wind/CSM, with a luminosity that is significantly smaller (by a factor comparable to the ratio of the dense and extended CSM densities) than that predicted for propagation in the compact dense CSM shell. It is important to note that for $R_\mathrm{dCSM}\lesssim3R_\mathrm{bo}$, the shock interaction with the dense CSM will be over before the spectrum evolves to X-ray energies (Note that the temporal dependence of the spectral luminosity emitted from the photosphere may also differ from our results since the extension of the diffusion time from the shock to the photosphere depends on $R_\mathrm{dCSM}$). Finally, in cases where $R_\star/R_\mathrm{bo}$ is not very small, the optical-UV emission will be dominated after the shock expands beyond $R_\mathrm{dCSM}$ by radiation escaping the expanding envelope, tracking models of envelope cooling emission.

Our results, that the spectrum shifts to the X-ray regime when the shock reaches $3R_\mathrm{bo}$ and that X-ray absorption in the upstream wind is insignificant, are in contrast with the results of some earlier works. As explained in \S~\ref{subsec:Previous_Works}, the discrepancy is due to incomplete treatment of relevant physical processes. These include, e.g., the effects of inelastic Compton scatterings, which we show are able to produce X-ray photons by unsaturated Comptonization (\S~\ref{subsec:piston_analytic}), the formation of CLS and its heating of the plasma to high temperatures (\S~\ref{subsec:Wind+CLS}, \S~\ref{sec:cooling_limiter}), and the ionization and heating of the upstream wind plasma (\S~\ref{sec:ionization}). The relatively low X-ray luminosities, typically $\lesssim10^{40}~$erg/s, inferred for several SNe that are believed to be associated with CSM breakouts \citep[e.g.][and more mentioned in \S~\ref{sec:intro}]{smith_sn_2007,ofek_x-ray_2013,irani_early_2024}, cannot be explained therefore as due to soft spectra of radiation escaping the shock or due to absorption in the dense CSM upstream. Rather, the low X-ray luminosity may suggest that the dense CSM does not extend beyond $\sim3R_\mathrm{bo}$, as is believed to be the case for SN~2023ixf (see \S~\ref{subsec:CSM}). We note that in such cases, as the X-ray luminosity produced by the CLS propagating in the lower-density extended wind is much weaker, it may be partially absorbed in the upstream, and may be consistent with observations of a decreasing neutral hydrogen column density with time.

The existence of X-ray-bright SNe, with $10^{41-43}~$erg/s X-ray luminosities (see \S~\ref{sec:intro}), may be accounted for by the presence of dense CSM extending beyond $\sim3R_\mathrm{bo}$. We note that the conclusion that in some cases the observed X-ray luminosity at times close to the optical maximum is too large to be accounted for by the wind breakout interpretation \citep{ofek_x-ray_2013}, is due to the results of earlier works suggesting that X-ray emission is suppressed. As explained above, we find that strong X-ray emission is expected as early as one breakout time past maximum optical luminosity time, consistent with the observed bright X-ray emission. 

Einstein Probe (EP), launched in 2024, carries a 3,600 deg$^{2}$ Wide-field of view X-ray Telescope (WXT) that reaches a $5\sigma$ sensitivity of $\approx 2.6\times10^{-11}\,{\rm erg\,cm^{-2}\,s^{-1}}$ in the 0.5-4 keV band for a single 1 ks survey exposure and sweeps roughly half the sky on a day timescale \citep{yuan_science_2025}. If CSM breakouts radiating $10^{42}\,{\rm erg\,s^{-1}}$ in WXT's band accompany a significant fraction of type II core-collapse SNe, a detection rate of $\approx 0.4$ events per year (at $7\sigma$, required for $<1\%$ false detection rate) is expected for the WXT survey (assuming a local type II SN volumetric rate of $\approx0.5\times10^{-4}\,{\rm Mpc^{-3}\,yr^{-1}}$; \citealp{li_nearby_2011}). If the X-ray emission remains bright on a day timescale, co-adding $\sim$10 daily snapshots improves the rate to a few per year. 

\begin{acknowledgments}
We thank Boaz Katz, Doron Kushnir, Ben Shenhar, and Jonathan Morag for insightful comments. This research was partially supported by ISF and IMOS grants.
\end{acknowledgments}

\appendix
    
\section{Cooling limiter}
\label{sec:cooling_limiter}

Here, we describe the cooling limiter that enables fast convergence with spatial grid resolution to the correct CLS-heated plasma temperature and the correct post-shock cooling profile.

\subsection{Shock Heating}
\label{subsec:shcok-heating}

\begin{figure}[ht]
    \centering
    \includegraphics[height=6.5cm]{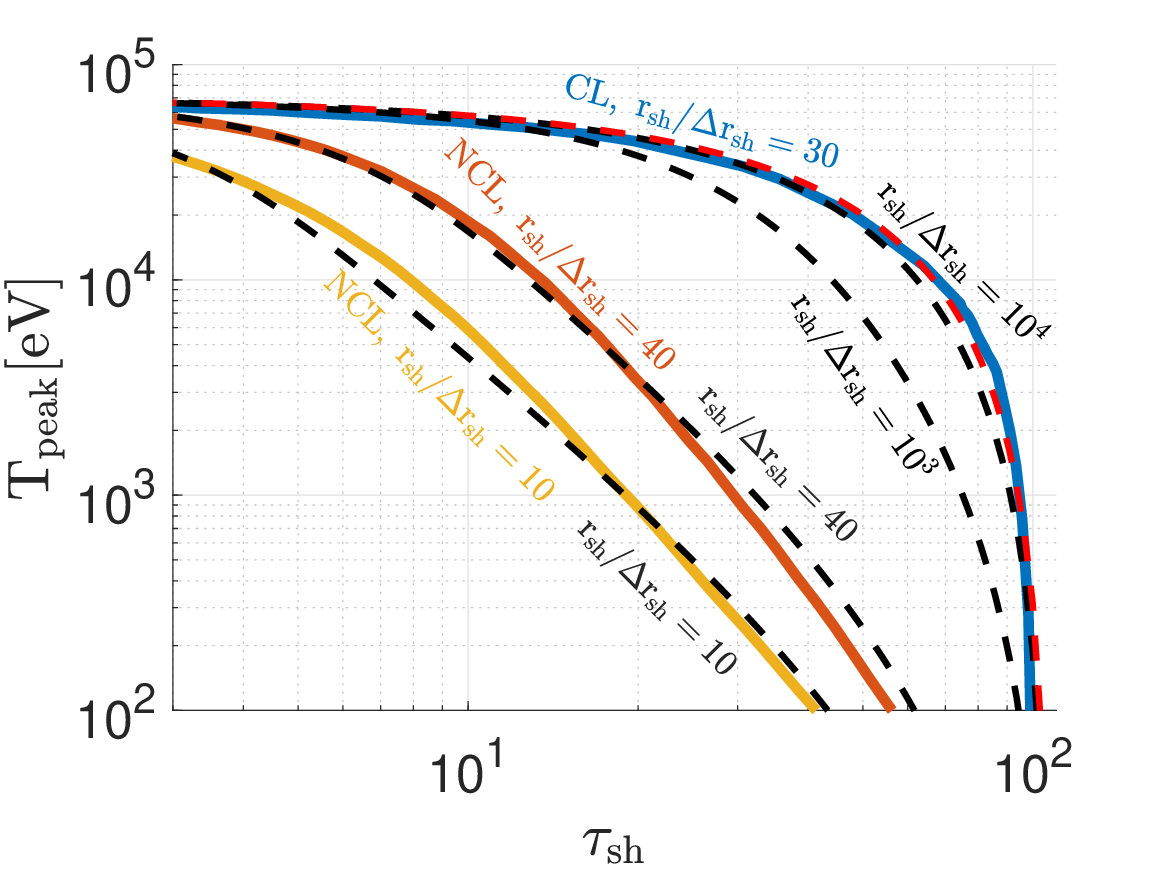}
    \caption{Solid lines show the peak plasma temperature at the CLS, obtained in numeric calculations with (CL) and without (NCL) a cooling limiter for various spatial grid resolutions, as a function of the optical depth ahead of the shock location, for $R_{14}=1$ and constant piston velocity $v_9=1$. $r_\mathrm{sh}/\Delta r_\mathrm{sh}$ is the ratio of the shock radius to the numeric spatial width of the shock. Black dashed lines show the analytic estimate for the peak temperature expected for different shock resolutions without a cooling limiter, eq. (\ref{eq:T_max}). The red dashed line shows the analytic estimate of the CLS temperature, eq. (\ref{eq:T_CLS}).}
    \label{fig:limiter}
\end{figure}

In Figure \ref{fig:limiter}, we show the peak plasma temperature achieved at the CLS during its propagation for different spatial grid resolution, characterized by the ratio of the shock radius to the numeric spatial width of the shock, $r_\mathrm{sh}/\Delta r_\mathrm{sh}$ (this ratio is time-independent when choosing a logarithmically spaced spatial grid). Using a cooling limiter, which we implement by turning off the radiation-plasma energy coupling of eq.~(\ref{eq:radmat}) over $\Delta r_\mathrm{sh}$, the correct temperature, $T_\mathrm{CLS}$ given by eq.~(\ref{eq:T_CLS}), is obtained for $r_\mathrm{sh}/\Delta r_\mathrm{s}=30$. Without applying the cooling limiter, a resolution $r_\mathrm{sh}/\Delta r_\mathrm{sh}>10^4$ would be required to obtain numerically the correct temperature for the RMS-CLS transition.

\begin{figure*}[ht]
    \centering
    \includegraphics[height=5.5cm]{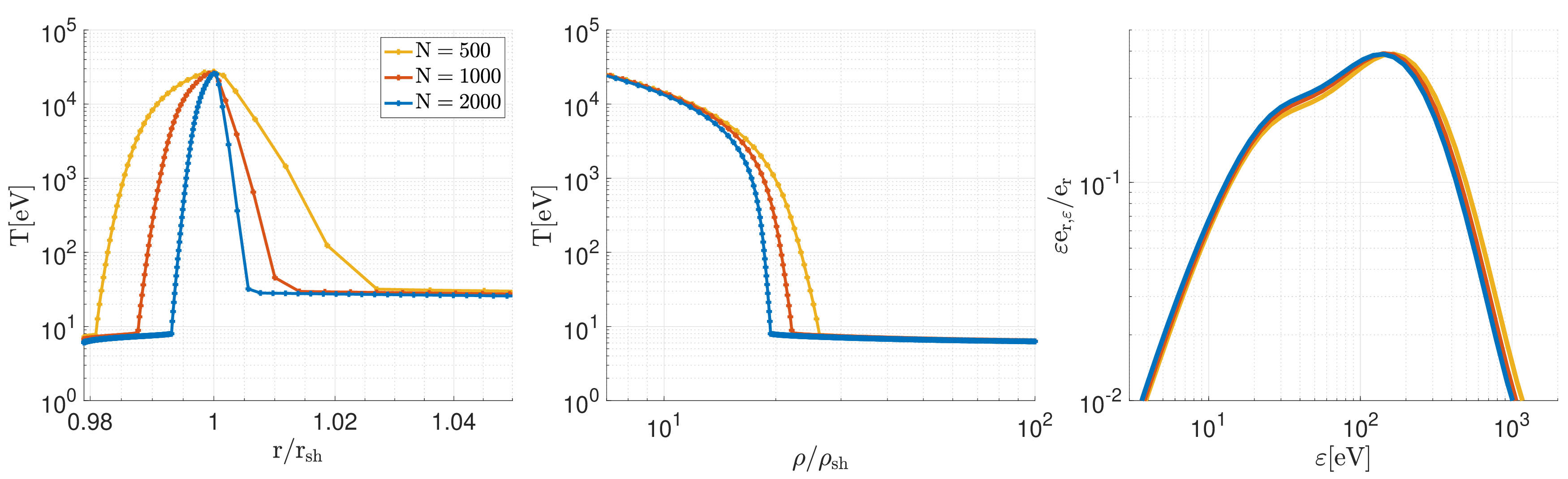}
    \caption{The convergence with spatial resolution (increasing by a factor 2 between different curve colors) of the radiation spectrum (\textit{right}) and of the thermodynamic trajectory of the cooling plasma (\textit{middle}) when using a cooling limiter as described in the text (results shown for the case of $R_{14}=1$ and a constant velocity $v_9=1$  piston-driven shock, at $\tilde{r}_\mathrm{sh}=1$). While the spatial width of the hot plasma layer, which extends over $\sim20$ grid points, changes with resolution (\textit{left}), the thermodynamic profiles and radiation spectrum converge.}
    \label{fig:limiter_comp}
\end{figure*}

We note that the resolution required for obtaining the correct temperature was not derived using brute-force, very high-resolution calculations, but rather by estimating analytically the temperatures that would be achieved in numeric calculations without a cooling limiter. The validity of this estimate, derived in the following paragraph and shown in dashed black lines in the plot, is demonstrated by the fact that it reproduces well the results obtained numerically at lower resolutions.

The relation between the numeric spatial resolution and $T_\mathrm{peak}$, the peak plasma temperature obtained numerically at the CLS, is estimated as follows. The temporal evolution of the temperature of a fluid element as it flows through the numeric shock transition region can be approximately described by
\begin{equation}
\label{eq:Tdot}
    \dot{T}=\frac{T_\mathrm{CLS}}{t_\mathrm{heat}}-\frac{T}{t_\mathrm{cool}},
\end{equation}
where $T_\mathrm{CLS}$ is given by eq. (\ref{eq:T_CLS}), the cooling time is given by eq. (\ref{eq:t_cool}), and the heating time is approximately given by $t_\mathrm{heat}=\Delta r_\mathrm{sh}/v_\mathrm{sh}$. With these approximations for the cooling and heating rates, we have
\begin{equation}
\label{eq:tc_th}
t_\mathrm{cool}/t_\mathrm{heat}\approx 10^{-3}\tilde{r}_\mathrm{sh}^2\times r_\mathrm{sh}/\Delta r_\mathrm{sh},
\end{equation}
and the solution of eq.~(\ref{eq:Tdot}) yields
\begin{equation}
\label{eq:T_max}
T_\mathrm{peak}/T_\mathrm{CLS}=\frac{t_\mathrm{cool}}{t_\mathrm{heat}}\left(1-e^{-t_\mathrm{heat}/t_\mathrm{cool}}\right).
\end{equation}

\subsection{Post-shock Cooling}
\label{subsec:shcok-cooling}

To properly describe the radiation emitted by the radiatively cooling post-CLS hot plasma, it may not be sufficient to choose a small time-step to enable following the plasma temperature temporal evolution. It may be required to spatially resolve the cooling region to enable following possible expansion/compression during the cooling phase. Since the cooling region width is very small compared to the other characteristic length scales of the problem, we artificially expand its width to a fixed number of spatial grid points in a manner that conserves the relative contributions of the different (heating and) cooling processes and thus enables rapid convergence with spatial grid resolution to the correct emitted radiation spectrum. 

The artificial expansion of the cooling region width is achieved by applying an adaptive cooling limiter (a variant of the burning limiter used in \citet{kushnir_accurate_2020} for detonation waves), reducing the rate of radiation-plasma energy transfer given by eq.~(\ref{eq:radmat}) by a (time- and space-dependent) factor that limits the cooling time, given by eq.~(\ref{eq:t_cool}), to be larger by a fixed factor, 20, than the sound crossing time of the spatial extent of the numeric cell. This leads to a widening of the cooling region width across $\sim20$ spatial grid points.

Figure \ref{fig:limiter_comp} demonstrates the convergence with spatial resolution of the radiation spectrum and of the thermodynamic trajectory of the cooling plasma. Note that while the widening of the cooling region increases the optical depth and Compton y-parameter of the hot plasma shell, this increase is compensated by the reduction of the strength of the radiation-plasma coupling.

\section{Upstream X-ray Absorption}
\label{sec:ionization}

In this section, we discuss the possible suppression of the X-ray, $>0.1$~keV, luminosity, that is emitted by the shocked plasma, by photo-ionization and Compton scattering as it propagates through the initially cold and neutral upstream plasma (for $>0.1$~keV, the contribution of bound-bound transitions to the absorption opacity is dominated by photo-ionization).

We first show that the evolution of the upstream plasma ionization and temperature follows four stages: ionization by the UV--X-ray photons, followed by a rapid thermalization of the ejected electrons (prior to any significant recombination), followed by evolution towards a quasi-equilibrium ionization and temperature at which a balance is maintained between cooling and heating (by ionization, recombination, Bremsstrahlung, and Compton interactions), followed by a gradual evolution of the quasi-steady state with the evolution of the ionizing radiation spectrum (note that the luminosity does not vary strongly with time). At any time $t$, corresponding to shock location $\tilde{r}_\mathrm{sh}$, the plasma is in (or close to) a quasi-equilibrium up to a radius $\tilde{r}_\mathrm{eq}$ that grows with time, and still evolving towards equilibrium at larger radii.

We show below that $\tilde{r}_\mathrm{eq}$ is always larger than the radius $\tilde{r}_\mathrm{Comp}$ up to which energy loss by Compton scattering may be significant, $\tilde{r}_\mathrm{Comp}\approx 4\varepsilon_\mathrm{r,10}^{1/2}v_9^{-1}$ for photons of energy $\varepsilon_\mathrm{r,10}=\varepsilon_\mathrm{r}/10$~keV (since inelastic energy loss to cold electrons is significant for $\varepsilon_\mathrm{r}>m_ec^2/N=m_ec^2/\tau_T^2$ where $N=\tau_T^2$ is the number of scatterings that the photon undergoes). Thus, the suppression of the X-ray luminosity may be due to ionization losses by the non-equilibrium plasma beyond $\tilde{r}_\mathrm{eq}$, which is discussed in \S~\ref{subsec:ion_loss}, and by Compton and ionization losses by the equilibrium plasma at smaller radii, that is shown to be dominated by the ``immediate" upstream plasma in the vicinity of $\tilde{r}_\mathrm{sh}$, and is discussed in \S~\ref{subsec:compton_loss}. We find that the X-ray luminosity suppression by the equilibrium plasma at small radii is insignificant for $v_9>0.4$ -- for lower velocities, the X-ray flux is suppressed up to $0.1v_9^{-4}t_\mathrm{bo}$. For an infinite wind, the ionization suppression by plasma at very large radii (and mass) is insignificant for $v_9>0.8$ -- for lower velocities, the X-ray flux is suppressed up to $1.5v_9^{-3}t_\mathrm{bo}$.

To determine the evolution of the upstream plasma ionization and temperature at different radii, we compare the rates of the different relevant processes. The timescale for photo-ionization of an atom, $\tilde{t}_{\mathrm{r},\mathrm{ionize}}$, for Compton heating of an electron (by energetic photons), $\tilde{t}_{\mathrm{Comp}}$, for energetic electron energy loss by collisional ionization, $\tilde{t}_{e,\mathrm{ionize,loss}}$, by collisions with free electrons, $\tilde{t}_{e,e}$, or ions $\tilde{t}_{e,p}$, losses by Bremsstrahlung emission $\tilde{t}_{Brem}$, and the recombination time of a free electron, $\tilde{t}_{rec}$, are \citep{rybicki_radiative_1979,huba_nrl_2013}
\begin{equation}
\label{eq:ion_times}
\begin{split}
    \tilde{t}_{\mathrm{r},\mathrm{ionize}}&\approx
    \begin{cases}
    10^{-11}\varepsilon_{\mathrm{r},0.1}^4\tilde{r}^3,&\tilde{r}<\tilde{r}_\mathrm{ph},\\
    10^{-9.5}\varepsilon_{\mathrm{r},0.1}^4\tilde{r}^2v_9^{-1},&\tilde{r}_\mathrm{ph}<\tilde{r},\\
    \end{cases}\\
    \tilde{t}_{\mathrm{Comp}}&\approx
    \begin{cases}
    10^{-3.5}(10\varepsilon_{\mathrm{e}}/\varepsilon_{\mathrm{r}})\tilde{r}^3,&\tilde{r}<\tilde{r}_\mathrm{ph},\\
    10^{-2}(10\varepsilon_{\mathrm{e}}/\varepsilon_{\mathrm{r}})\tilde{r}^2v_9^{-1},&\tilde{r}_\mathrm{ph}<\tilde{r},\\
    \end{cases}\\
    % \tilde{t}_{e,\mathrm{ionize}}&\approx10^{-8}f^{-1}\varepsilon_{e,100}^{3/2}\tilde{r}^2v_9^2A/Z^5,\\
    \tilde{t}_{e,\mathrm{ionize,loss}}&\approx10^{-8.5}(1-f)^{-1}\varepsilon_{e,0.1}^{3/2}\tilde{r}^2v_9^2,\\
    \tilde{t}_{e,e}&\approx10^{-8.5}f^{-1}\varepsilon_{e,0.1}^{3/2}\tilde{r}^2v_9^2,\\
    \tilde{t}_{e,p}&\approx10^{-5}f^{-1}\varepsilon_{e,0.1}^{3/2}\tilde{r}^2v_9^2,\\
    \tilde{t}_{\rm Brem}&\approx10^{-2.5}f^{-1}\varepsilon_{e,0.1}^{1/2}\tilde{r}^2v_9^2,\\
    \tilde{t}_{\rm rec}&\approx
    \begin{cases}
        10^{-1.5}f^{-1}\varepsilon_{e,0.1}^{3/2}\tilde{r}^2v_9^2&I_H<\varepsilon_e\\
        10^{-3.5}f^{-1}\varepsilon_{e,0.001}^{1/2}\tilde{r}^2v_9^2&\varepsilon_e<I_H\\
    \end{cases}.
\end{split}
\end{equation}
These timescales are obtained for a hydrogen plasma, with density given by eq.~(\ref{eq:density}) and radiation energy density given by eq.~(\ref{eq:e_r}), with photon (electron) energy $\varepsilon_{\mathrm{r}(e),X}\equiv\varepsilon_{\mathrm{r}(e)}/X~$keV, hydrogen ionization fraction $f$ and binding energy $I_H$, neglecting logarithmic corrections and the small corrections due to ionization/recombination from/to excited states.

In what follows, we address the impact of the presence of ``metals" on the results. For a solar (or lower) abundance, the ratio of the densities of nuclei with atomic number $Z$ to the density of hydrogen satisfies $n_Z/n_H<Z^{-3}$. Only the ionization and recombination rates, which may increase with $Z$ faster than $Z^3$, are significantly affected. In contrast, the number density of free electrons, which is determined by the ionization fraction of hydrogen (and helium), as well as the electron energy loss rates by collisions with ions ($\propto Z_\mathrm{eff}^2/A$) and the Bremsstrahlung emission ($\propto Z_\mathrm{eff}^2$), are not significantly affected ($Z_\mathrm{eff}$ is the properly averaged effective ion charge, which is smaller than $Z$ for partial ionization and for ionization of electrons from energy levels above the lowest). We show below that the presence of solar metal abundance does not affect the results significantly.

Comparing the different timescales, we find that electrons emitted by ionization collisionally ionize other atoms or thermalize with other free electrons (and then also with the protons) prior to significant Bremsstrahlung energy loss or recombination (for incident electrons with energy smaller than the field electron temperature $T$, the thermalization timescale scaling changes to $\varepsilon^{3/2}\rightarrow T^{1/2}\varepsilon$ with the same prefactor, such that the above statement remains valid). This is true also in the presence of solar abundance of metals, since considering recombination with fully ionized ions (for which the recombination cross-section is largest), the recombination rate scales only as $Z^2$ (and is hence dominated by hydrogen) for low energy  ($<I_HZ^2$) electrons, while the recombination cross-section for higher energy electrons scales as $Z^4$, implying $\tilde{t}_{e,e}/\tilde{t}_\mathrm{rec}\approx10^{-7}(n_Z/n_e)Z^{4}\ll1$. Since the photo-ionization is much faster than the collisional ionization electron energy loss rate, the hydrogen ionization fraction increases significantly before significant collisional ionization energy loss, and the free electron thermalization rate quickly becomes much faster than the collisional ionization loss rate. Thus, at any radius, an initially neutral plasma is first almost fully ionized to $f_\mathrm{eq}\approx1-\tilde{t}_\mathrm{r,ionize}/\tilde{t}_\mathrm{rec}$, and then the free electrons thermalize prior to significant secondary collisional ionization, recombination or Bremsstrahlung energy loss. For photons' energy that is significantly larger than $I_H$, the ionized electron temperature will initially (after thermalization with protons) be approximately given by $\varepsilon_{\mathrm{r}}/3$. During and following the breakout, the ionizing radiation is characterized by $\varepsilon_{\mathrm{r}}\gtrsim100$~eV, implying that the ionized electrons promptly thermalize to a temperature exceeding 10's of eV.

Examining eq. (\ref{eq:ion_times}), the slowest process affecting the promptly ionized electrons is Bremsstrahlung emission (since the ionization time is very short, the hydrogen ionization fraction of the promptly heated plasma will be close to equilibrium at the time-dependent temperature). This implies that the plasma will be close to equilibrium at radii below which the Bremsstrahlung time is short compared to $t$, i.e. that $\tilde{r}_\mathrm{eq}$ is determined by $\tilde{t}_\mathrm{Brem}(\tilde{r}_\mathrm{eq})=\tilde{r}_\mathrm{sh}$. A lower limit to $\tilde{r}_\mathrm{eq}$ is obtained by assuming the electron temperature to equal the ionizing radiation photon energy, which yields an upper limit to the Bremsstrahlung cooling time. For this limit, we obtain $\tilde{r}_\mathrm{eq}\approx10\varepsilon_\mathrm{r,10}^{-1/4}\tilde{r}_\mathrm{sh}^{1/2}v_9^{-1}$ and $\tilde{r}_\mathrm{Comp}/ \tilde{r}_\mathrm{eq}\approx0.4\varepsilon_\mathrm{r,10}^{3/4}\tilde{r}_\mathrm{sh}^{-1/2}$, satisfying $\tilde{r}_\mathrm{Comp}/ \tilde{r}_\mathrm{eq}<1$ as mentioned above (recall that the shocked plasma emission is dominated by X-rays only at $3R_\mathrm{bo}$). Note that for $\tilde{r}_\mathrm{sh}\gtrsim100\varepsilon_\mathrm{r,10}^{-1/2}v_9^{-2}$ the plasma lying ahead of the shock does not reach equilibrium.

\subsection{$r>r_\mathrm{eq}$}
\label{subsec:ion_loss}

As the photo-ionization timescale is orders of magnitude shorter than the breakout time and the recombination time, eq. (\ref{eq:ion_times}), the plasma may first reach nearly full ionization at all radii (metals will be ionized up to the photons' energy threshold\footnote{Typically, the most bound electron will be ejected first, followed by a quick de-excitation of the ion.}).
Beyond $\tilde{r}_\mathrm{eq}$, the Bremmstrahlung time, and thus the recombination time, is longer than the dynamical time\footnote{This is true for a hydrogen plasma, eq. (\ref{eq:ion_times}), as well as for a solar abundance plasma. The shortest recombination time, which is obtained for fully ionized ions, is longer than or equal to the Bremsstrahlung time for all electron energies or ion charges, as implied by the scalings discussed above.}, such that after the plasma is ionized a single time, it does not recombine (and does not get re-ionized multiple times) over a dynamical time. Thus, it is sufficient to consider a single ionization of the plasma above $\tilde{r}_\mathrm{eq}$ (multiple ionization/recombination of plasma lying below this radius is discussed in the next subsection).

The CLS deposits $\approx110v_9^2$~keV per proton in the plasma thermal energy, which is radiated mostly in ionizing photons. The energy radiated as the shock propagates to $\tilde{r}_\mathrm{sh}$ is sufficient, therefore, for ionizing a plasma of mass larger by a factor $\approx10^4v_9^2$ than the accumulated shocked mass $M_\mathrm{sh}$\footnote{The average ionization energy for large $Z$ is $\approx10 ZI_H$ \citep{segre_nuclei_1982}, so that the energy required for ionization is dominated by the hydrogen atoms also for solar abundance.}. As the wind mass increases linearly with radius, the radiation energy emitted as the shock reaches $\tilde{r}_\mathrm{sh}$ is sufficient to ionize the plasma up to radius $\approx10^4\tilde{r}_\mathrm{sh}v_9^2 R_\mathrm{bo}$. Noting that the ionization opacity for solar abundance is $\kappa_\mathrm{eff}\approx300\varepsilon_\mathrm{r,1}^{-8/3}\kappa_\mathrm{T}$ \citep[higher than for hydrogen $\kappa_{\mathrm{ionize},H}\approx30\varepsilon_\mathrm{r,1}^{-3}\kappa_\mathrm{T}$][]{cruddace_opacity_1974,longair_high_2011}, and that the Thomson optical depth is $c/v\approx30v_9^{-1}$ at $R_{\rm bo}$, we find that the energy emitted at breakout is sufficient to ionize the plasma up to the radius at which the ionization optical depth for 1~keV photons drops to unity, $\approx10^4v_9^{-1} R_\mathrm{bo}$. Thus, if the dense CSM wind profile extends to very large radii, $>10^4v_9^{-1} R_{\rm bo}$, with very large CSM mass, $>10^2 R_{\rm 14}^2M_\odot$, the luminosity of the $>1$~keV radiation will be strongly suppressed over roughly $1.5v_9^{-3}t_\mathrm{bo}$ (up to shock radius $1.5v_9^{-3}R_\mathrm{bo}$), rendering a nonsignificant X-ray absorption by ionization starting from the X-ray radius of $3R_\mathrm{bo}$, for $v_9>0.8$. For a finite dense CSM radius $\tilde{r}_\mathrm{dCSM}$, the X-ray luminosity emitted by the shock is suppressed up to a shock radius
\begin{equation}
    \tilde{r}_\mathrm{sup,ion1}\approx\min(1.5v_9^{-3},(\tilde{r}_\mathrm{dCSM}/10^4)v_9^{-2}).
\end{equation}
At later times, the plasma will be fully ionized up to the ``ionization photosphere" (or the dense CSM radius) and $>1$~keV photons will be able to escape. We note that the luminosity in lower energy, $\sim0.1$~keV, photons may be suppressed over a longer timescale if the dense wind extends to even larger radii.

\subsection{$r<r_\mathrm{eq}$}
\label{subsec:compton_loss}

In this region, the timescales for ionization, recombination, Bremsstrahlung, and Compton interactions to reach the equilibrium values are short compared to the dynamical time; hence, the plasma ionization and temperature reach a steady state balancing cooling and heating by ionization, recombination, Bremsstrahlung, and Compton interactions. The ionization fraction and temperature depend on the incident radiation spectrum (recall that the luminosity is nearly time-independent following breakout), which evolves on the breakout timescale. This evolution leads to a corresponding evolution of the plasma steady state, to which we therefore refer as a quasi-steady state. We show below that energy absorption (and emission) by the quasi-equilibrium plasma is dominated by the ``immediate" upstream plasma near the shock and is negligible compared to the shock-generated X-ray luminosity for $v_9>0.4$.

The plasma (hydrogen) ionization fraction and temperature evolve toward an equilibrium according to
\begin{equation}
\label{eq:ion_evol}
\begin{split}
        \dot{f}&=\frac{1-f}{t_\mathrm{r,ionize}}-\frac{f}{t_\mathrm{rec}},\\
        \dot{e}_\mathrm{pl}&=\dot{e}_\mathrm{ionize}+\dot{e}_\mathrm{Comp}-\dot{e}_\mathrm{rec}-\dot{e}_\mathrm{Brem},
\end{split}
\end{equation}
where $\dot{e}_\mathrm{Comp},\dot{e}_\mathrm{Brem}$ are given by eq. (\ref{eq:bolradmat}) with an additional multiplication by $f$ and $f^2$ correspondingly, and the heating and cooling of (monochromatic) photo-ionization and recombination (from/to the ground state of hydrogen plasma) are \citep{rybicki_radiative_1979}
\begin{equation}
\begin{split}
    \dot{e}_\mathrm{ionize}&=(1-f)\rho\kappa_{\mathrm{ionize},H}ce_\mathrm{r},\\
    \dot{e}_\mathrm{rec}&=f^2\rho\kappa_\mathrm{T}c\frac{8}{\sqrt{3\pi}}\frac{\rho}{m_p}\sqrt{\frac{I_H}{T}}I_H.
\end{split}
\end{equation}
We assume in eq.~(\ref{eq:ion_evol}) that the photons emitted by recombination promptly escape the plasma. This leads to an overestimate of the effective recombination rate, yielding, therefore, an upper limit to the fraction of the incident luminosity that is absorbed (and re-emitted) by the upstream plasma. Note that in deriving the fraction of the X-ray luminosity that is ``reprocessed", i.e., absorbed and re-emitted at lower energy, we consider all the energy emitted by the heated plasma (including by recombination) as ``reprocessed".

\begin{figure}[ht]
    \centering
    \includegraphics[height=6.5cm]{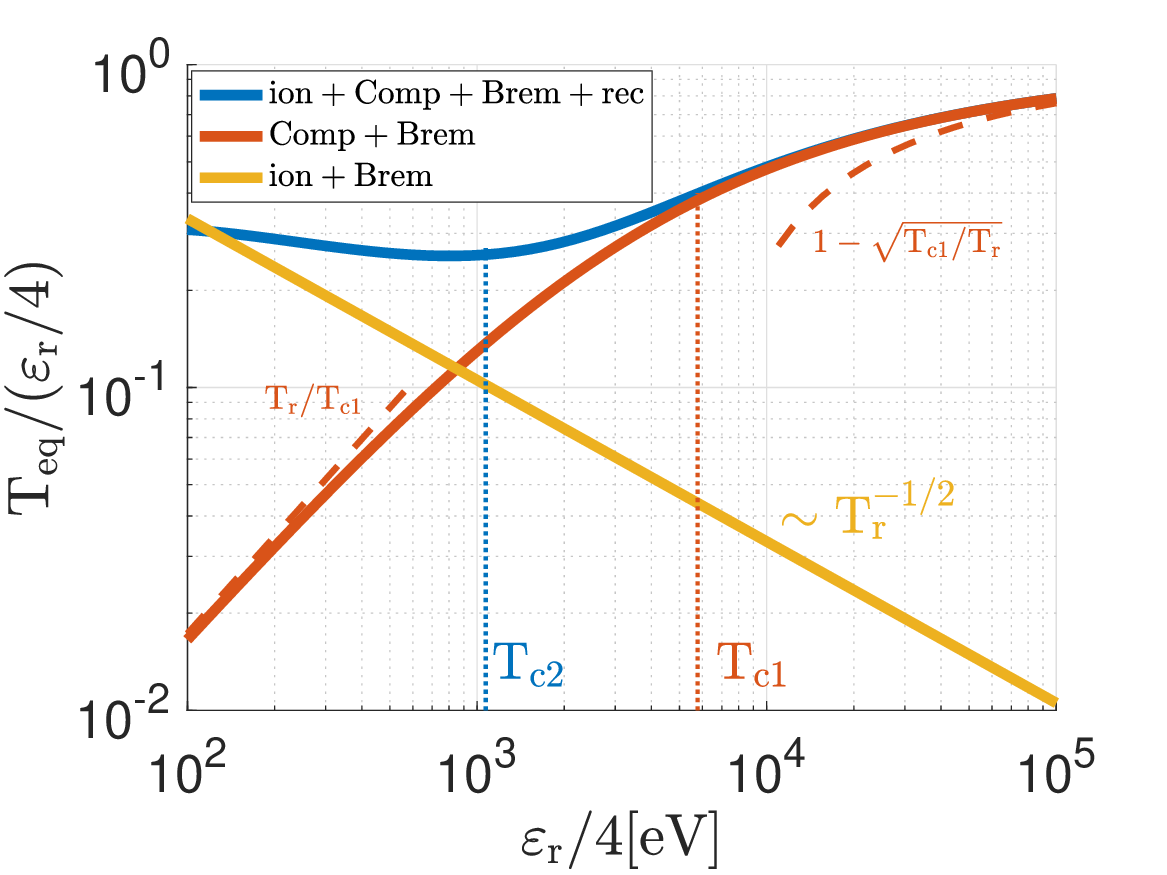}
    \caption{The equilibrium upstream plasma temperature $T_\mathrm{eq}$, obtained by numerically solving the steady solution of eq.~ (\ref{eq:ion_evol}), as a function of the photon energy $\varepsilon_r$ of a monochromatic incident radiation, for $\tilde{r}=3$, $v_9=0.4$. We also present the Compton-Bremsstrahlung equilibrium, eq. (\ref{eq:T_US}) (with its limiting behaviors for radiation temperature higher or lower than $T_\mathrm{c1}$), and the photoionization-Bremsstrahlung equilibrium, eq. (\ref{eq:T_eq}). $T_\mathrm{c2}$ separates regions where Compton or photo-ionization heating dominates.}
    \label{fig:T_eq}
\end{figure}

Figure \ref{fig:T_eq} shows the equilibrium upstream plasma temperature $T_\mathrm{eq}$, obtained by numerically solving for the steady solution of eq.~ (\ref{eq:ion_evol}), as a function of the photon energy $\varepsilon_r$ of a monochromatic incident radiation. While the resulting $T_\mathrm{eq}$ varies in the range $10~\mathrm{eV}<T_\mathrm{eq}<\varepsilon_r/4$ (recalling that Compton equilibrium is obtained for $\varepsilon_r= 4T$), the ionization fraction is always very close to unity, eq. (\ref{eq:ion_times}). In the resulting equilibrium temperature range, the Bremsstrahlung emissivity dominates over the recombination emissivity (the ratio is $\approx0.1T_\mathrm{eq,0.1}^{-1}$, eq. (\ref{eq:ion_times}), valid also for a solar abundance). Hence, $T_\mathrm{eq}$ is determined by a balance of Compton heating and Bremsstrahlung cooling (that are included in the numeric calculations, and for which we derived an analytic expression, eq. (\ref{eq:T_US}), Figure \ref{fig:T_eq}), or by a balance of photo-ionization heating (which is not included in our numeric calculations for fully-ionized plasma) and Bremsstrahlung cooling. The dominant heating process depends on the radiation photon energy. For radiation temperatures lower than $T_\mathrm{c2}\equiv\frac{2^{3/2}}{\pi^2 3^{1/6}}\left(\frac{m_e c^2}{e_\mathrm{r}/n_e}\right)^{4/3}I_H\approx15\tilde{r}^{4/3}v_9^{-8/3}~$eV, the ionization heating dominates, leading to a radius independent equilibrium temperature of 
\begin{equation}
\label{eq:T_eq}
    T_\mathrm{eq}\approx\sqrt{\frac{I_H\varepsilon_\mathrm{r}}{2^{3/2}3^{1/2}}}\approx17\varepsilon_\mathrm{r,0.1}^{1/2}~\mathrm{eV}.
\end{equation}
This behavior is evident in Figure \ref{fig:T_eq}.

Using the above determination of $T_\mathrm{eq}$, we may proceed to calculate the luminosity emitted by the quasi-equilibrium upstream plasma. Since the emissivity is dominated by Bremsstrahlung emission, which scales as $\rho^2T_\mathrm{eq}^{1/2}$, the upstream luminosity is dominated by plasma at the immediate upstream of the shock, $\dot{E}_\mathrm{US}\approx4\pi r_\mathrm{sh}^3\dot{e}_\mathrm{Brem,sh}({r}_\mathrm{sh})\propto r_\mathrm{sh}^{-1}T_\mathrm{eq}^{1/2}({r}_\mathrm{sh})$ (the highest equilibrium temperature is obtained at ${r}_\mathrm{sh}$). For low shock velocities, the radiation temperature is below $T_\mathrm{c2}$, and $T_\mathrm{eq}$ is well approximated by eq. (\ref{eq:T_eq}) (see Figure \ref{fig:T_eq}), and the radius beyond which $\dot{E}_\mathrm{US}$ is smaller than the shock luminosity is
\begin{equation}
\label{eq:r_rep,ion}
\tilde{r}_\mathrm{sup,ion}\approx0.1\varepsilon_\mathrm{r,0.1}^{1/4}v_9^{-4}.
\end{equation}
This suppression radius matches very closely the radius calculated with the exact $T_\mathrm{eq}$ solution of eq. (\ref{eq:ion_evol}), and is shown in Figure \ref{fig:xray}. For $v_9>0.4$, the equilibrium upstream energy absorption rate can be neglected at radii at which X-rays dominate the shocked plasma emission.

In the presence of metals, the increased recombination and ionization rates imply a higher $T_\mathrm{eq}$ and hence larger Bremsstrahlung emission and larger suppression radius $\tilde{r}_\mathrm{sup,ion}$. For solar abundance, this leads to only a small modification. Assuming conservatively that ionization heating scales as $Z^4$ we find that $\tilde{r}_\mathrm{sup,ion}$ is increased by a factor $\approx1.5(2)$ for $\varepsilon_\mathrm{r}=1(10)~$keV (assuming full ionization of ions up to Ne(Fe)).

Note that in our numeric calculations of fully-ionized plasma, the upstream equilibrium is determined by a Compton-Bremsstrahlung balance, yielding an equilibrium temperature given by eq. (\ref{eq:T_US}) instead of eq. (\ref{eq:T_eq}) (see Figure \ref{fig:T_eq}), which implies a suppression radius of
\begin{equation}
\label{eq:r_rep,comp}
    \tilde{r}_\mathrm{sup,Comp}\approx0.4\varepsilon_\mathrm{r,0.1}^{3/8}v_9^{-1}.
\end{equation}
This fits the break in the numerically derived radius beyond which X-ray dominates the emission, Figure \ref{fig:xray}.

Finally, the energy required for heating the upstream plasma to its quasi-equilibrium temperature is always very small compared to the shock-generated energy. This is due to the fact that only for high velocities, and only at the ``immediate" upstream of the shock, the plasma is heated to a high temperature comparable to the radiation temperature (see Figure \ref{fig:consttemp}). Note that the upstream heating guarantees that our comptonization treatment is valid and energetic photons are not absorbed in the immediate upstream (when they scatter back and forth to the hot layer).

\section{The validity of the Electron-Proton Equipartition Approximation}
\label{sec:electron-proton}
We present a similar argument to that of \citet{katz_x-rays_2011}. When Compton cooling dominates ($Q_{\rm brem}<1$, eq. (\ref{eq:Q})), the temperature of the electrons and protons may decouple. The CLS initially heats the protons on a timescale of $\omega_p^{-1}$ to $T_{p0}=2T_\mathrm{CLS}\approx2\times56v_9^{-2}~$keV, eq. (\ref{eq:T_CLS}). The electron temperature evolution depends on the unknown amount of collisionless heating. A lower limit for the achieved electron temperature can be obtained by assuming that there is no collisionless heating. Then, the electron temperature evolution is affected by collisional heating and inverse Compton cooling
\begin{equation}
    \label{T_ed ot}
    \dot{T_e}=\frac{T_p-T_e}{t_\mathrm{e-p}}-\frac{T_e}{t_\mathrm{cool}},
\end{equation}
where the Compton cooling time is given by eq. (\ref{eq:t_cool}) (without factor $2$ for equipartition), and the electron-proton equipartition time is given by \citep{huba_nrl_2013}
\begin{equation}
\begin{split}    
    t_{e-p}&=(\rho\kappa_Tc)^{-1}\sqrt{\frac{\pi}{2}}\frac{m_p}{m_e}\lambda_\mathrm{C}^{-1}\left(\frac{T_e}{m_e c^2}+\frac{T_p}{m_p c^2}\right)^{3/2}\\
    & \approx 60\left(\frac{T_e}{T_\mathrm{CLS}}\right)^{3/2}\tilde{r}_\mathrm{sh}^2R_{14}v_9~\mathrm{s},
    \end{split}
\end{equation}
where $\lambda_\mathrm{C}\approx30$ is the Coulomb logarithm. If $t_{e-p}\ll t_\mathrm{cool}$, the electrons will heat quickly to $T_{p0}/2=T_\mathrm{CLS}$. Otherwise, the electron temperature will only reach a fraction $t_\mathrm{cool}/t_{e-p}$ of the proton temperature. Solving for $T_e$ we obtain
\begin{equation}
\label{eq:TeTp}
    T_e/T_\mathrm{CLS}\approx \mathrm{min} \left(1, 1.7\tilde{r}_\mathrm{sh}^{2/5} v_9^{-2}\right).
\end{equation}
The radius beyond which the electrons are in equipartition with protons is (see Figure \ref{fig:xray})
\begin{equation}
\label{eq:r_ep}
    \tilde{r}_\mathrm{e-p}\approx 0.1 v_9^5
\end{equation}
(at shock radius $\tilde{r}_\mathrm{e-p}$ the electron temperature is $0.7 T_\mathrm{CLS}$). Only for $v_9\gtrsim 2$ might the electron temperature fall significantly below $T_\mathrm{CLS}$ within the first few breakout radii. This means that the approximation of a single temperature is valid for most of our parameter range. Note that this is an upper bound for $\tilde{r}_{e-p}$, as collisionless heating has been neglected.

\section{Expanding Envelope Radiation}
\label{sec:luminosity_contributions}

The emission of radiation from the expanding stellar envelope can be derived analytically for $r_\mathrm{sh}\gg R_\star$ \citep{nakar_early_2010,rabinak_early_2011}.

We treat the expanding envelope as adiabatic up to where radiation escapes efficiently, $\tau=c/v$. The  luminosity at that point is then given by
\begin{equation}
\label{eq:L_*}
    L_\star(t)\approx 4\pi r_\mathrm{ej}(m,t)^2\frac{- \partial_r e_\mathrm{ej}(m,t)c}{3\rho_\mathrm{ej}(m,t)\kappa_\mathrm{T}}\Bigr\rvert_{\tau_\mathrm{ej}(m) = c/v_\mathrm{ej}(m)},
\end{equation}
where $r_\mathrm{ej}=v_\mathrm{ej}(m)t$, the ejecta velocity is given by eq. (\ref{eq:v_ej}), and the ejecta density is
\begin{equation}
\label{eq:rho_ej}
    \rho_\mathrm{ej}=\frac{dm}{4\pi r_\mathrm{ej}^2(-t dv)}=\frac{n+1}{4\pi\lambda}\frac{m}{r_\mathrm{ej}^3}.
\end{equation}
The optical depth is determined by this density profile, and the radiation energy density $e_\mathrm{ej}$ is derived from adiabatic suppression of the initial shocked envelope profiles
\begin{equation}
\label{eq:e_ej}
    \frac{e_\mathrm{ej}(m,t)}{\frac{18}{7}\rho_0(m)v_\mathrm{ej}(m)^2}=\left(\frac{\rho_\mathrm{ej}(m,t)}{7\rho_0(m)}\right)^{4/3},
\end{equation}
where $\rho_0$ is given by eq. (\ref{eq:rho_env}). Plugging eq. (\ref{eq:rho_ej}) and (\ref{eq:e_ej}) into (\ref{eq:L_*}), we find
\begin{equation}
    \frac{L_\star}{L_\mathrm{sh}}\approx0.1\left(\frac{R_{\star,13}}{R_{14}}\right)\left(\frac{R_{14}^2v_9^{-1}}{M_{\star,10}}\right)^{0.13}\tilde{t}^{0.14},
\end{equation}
where $L_\mathrm{sh}$ includes the temporal decay discussed in \S~\ref{sec:ejecta_results}. $L_\star/L_\mathrm{sh}$ depends mainly on the ratio of the progenitor radius to the breakout radius. Even in the limiting case of equal ejecta and wind masses ($M_\star=0.14R_{14}^2v_9^{-1}M_\odot$), the ejecta luminosity remains lower than the shock luminosity up to the wind photosphere for $R_\star/R_\mathrm{bo}\lesssim0.5$.

Note that for cases where the breakout radius is very close to the stellar radius, further analysis is necessary to determine the radius (if it exists) of the RMS-CLS transition (\S~\ref{subsec:Wind+CLS}), as the assumption of wind energy domination may not be valid. Care must be taken to account for the correct amount of radiation diffusing from the envelope that may contribute to the upstream acceleration and may affect the shock structure. %We leave this to future work.

\section{Static Radiation Calculations}
\label{sec:static}

The validity of our analytic interpretation and derivation of the various spectral components' contribution to the radiation field (\S~\ref{subsec:piston_analytic}) is supported by the results of static numeric calculations. In these calculations, we start with a snapshot of the dynamic calculation at a chosen time (or shock radius) and allow the radiation to evolve while holding the plasma density and temperature spatial distributions constant in time. The goal is to calculate the steady-state radiation spectrum solution at late times (typically converges after a few dynamical times). Additionally, we examine the effects of disabling the contribution of different regions of the plasma to Bremsstrahlung and/or Compton interactions.

\begin{figure*}[ht]
    \centering
    \includegraphics[height=5.5cm]{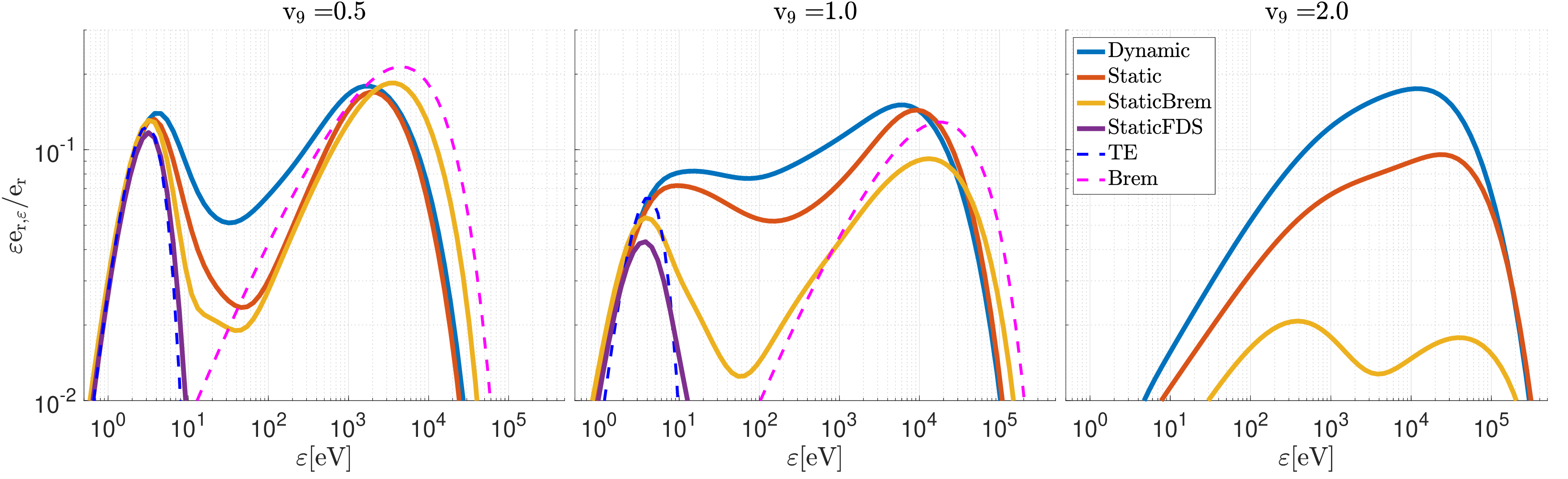}
    \caption{Comparison of radiation spectral energy density from dynamic and different static calculations, normalized to the bolometric energy density at the shock position, for $\tilde{r}_\mathrm{sh}=6$, $R_{14}=1$ and $v_9=0.5,1,2$ for constant velocity piston-driven shocks. Solid lines represent numeric calculations: dynamic calculation (blue), static calculation (i.e., No Hydrodynamics, orange), static calculation with only Bremsstrahlung (yellow), and static calculation with only Bremsstrahlung in the far downstream near the dense shell (purple). Dashed lines represent analytic approximations: the reprocessed component of thermal radiation (blue, eq. (\ref{eq:T_TE})) and Bremsstrahlung cooling (magenta, eq. (\ref{eq:brem})).}
    \label{fig:static}
\end{figure*}

In Figure \ref{fig:static}, we compare radiation spectra obtained in dynamic and different static calculations. Bremsstrahlung interactions dominate the spectrum for low and intermediate velocities, while Compton scatterings dominate at higher velocities. The shape of the Bremsstrahlung cooling spectrum and the thermal radiation originating from the dense shell far downstream are also consistent with our analysis (\S~\ref{subsec:piston_analytic}).

\begin{figure}[ht]
    \centering
    \includegraphics[height=6.5cm]{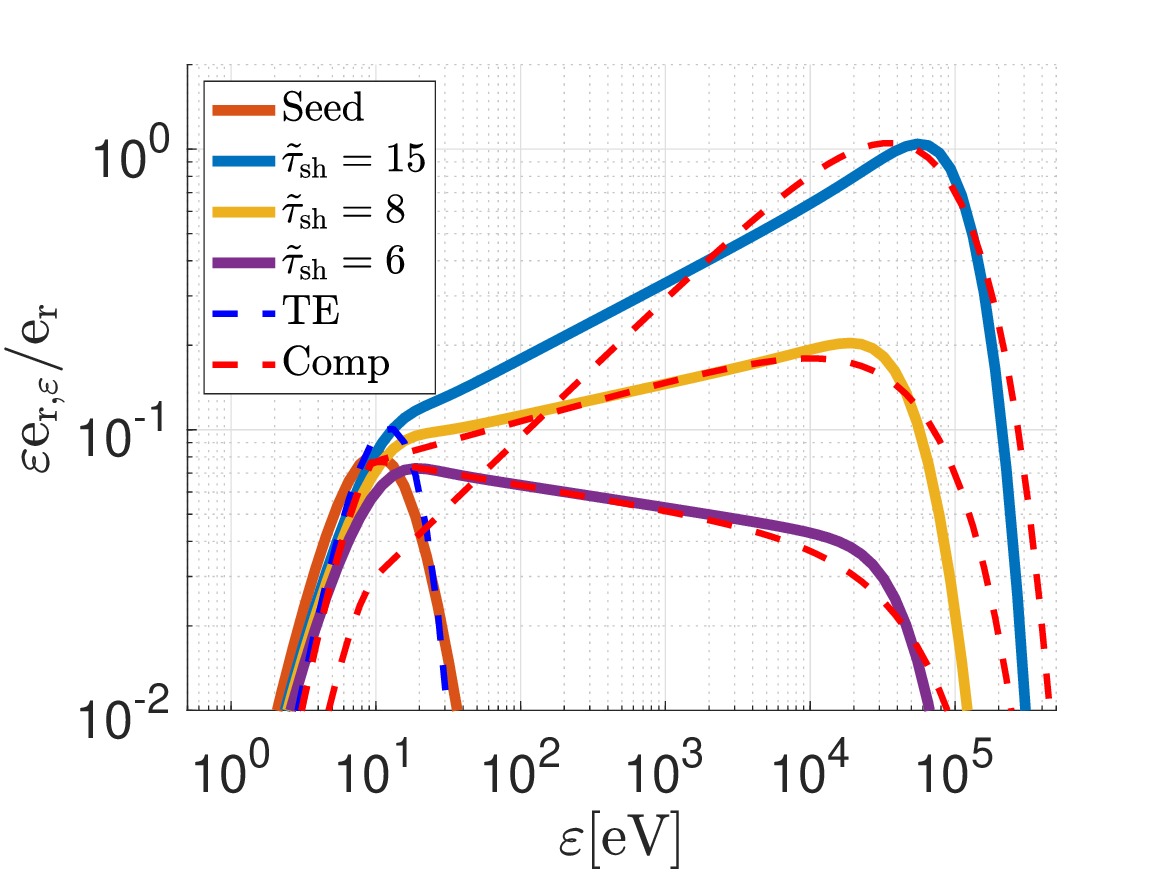}
    \caption{Comparison of radiation spectral energy density from static calculations, normalized to the bolometric energy density at the shock position, for $\tilde{r}_\mathrm{sh}=2$, $R_{14}=1$ and $v_9=1$ for a constant velocity piston-driven shock. Solid lines represent static numeric calculations with varying optical depths (manually adjusted) for Compton scattering of the thermal radiation seed emitted far downstream near the dense shell. Dashed lines represent analytic approximations: unsaturated Comptonization
    (red dashed, eq.(\ref{eq:unsaturated})) and thermal radiation (blue, eq. (\ref{eq:T_TE})).}
    \label{fig:compstat}
\end{figure}

In Figure \ref{fig:compstat}, we compare radiation spectra from static calculations where parts of the outer material are manually ``cut out" to reduce the optical depth and Compton $y$-parameter. The static numeric results match the analytic results obtained for different $y$-parameter values.

\bibliography{refs}

\end{document}